\documentclass[12pt]{article}
\pdfoutput=1

\usepackage{jheppub}
\usepackage{amsmath,amssymb,euscript,array,mathrsfs}

\def\a{\alpha}
\def\b{\beta}
\def\c{\gamma}
\def\d{\delta}
\def\e{\epsilon}

\def\k{\kappa}
\def\l{\lambda}
\def\m{\mu}
\def\n{\nu}

\def\r{\rho}
\def\s{\sigma}

\def\u{\upsilon}

\def\L{\Lambda}

\def\hh{{\mathfrak h}}

\def\tr{{\rm tr}}

\def\Dbarslash{\,\,{\raise.15ex\hbox{/}\mkern-12mu {\bar D}}}
\def\Dslash{\,\,{\raise.15ex\hbox{/}\mkern-12mu D}}
\def\delslash{\,\,{\raise.15ex\hbox{/}\mkern-9mu \partial}}
\def\delbarslash{\,\,{\raise.15ex\hbox{/}\mkern-9mu {\bar\partial}}}

\def\half{\frac{1}{2}}

\def\rta{\rightarrow}
\def\tr{{\rm tr}}

\title{\begin{center}Memory, Penrose Limits and the Geometry of
    Gravitational Shockwaves and Gyratons  \\
\end{center}}
\author{Graham M. Shore}

\affiliation{Department of Physics,\\ 
College of Science,\\
Swansea University,\\
Swansea,\\ 
SA2 8PP, UK.}

\emailAdd{g.m.shore@swansea.ac.uk}

\abstract{ The geometric description of gravitational memory for strong gravitational waves
is developed, with particular focus on shockwaves and their spinning analogues, gyratons.
Memory, which may be of position or velocity-encoded type, characterises the residual separation
of neighbouring `detector' geodesics following the passage of a gravitational wave burst,
and retains information on the nature of the wave source. 
Here, it is shown how memory is encoded in the Penrose limit of the original gravitational wave spacetime
and a new `timelike Penrose limit' is introduced to complement the original plane wave limit
appropriate to null congruences. A detailed analysis of memory is presented for timelike and null
geodesic congruences in impulsive and extended gravitational shockwaves of Aichelburg-Sexl type,
and for gyratons. 
Potential applications to gravitational wave astronomy and to quantum gravity, 
especially infra-red structure and ultra-high energy scattering, are briefly mentioned. }

\begin{document}

\maketitle

\setlength{\parskip}{10pt}


\textheight=600pt

\section{Introduction}\label{sect 1}

Gravitational memory is becoming an increasingly important topic in gravitational wave physics, not only 
because of its potential observation in gravitational waves from astronomical sources, but also for its importance
in theoretical issues in quantum gravity, including notably soft-graviton theorems, quantum loop effects and 
Planck energy scattering.

In this paper, we develop a geometric formalism for the description of gravitational memory which goes beyond 
the conventional weak-field analysis and is applicable to strong gravitational waves, especially gravitational
shockwaves and their spinning generalisations, gyratons. 

Gravitational memory refers to the residual separation of `detectors' following the passage of a gravitational 
wave burst. This may take the form of a fixed change in position, or a constant separation velocity, or both.
We refer to these as `position-encoded' \cite{Zeldovich,Braginsky:1986ia} 
and `velocity-encoded' \cite{Bondi:1957dt, Grishchuk:1989qa} memory respectively. 
From a geometric point of view, such idealised detectors are represented as neighbouring geodesics in
a timelike congruence. The description of memory is therefore part of the more general geometric 
analysis of geodesic deviation. To be precise, the separation of nearby detectors is identified as the 
connecting vector $\Delta z^i$ of neighbouring geodesics, which for a null congruence is given in
suitable Fermi normal coordinates as
\begin{equation}
\Delta z^i = \int_{-\infty}^{u} du\,\hat{\Omega}^i{}_j(u)\, z^j \ ,
\label{a1}
\end{equation}
where $\hat{\Omega}_{ij} = \half \hat{\theta} + \hat{\s}_{ij} + \hat{\omega}_{ij}$ defines the expansion, 
shear and twist optical tensors which characterise the congruence.  A similar expression holds for 
timelike congruences with the lightlike coordinate $u$ replaced by time $t$.  Memory resides in the value of
$\Delta z^i$, and $\Delta \dot{z}^i$, in the future region following the interaction with the gravitational 
wave burst, and is determined by integration of the optical tensors through the interaction region.
Here, we develop the theory of geodesic deviation and memory for both null and timelike geodesic
congruences in strong gravitational waves.

Central to this analysis is the observation that the geometry of geodesic deviation around a chosen null geodesic 
$\gamma$ in a given background spacetime is encoded in its Penrose limit 
\cite{Penrose:1965rx, Penrose, Blau:2006ar, Hollowood:2009qz}. 
This limit is a plane wave \cite{Stephani:2003tm, Griffiths:2009dfa},
so the description of memory for null observers in a general spacetime can be reduced to that in
an equivalent gravitational plane wave. For timelike observers, we define here a new `timelike Penrose
limit' with the same property. Moreover, we show that if the original spacetime is itself in the general
class of pp waves, the transverse geodesic equations defining memory are in fact the same for both
timelike and null congruences. 

We set out this general theory in section \ref{sect 2}, defining the null and timelike Penrose limits and 
relating our approach to the conventional analysis of weak gravitational waves considered so far
in astrophysical applications \cite{Gibbons:1972fy, Favata:2010zu, Lasky:2016knh, McNeill:2017uvq, Talbot:2018sgr}. 
The important, and very general, r\^ole of gravitational plane waves in encoding memory is discussed in 
some detail.  Closely related work on geodesics and gravitational plane waves, including memory effects,
may be found in \cite{Hollowood:2009qz, Harte:2012jg, Harte:2015ila, 
Hollowood:2015elj, Duval:2017els, Zhang:2017rno, Zhang:2017geq,
Shore:2017dqx, Zhang:2017jma, Zhang:2018srn, Zhang:2018gzn, Zhang:2018upz}) .

Motivated primarily by issues in quantum gravity, our focus in this paper then turns, in section \ref{sect 3}, 
to gravitational shockwaves.
These are described by generalised Aichelburg-Sexl metrics of the form \cite{Aichelburg:1970dh, Dray:1984ha},
\begin{equation}
ds^2 = 2 \,du\, dv + f(r) \chi_F(u)\, du^2 + dr^2 + r^2 d\phi^2    \ ,
\label{a2}
\end{equation}
where the potential $f(r)$ is fixed by the Einstein equations through the relation 
$R_{uu}= 8\pi G T_{uu} =  -\half \Delta f(r) \chi_F(u)$. (Here, $\Delta$ denotes the two-dimensional
Laplacian.) With the profile function $\chi_F(u)$ chosen to be 
impulsive, $\chi_F(u) = \d(u)$, this is the original Aichelburg-Sexl metric describing the spacetime around
an infinitely-boosted source localised on the surface $u=0$. 

For a particle source, $f(r) = -4 G E \log\left(r/r_0\right)^2$, and metrics of this type are important
in analyses of Planck energy scattering. At such ultra-high energies, scattering is dominated
by gravitational interactions and the leading eikonal behoviour of the scattering amplitude, generated 
by ladder diagrams representing multi-graviton exchange, can be reproduced by identifying
the corresponding phase shift with the discontinuous lightcone coordinate jump $\Delta v$ of 
test geodesics as they interact with the shockwave \cite{tHooft:1987vrq, Muzinich:1987in, 
Amati:1987wq}. While this simply requires the solution 
for a single null geodesic in the Aichelburg-Sexl background, ultra-high energy scattering
in an interacting quantum field theory including loop contributions depends on the geometry
of the full congruence. These QFT effects, the geometry of the relevant Penrose limits, and their
importance in resolving fundamental issues with causality and unitarity, have been studied extensively
in the series of papers \cite{Hollowood:2007ku, Hollowood:2008kq, Hollowood:2011yh, Hollowood:2015elj,
Hollowood:2016ryc}.

Several generalisations are also of interest, giving rise to different potentials $f(r)$ and distinguishing
between an extended profile $\chi_F(u)$ typical of a sandwich wave and its impulsive limit $\d(u)$. 
For example, an infinitely-boosted Schwarzschild black hole \cite{Aichelburg:1970dh} gives a shockwave 
metric with $f(r) \sim \log r$ and $\chi_F(u) \rta \d(u)$, while other black holes such as 
Reissner-Nordstr\"om \cite{Lousto:1988ej}, 
Kerr \cite{Ferrari:1990, Balasin:1994tb, Balasin:1995tj, Hayashi:1994rf}, 
Kerr-Newman \cite{Lousto:1989ha, Lousto:1992th, Yoshino:2004ft} and dilatonic \cite{Cai:1998ii}
also give impulsive shockwaves with modified potentials of the form $f(r) \sim \log r + 1/r + O(1/r^2)$.
It would be interesting if such shockwaves from extremely fast-moving black holes have an
important r\^ole in astrophysics.
Also note that in certain higher-dimensional theories of gravity, the Planck scale can be lowered to 
TeV scales, in which case the formation of trapped surfaces \cite{Eardley:2002re, Yoshino:2005hi,
Yoshino:2006dp, Yoshino:2007ph} in the scattering of 
such shockwaves becomes a model for black hole production at the LHC or FCC. 

A natural extension of the Aichelburg-Sexl shockwave metric is to `gyratons' \cite{Frolov:2005zq, 
Frolov:2005in, Podolsky:2014lpa}. These are a special
class of gravitational pp waves with metric,
\begin{equation}
ds^2 = 2\,du\,dv + f(r) \chi_F(u)\,du^2 - 2 J \chi_J(u)\,du\,d\phi + dr^2 + r^2 d\phi^2 \ .
\label{a3}
\end{equation}
These describe the spacetime generated by a pulse of null matter carrying an angular momentum,
related to $J$. They are the simplest models in which to study the gravitational effect of spin
in ultra-high energy scattering.\footnote{Note that this is not achieved by, for example, infinitely boosting
a black hole with spin (the Kerr metric), since as mentioned above this simply modifies $f(r)$ in the 
Aichelburg-Sexl metric while retaining the impulsive profile $\d(u)$.}
In this case, however, it is necessary to choose the spin profile $\chi_J(u)$ to be extended. 
This is because the curvature component $R_{ru\phi u}$ from (\ref{a3}) involves $\chi'_J(u)$, so an 
impulsive profile $\chi_J(u) \sim \d(u)$ would give an unphysically singular curvature. 
This also allows the spin in the metric time to act on the scattering geodesic (detector) imparting 
an angular momentum. In section \ref{sect 4}, we study these orbiting geodesics and the associated null
and timelike congruences in detail, determining the optical tensors, the relevant Penrose limits, 
and the eventual gravitational memory. A particular question is whether the gyraton spin gives
rise to a `twist memeory' in which the final $\Delta z^i$ would be determined by a non-vanishing
twist $\hat{\omega}_{ij}$ in the optical tensors characterising the congruence.

We include four appendices. In Appendix A, we review the relation of the scattering amplitude 
${\cal A}(s,t)$ for Planck energy scattering to the lightlike coordinate shift $\Delta v$ for a null
geodesic in an Aichelburg-Sexl spacetime, illustrating the origin of the poles at complex integer 
values of the CM energy $s$, and calculate the leading corrections arising from an extended 
profile $\chi_F(u)$.  In Appendix B, we describe the symmetries associated with the
shockwave and corresponding plane wave metrics, in particular considering potential enhanced 
symmetries for impulsive profiles. In Appendix C, we consider more general gyraton metrics
showing especially how the curvature constrains the the form of the profiles $\chi_F(u)$ and 
$\chi_J(u)$ and motivating the particular choice of metric (\ref{a3}) considered here.
Finally, the related phenomenon of gravitational spin memory \cite{Pasterski:2015tva} 
is described for gyratons in Appendix D.

\section{Memory, Optical Tensors and Penrose limits}\label{sect 2}

Gravitational memory concerns the separation of neighbouring geodesics following the passage 
of a gravitational wave burst, either an extended (sandwich) wave or, in the impulsive limit, a shockwave.
The appropriate mathematical description of memory is therefore the geometry of geodesic congruences,
in particular geodesic deviation characterised by the optical tensors in the Raychoudhuri equations.

In this section, we describe in quite general terms the geometry of geodesic congruences for the class
of gravitational waves of interest. We focus particularly on two examples of pp waves -- the Aichelburg-Sexl
shockwave and its non-impulsive extension, and gyratons. We consider both timelike geodesics, relevant
for the interpretation in terms of detectors for astrophysical gravitational waves, and null geodesics, 
which will also be appropriate for more foundational questions involving shockwaves and Planck energy 
scattering. We also discuss the difference in the origin of position-encoded memory, in which neighbouring
geodesics acquire a fixed separation after the gravitational wave has passed, and velocity-encoded memory,
in which they separate or focus with fixed velocity.

A key observation is that the geometry of geodesic deviation around a given null geodesic in a curved
spacetime background is encoded in the corresponding Penrose plane wave limit. This implies the
remarkable simplification that the properties of memory for a general background spacetime may be 
entirely described by studying congruences in an appropriate plane wave background. 
We also describe here a generalisation of the Penrose limit construction for the case of timelike geodesics.

\subsection{Geodesic deviation}\label{sect 2.1}

Consider a congruence centred on a chosen (null or timelike) geodesic $\c$ with tangent vector $k^\m$.
Let $z^\m$ be the `connecting vector' specifying the orthogonal separation to a neighbouring geodesic.
By definition, the Lie derivative of $z^\m$ along $\c$ vanishes, {\it i.e.}
\begin{equation}
{\cal L}_k z^\m = k.D z^\m - (D_\n k^\m) z^\m = 0 \ ,
\label{b1}
\end{equation}
where $D_\m$ is the covariant derivative. It follows that
\begin{equation}
k.D z^\m = \Omega^\m{}_\n z^\n \ ,
\label{b2}
\end{equation}
where we define the tensor $\Omega_{\m\n} = D_\n k_\m$ which will be fundamental to our analysis.
Differentiating (\ref{b2}), and using the geodesic equation $k.D k^\m = 0$, we find 
\begin{equation}
(k.D)^2 z^\m = - R^\m{}_{\r\n\s} k^\r k^\s z^\n \ ,
\label{b3}
\end{equation}
which is the Jacobi equation for geodesic deviation. In more familiar form, if the geodesic is affine parametrised
as $x^\m(\l)$ and the tangent vector is given by $k^\m = dx^\m/d\l$, this is written in terms of the intrinsic
derivative along $\c$ as
\begin{equation}
\frac{D^2 z^\m}{D\l^2} = - R^\m{}_{\r\n\s} \,\dot{x}^\r\, \dot{x}^\s \,z^\m \ ,
\label{b3a}
\end{equation}
where the dot denotes a derivative w.r.t.~$\l$.
The consistency of (\ref{b2}), (\ref{b3}) is ensured by the identity,
\begin{equation}
k.D \Omega^\m{}_\n + \Omega^\m{}_\l \Omega^\l{}_\n = - R^\m{}_{\r\n\s} k^\r k^\s \ ,
\label{b4}
\end{equation}
which holds in general given only that $k^\m$ satisfies the geodesic equation. 
This is in essence the Raychoudhuri equation.

The next step is to establish a frame adapted to the chosen congruence. That is, we choose a pseudo-orthonormal frame 
${\bf e}^A$ which is parallel-propagated along $\c$. This will define Fermi normal coordinates
(FNCs) in the neighbourhood of $\c$. For lightlike $\c$, we choose a frame such that the metric in the
neighbourhood of $\c$ is\footnote{At a point, this is identified
with a Newman-Penrose (null) basis $(\ell^\m,n^\m,m^\m, \bar{m}^\m)$ through 
\begin{equation*}
\ell^\m = k^\m = e^{u\m} \ , ~~~~ n^\m = - e^{v\m} \ , ~~~~ m^\m = \tfrac{1}{\sqrt{2}}(e^{1\m} \pm i e^{2\m}) \ ,
\end{equation*}
with the usual contractions $\ell.n = -1$, $m.\bar{m} = 1$, $\ell^2 = n^2 = m^2 = \bar{m}^2 = 0$.
The FNC basis is just this $NP$ basis parallel-propagated along $\c$, {\it i.e.}~we impose $k.D e^{A\m} = 0$ for all $A = u,v,1,2$.
In our previous work on Penrose limits \cite{Hollowood:2009qz, Shore:2017dqx}, we used this NP notation extensively.}
\begin{align}
g_{\m\n}\big|_\c  &= \eta_{AB}\, e^A{}_\m  \,  e^B{}_\n \ ,       ~~~~~~~~~~~~~~~~
\eta_{AB} = \begin{pmatrix} 0&~1&~0&~0\\1&~0&~0&~0\\0&~0&~1&~0\\0&~0&~0&~1\end{pmatrix}   \nonumber \\
&=  e^u{}_\m\,e^v{}_\n + e^v{}_\m\,e^u{}_\n + \d_{ij}\,e^i{}_\m\,e^j{}_\n\ , ~~~~~~~~i,j = 1,2 
\label{b5}
\end{align}
with $e^{u\m} = k^\m$ chosen to be tangent to the geodesic $\c$, and where the basis vectors satisfy 
$k.D e^A{}_\m = 0$.  This defines null FNCs $(u,v,x^i)$.

For timelike $\c$, we choose 
\begin{align}
g_{\m\n}\big|_\c &= \eta_{AB}\, e^A{}_\m\,e^B{}_\n \ ,
          ~~~~~~~~~~~~~~~~~ 
\eta_{AB} = {\rm diag}~(-1,1,1,1) 
\nonumber \\
&= - e^0{}_\m\,e^0{}_\n  + \d_{rs} \, e^r{}_\m\,e^s{}_\n \ , ~~~~~~~~r,s = 1,2,3 
\label{b6}
\end{align}
with $e^{0\m} = k^\m$ and $k.D e^A{}_\m = 0$, defining timelike FNCs $(t, x^r)$.

In terms of these coordinates, where by the definition of FNCs the Christoffel symbols vanish locally along $\c$, 
the Jacobi equations (\ref{b3a}) become simply
\begin{equation}
\ddot{z}^i  = - R^i{}_{u j u}  \,\dot{u} \,\dot{u} \,z^j \ , ~~~~~~~~~~~~
\ddot{z}^r = - R^r{}_{0 s 0} \,\dot{t} \,\dot{t}  \,z^s \ ,
\label{b7}
\end{equation}
for null, timelike congruences respectively.

\subsection{Optical tensors}\label{sect 2.2}

Geodesic deviation, and therefore gravitational memory, is described in terms of the optical tensors -- expansion, 
shear and twist -- characterising the congruence. For a null congruence, the transverse space spanned 
by the connecting vector is two-dimensional. Taking a cross-section through the congruence, those geodesics at 
fixed separation from $\c$ form a ``Tissot ring'' \cite{Zhang:2017geq} -- initially a circle, this distorts as the gravitational
wave burst passes displaying clearly the effects of expansion, shear and twist. For a timelike congruence,
the transverse space and optical tensors are in general three-dimensional although, as we shall see, the 
special symmetry characterising pp waves means that this space remains effectively two-dimensional
and the optical tensors are identical to the null case.

The optical tensors are defined from the projections of $\Omega_{\m\n}$ onto the appropriate transverse subspace
(see {\it e.g.}~\cite{Poisson}). For a null congruence, we have the projection matrix
\begin{equation}
\hat{g}_{\m\n} \equiv g_{\m\n} - e^u{}_\m e^v{}_\n - e^v{}_\m e^u{}_\n ~=~ \d_{ij}\,e^i{}_\m e^j{}_\n  \ ,
\label{b8}
\end{equation}
and define
\begin{equation}
\hat{\Omega}_{\m\n} = \left( \hat{g} \,\Omega \,\hat{g} \right)_{\m\n}  \ .
\label{b9}
\end{equation}
It is readily checked that $\hat{\Omega}_{\m\n} e^{u\n} = 0$ and $\hat{\Omega}_{\m\n} e^{v\n} = 0$, 
so $\hat{\Omega}$ is effectively two-dimensional. We define the optical tensors from the decomposition
of
\begin{equation}
\hat{\Omega}^{ij} \equiv  e^i{}_\m \hat{\Omega}^{\m\n} e^j{}_\n  ~=~  e^i{}_\m \Omega^{\m\n} e^j{}_\n
\label{b10}
\end{equation}
as
\begin{equation}
\hat{\Omega}_{ij} = \tfrac{1}{2} \hat{\theta} \,\d_{ij}  + \hat{\s}_{ij} + \hat{\omega}_{ij} \ .
\label{b11}
\end{equation}
Here, the shear $\hat{\s}_{ij}$ is symmetric and traceless, the twist $\hat{\omega}_{ij}$ is antisymmetric,
while the expansion $\hat{\theta} = {\rm tr}\, \hat{\Omega}$. 
The relation (\ref{b4}) is then seen to be equivalent to the Raychoudhuri equation for the optical tensors,
since for null FNCs it is simply,
\begin{equation}
\frac{d}{du} \hat{\Omega}_{ij} = - (\hat{\Omega}^2)_{ij} - R_{iuju} \ .
\label{b12}
\end{equation}

\begin{figure}[h]
\centering
\includegraphics[scale=0.4]{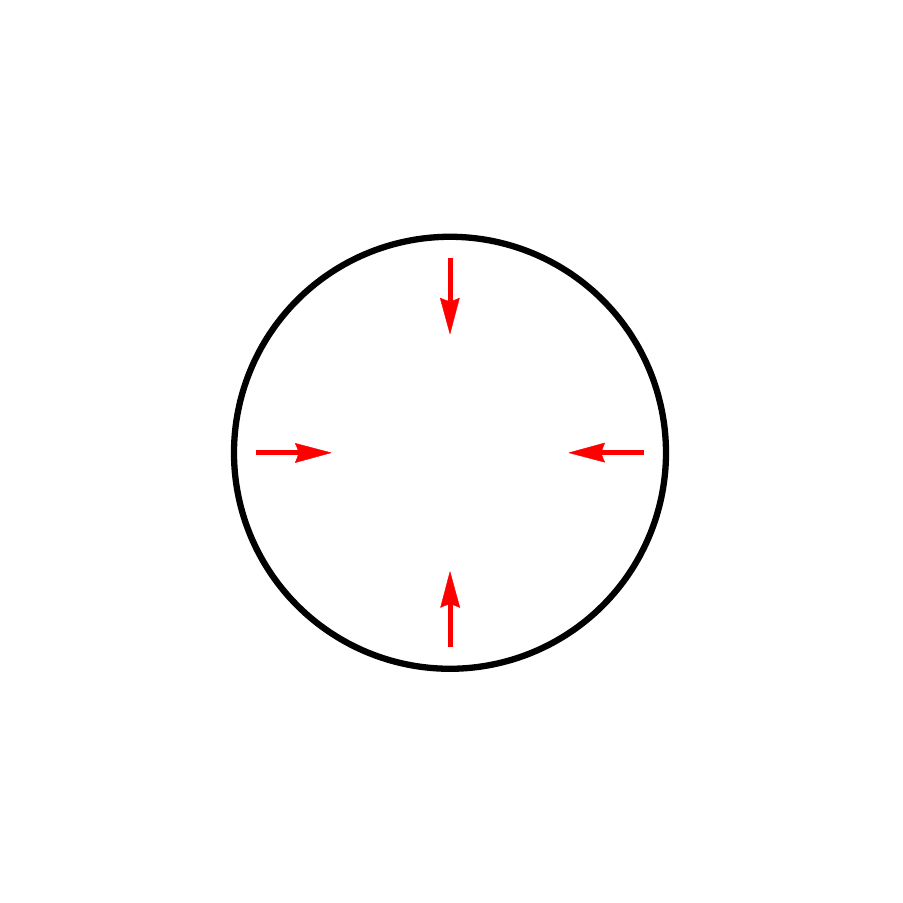} \hskip0cm
\includegraphics[scale=0.4]{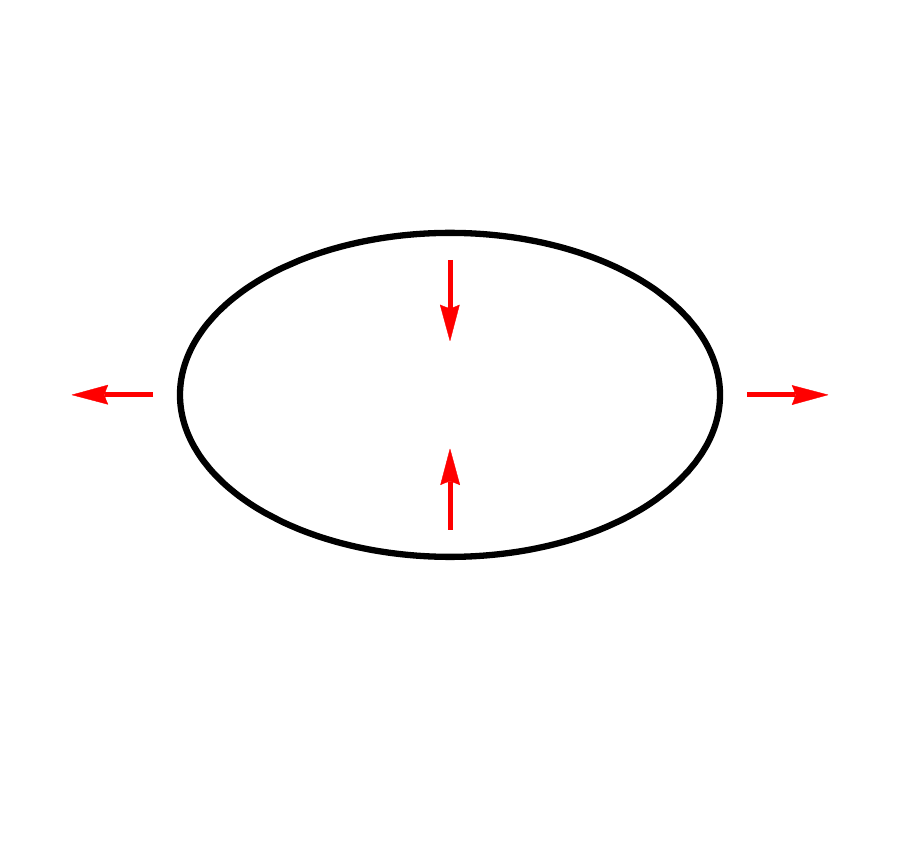} \hskip0.3cm
\includegraphics[scale=0.4]{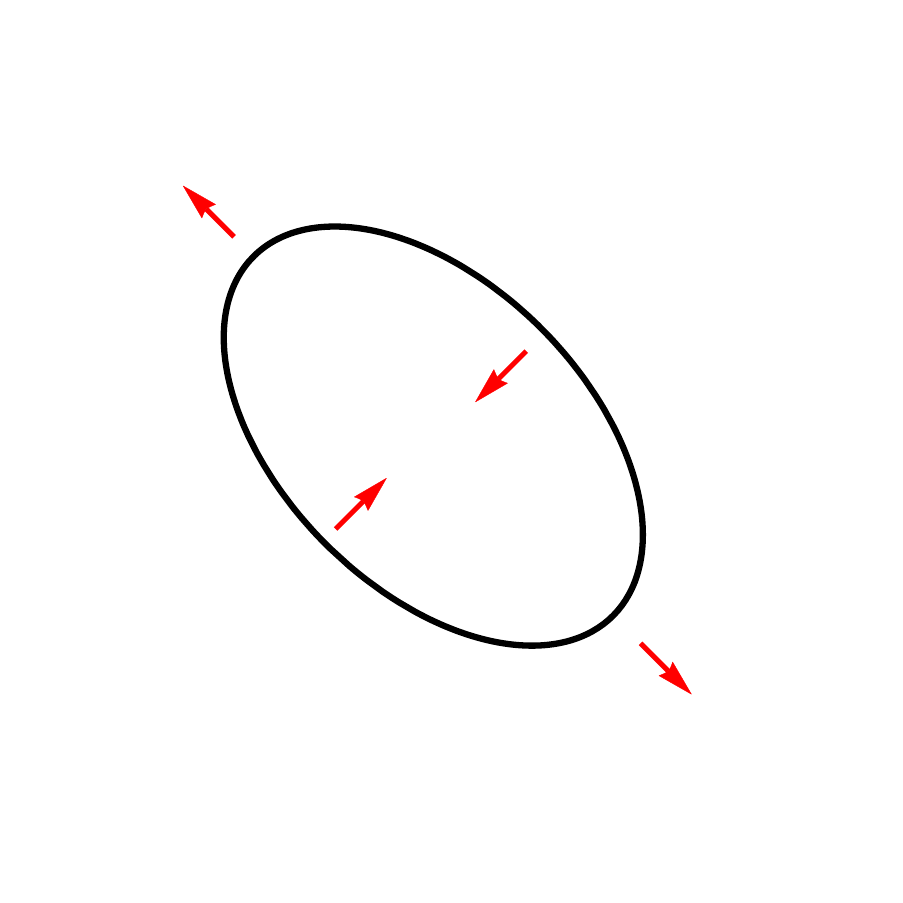} \hskip0cm
\includegraphics[scale=0.4]{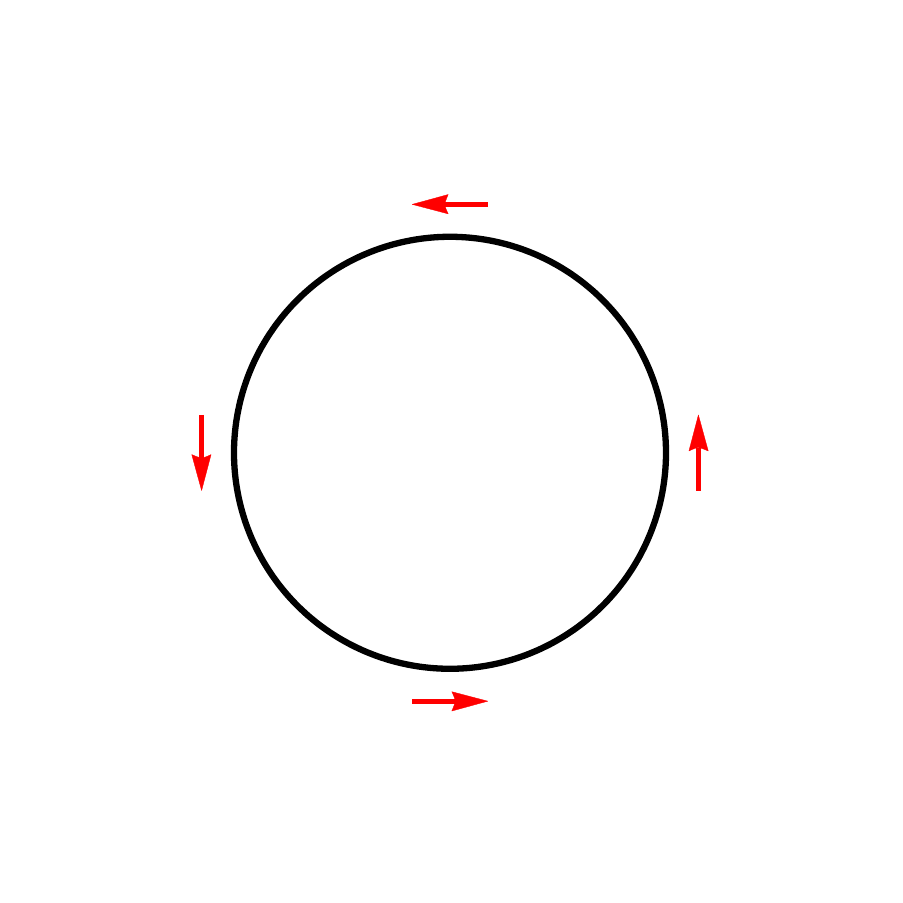} \hskip0cm
\caption{Illustration of the effect of the optical tensors on the Tissot circle. From left to right, the figures
show the expansion $\hat{\theta}$, $+$ oriented shear $\hat{\s}_+$,  $\times$ oriented shear 
$\hat{\s}_{\times}$, and the twist $\hat{\omega}$. }
\label{Tissot}
\end{figure}

Since the transverse space is two-dimensional, we can further simplify this description by writing 
$\hat{\Omega}_{ij}$ as
\begin{equation}
\hat{\Omega}_{ij} = \begin{pmatrix} \tfrac{1}{2} \hat{\theta} + \hat{\s}_+  &~~ \hat{\s}_\times + \hat{\omega} \\
\hat{\s}_\times - \hat{\omega} &~~ \tfrac{1}{2} \hat{\theta} - \hat{\s}_+ \end{pmatrix} \ ,
\label{b13}
\end{equation}
defining the optical scalars $\hat{\theta}$ (expansion), $\hat{\s}_+$ and $\hat{\s}_\times$ (shear with 
$+$ and $\times$ oriented axes), and $\hat{\omega}$ (twist).
Their action on the Tissot ring is indicated schematically in Fig.~\ref{Tissot}.

For a timelike congruence, the analogous projection matrix is
\begin{equation}
\hat{g}_{\m\n} \equiv  g_{\m\n} + \hat{k}_\m \hat{k}_\n ~=~ \d_{rs} \,e^r{}_\m \, e^s{}_\n\ ,
\label{b14}
\end{equation}
where we normalise $\hat{k}^2 = -1$.  In this case,
\begin{equation}
\hat{\Omega}_{\m\n} = (\hat{g} \, \Omega \, \hat{g})_{\m\n} \ ,
\label{b15}
\end{equation}
defines three-dimensional optical tensors through 
\begin{equation}
\hat{\Omega}^{rs} \equiv e^r{}_\m \hat{\Omega}^{\m\n} e^s{}_\n = e^r{}_\m \Omega^{\m\n} e^s{}_\n \ ,
\label{b15a}
\end{equation}
 as
\begin{equation}
\hat{\Omega}_{rs} = \tfrac{1}{3} \hat{\theta}\, \d_{rs} + \hat{\s}_{rs} + \hat{\omega}_{rs} \ .
\label{b16}
\end{equation}
The timelike Raychoudhuri equations follow straightforwardly. Again, however, note that with the
defining symmetry of the pp waves considered here, we will find only the components $\hat{\Omega}_{rs}$
with $r,s = 1,2$ are non-vanishing, so the optical tensors remain effectively two-dimensional and can also be 
visualised with a Tissot ring.

\subsection{Penrose limits for null and timelike congruences}\label{sect 2.3}

For null congruences, the geometry of geodesic deviation is encoded in the 
Penrose limit of the background geometry with the chosen geodesic $\c$.
An elegant construction of the Penrose limit in terms of Fermi null coordinates is given in \cite{Blau:2006ar}.

In the neighbourhood of a null geodesic $\c$, and choosing null FNCs according to the construction described
above, we can expand the metric as follows \cite{Blau:2006ar, Poisson}
\begin{align}
ds^2~=~ &2 du dv + \d_{ij}dx^i dx^j \\
&- \left( R_{\a u \b u} \big|_\c x^\a x^\b du^2 + \tfrac{4}{3} R_{\a\c \b u}\big|_\c x^\a x^\b du dx^\c + 
\tfrac{1}{3} R_{\a\c\b\d}\big|_\c x^\a x^\b dx^\c dx^\d \right)  + O(x^3) \ ,
\label{b17}
\end{align}
where $x^\a \equiv (v, x^i)$ here.  Note that the curvatures are evaluated on the geodesic $\c$ and are therefore
functions of $u$ only.

The conventional (null) Penrose limit follows from the rescaling
$u\rta u$, $v\rta \k^{-2} v$, $x^i \rta \k^{-1} x^i$, \cite{Penrose:1965rx, Penrose}. 
Keeping only those terms in $ds^2$ which scale  as $\k^{-2}$,
{\it i.e.}~neglecting $O(\k^{-3}, \k^{-4})$, leaves the following {\it truncation} of (\ref{b17}):
\begin{equation}
ds_P^2 = 2 du dv + \d_{ij} dx^i dx^j - R_{iuju}\big|_\c x^i x^j du^2 \ .
\label{b18}
\end{equation}
We immediately see that this truncation leaves only the curvature components $R_{iuju}$, precisely
those that determine geodesic deviation through the Jacobi equation (\ref{b7}).
The second key property of the Penrose limit metric (\ref{b18}) is that it describes a gravitational
{\it plane wave} expressed in Brinkmann coordinates, {\it i.e.}
\begin{equation}
ds_P^2 = 2 du dv + h_{ij}(u) x^i x^j du^2 + \d_{ij} dx^i dx^j \ ,
\label{b19}
\end{equation}
with the profile function $h_{ij}(u)$ identified in terms of the curvature tensor of the original spacetime
evaluated on $\c$ as $h_{ij}(u) = - R_{iuju}\big|_\c$. 

The geodesic equation for the transverse Brinkmann coordinates $x^i$ in the plane wave metric (\ref{b18})
is well known:
\begin{equation}
\frac{d^2 x^i}{d\l^2} - h^i{}_j(u) \left(\frac{du}{d\l}\right)^2 x^j = 0 \ .
\label{b20}
\end{equation}
This is identical to the geodesic deviation equation (\ref{b7}) around $\c$ in the original metric, where
we identify the connecting vector $z^i$ in FNCs with the Brinkmann $x^i$ in the plane wave.

This confirms the claim that the Penrose limit captures precisely the geometry of geodesic deviation.
The ability to analyse physical effects controlled by geodesic deviation (such as quantum loop corrections
in QFT in curved spacetime \cite{Hollowood:2007ku, Hollowood:2008kq, Hollowood:2011yh, Hollowood:2015elj,
Hollowood:2016ryc}) entirely in the simpler and well-studied case of plane waves 
has proved to be extremely powerful. Here, we demonstrate this in the context of gravitational memory.

Given this description of geodesic deviation for null congruences, it is now natural to repeat the
construction for timelike congruences, defining what we may call the ``timelike Penrose limit''.
Using the timelike FNCs defined above, we expand the original background metric in the neighbourhood of 
a chosen timelike geodesic $\c$ as \cite{Poisson}:
\begin{align}
ds^2 ~=~ &-dt^2  + \d_{rs}dx^r dx^s \\
&- \left( R_{r 0 s 0} \big|_\c x^rx^s dt^2 + \tfrac{2}{3} R_{rps0}\big|_\c x^r x^s dt dx^p + 
\tfrac{1}{3} R_{rpsq}\big|_\c x^r x^s dx^p dx^q \right)  + O(x^3) \ .
\label{b21}
\end{align}
Without invoking a scaling argument as in the original Penrose limit derivation, we may simply make 
an analogous truncation of (\ref{b21}) keeping only the curvature terms which enter the Jacobi equation.
This leaves
\begin{equation}
ds_P^2 = -dt^2 - R_{r0s0}\big|_\c x^r x^s dt^2 + \d_{rs} dx^r dx^s \ .
\label{b22}
\end{equation}
In general, therefore, we define the timelike Penrose limit as a metric of the form
\begin{equation}
 ds_P^2 = -\bigl( 1 - h_{rs}(t) x^r x^s \bigr) dt^2 + \d_{rs} dx^r dx^s \ ,
\label{b23}
\end{equation}
with $h_{rs}(t) = - R_{r0s0}\big|_\c$. 

The geodesic equation for the coordinates $x^r$ derived from  the metric (\ref{b23}) is
\begin{equation}
\frac{d^2 x^r}{d\l^2} - h^r{}_s (t) \left(\frac{dt}{d\l}\right)^2 x^s = 0 \ ,
\label{b24}
\end{equation}
which is identical to the timelike geodesic deviation equation (\ref{b7}) for the 
connecting vector $z^r$ in the original spacetime.  This confirms that the timelike Penrose limit metric (\ref{b23})
fully captures the geometry of geodesic deviation. Moreover, for the pp waves of interest here,
we find that only the two-dimensional transverse components $h_{ij}(t)$ with $i,j = 1,2$ are non-zero,
since for these backgrounds we have $R_{3 0 3 0} = R_{3 0 i 0} = 0$.\footnote{To complete the demonstration 
that the two-dimensional optical tensors are the same for the null and timelike cases, we need the further observation
that whereas in the null case (\ref{b19}) we have $\ddot{u} = 0$ and can simply take $u = \l$ as the affine parameter
(so $\dot{u} = 1$ in (\ref{b7})),
for the geodesics in the metric (\ref{b23}) we only have $\ddot{t} = O(x^r)^2$ and can at best parametrise
such that $\dot{t} = 1 + O(x^r)^2$ in (\ref{b7}). However, this is sufficient to establish the equivalence
of the optical tensors defined in the neighbourhood of $\c$ and given by $\Omega_{ij}$.\label{timelike}}

Of course, introducing the Penrose limit metric (\ref{b19}) does not in principle give any information that is not
already present in the original derivation of the optical tensors from $\Omega_{ij}$. However, it does allow us
to exploit the whole body of knowledge on the geometry of gravitational plane waves,
and to expose a large measure of universality in phenomena controlled by geodesic deviation.
In particular, the enhanced symmetries of plane waves (expressed as an extended Heisenberg algebra
\cite{Blau:2006ar, Shore:2017dqx} or Carroll symmetry \cite{Duval:2017els}), and their classification, brings considerable
insight into the nature of the geodesic solutions and congruences and, by extension, into the form 
of gravitational memory.  The symmetries of shockwaves and their plane wave Penrose limits are 
described in Appendix \ref{Appendix B}.
The same benefits should also arise for the timelike Penrose limit (\ref{b23}) although, to our knowledge,
metrics of this form have not been so widely studied in the general relativity literature.

\subsection{Gravitational plane waves}\label{sect 2.4}

The discussion above shows that memory for null observers in a general curved spacetime background 
can be reduced to the simpler case of the Penrose limit plane wave.  The geometry of geodesic congruences
in plane waves is well understood and we present here only a brief summary of some key results.
Of course, gravitational plane waves are an important physical example in their own right.

The full set of geodesic equations for the metric {\ref{b19}) are
\begin{align}
&\ddot{u} = 0 \ ,  \nonumber \\
&\ddot{v} + \frac{1}{2} h_{ij} x^i x^j \,\dot{u}^2 + 2 h_{ij} x^j \,\dot{x}^i \,\dot{u} =0 \ ,  \nonumber \\
&\ddot{x}^i - h^i{}_j x^j \,\dot{u}^2 = 0  \ .
\label{b25}
\end{align}
This allows us to immediately take $u = \l$ as an affine parameter, simplifying (\ref{b25}).

The solutions are then written in terms of a zweibein $E^i{}_a(u)$, $a=1,2$, as
\begin{align}
v &= V + \eta u - \frac{1}{2} {\bf \Omega}_{ab}(u) X^a X^b \ , \nonumber \\
x^i &= E^i{}_a(u) X^a \ ,  
\label{b26}
\end{align}
where the integration constants $V, X^a$ label the geodesic and $\eta=0$ ($\eta <0$) for 
null (timelike) geodesics.
The zweibein satisfies the key `oscillator equation',
\begin{equation} 
\ddot{E}^i{}_a(u) - h^i{}_j(u) E^j{}_a(u) = 0 \ .
\label{b27}
\end{equation}
In (\ref{b26}), ${\bf \Omega}_{ab} = (E^{T} \,\Omega\, E)_{ab}$ with $\Omega_{ij}$ defined as
$\Omega_{ij} = (\dot{E} E^{-1})_{ij}$  (the dot now signifying $d/du$). It follows immediately that
\begin{equation}
\dot{x}^i = \dot{E} E^{-1} x = \Omega^i{}_j\, x^j
\label{b28}
\end{equation}
so we see that the definition of $\Omega_{ij}$ here precisely matches that given in (\ref{b2}).
We therefore use the same notation for economy.  Using (\ref{b27}), these $\Omega_{ij}$ are readily seen
to satisfy
\begin{equation}
\dot{\Omega}_{ij} + (\Omega^2)_{ij} = h_{ij}  \ , 
\label{b29}
\end{equation}
to be compared with (\ref{b4}).

With $\eta=0$, the expressions (\ref{b26}) give the change of variables
from Brinkmann coordinates $(u, v, x^i)$ to Rosen coordinates $(u, V, X^a)$ (referred to as ``BJR'' coordinates 
in \cite{Duval:2017els, Zhang:2017rno, Zhang:2017geq}), in terms of which the plane wave metric takes the form,
\begin{equation}
ds^2 = 2\, du \,dV +  {\bf C}_{ab}(u) \,dX^a\, dX^b \ .
\label{b29aa}
\end{equation}
The metric components ${\bf C}_{ab}= (E^T E)_{ab}$ are used to contract the transverse Rosen 
indices.

The nature of the congruences is determined by the particular solutions of (\ref{b27}) for the 
zweibein $E^i{}_a(u)$ for specified boundary conditions. This is discussed in detail in
\cite{Blau:2002js,Shore:2017dqx}, the former reference focusing on geodesics exhibiting twist.
It is convenient to consider the complete set of solutions $f^i_{(r)}$ and $g^i_{(r)}$ $(r=1,2)$ 
defined with canonical `parallel' and `spray' boundary conditions respectively, as given in 
(\ref{bb10}) and (\ref{bb11}) in Appendix \ref{Appendix B}. In general, the zweibein is a linear combination
of these $f^i_{(r)}$ and $g^i_{(r)}$ solutions. The choice of zweibein corresponding to an initially parallel 
congruence, as appropriate in the flat spacetime region before an encounter with a shockwave,
is therefore $E^i{}_a(u) = f^i_{(r)}(u) \d_{ra}$.

Now, it is shown in \cite{Shore:2017dqx} that the Wronskian associated with a particular choice of 
zweibein is 
\begin{align}
W_{ab} &= \left(E^T \dot{E} - \dot{E}^T E \right)_{ab} \nonumber \\
&= \left(E^T (\Omega - \Omega^T)E\right)_{ab} \nonumber \\
&= \left({\bf \Omega} - {\bf \Omega}^T\right)_{ab} \nonumber \\
&= 2 \hat{\boldsymbol \omega}_{ab} \ , 
\label{b29a}
\end{align}
where $\hat{\boldsymbol \omega}_{ab}= E^T \hat{\omega} E$ is the twist in Rosen coordinates.
It follows that for a congruence to exhibit twist, the Wronskian of the zweibein must 
not vanish. However, noting that the Wronskian is $u$-independent and can therefore be evaluated
at any value of $u$, it follows from (\ref{b29a}) and the boundary conditions $f^i_{(r)}(u_0) = \d^i_r$, 
$\dot{f}^i_{(r)}(u_0) = 0$, that $W_{ab}(u_0) = 0$, so an initially parallel congruence can never develop 
a non-vanishing twist. 

Indeed, this is already apparent from expanding (\ref{b29}) into the individual Raychoudhuri equations
for expansion, shear and twist, {\it viz}.
\begin{align}
\frac{d}{du}\, \hat{\theta} ~~&= -\tfrac{1}{2}\hat{\theta}^2 - \tr\, \hat{\s}_{ij}^2 
- \tr\, \hat{\omega}_{ij}^2 - R_{uu} \ ,\nonumber \\
\frac{d}{du\,} \hat{\sigma}_{ij} &= - \hat{\theta}\, \hat{\sigma}_{ij} - C_{iuju} \ , \nonumber \\
\frac{d}{du}\, \hat{\omega}_{ij} &= - \hat{\theta}\, \hat{\omega}_{ij} \ ,
\label{b29b}
\end{align}
where $R_{uu} = -\tr\, h_{ij}$ and the Weyl tensor is $C_{iuju} = - h_{ij} + \tfrac{1}{2} \tr\, h \,\d_{ij}$.
It follows that while a non-vanishing expansion and shear can be induced as the congruence encounters 
a region of non-vanishing curvature such as a shockwave, the twist remains zero by virtue of the
last equation of (\ref{b29b}).  Similar considerations apply to timelike congruences, following the discussion 
in section \ref{sect 2.3}.

The implications for gravitational memory are that since we start with detectors forming a twist-free
congruence in flat spacetime, and since their subsequent evolution is governed by the appropriate
Penrose limit spacetime, the congruence will remain twist-free during and after its encounter with
the impulsive gravitational wave. This rules out gravitational twist memory, showing that the evolution 
of the Tissot ring is always determined by expansion and shear alone.

As discussed in refs.~\cite{Hollowood:2007ku, Hollowood:2008kq, Hollowood:2011yh, Hollowood:2015elj,
Hollowood:2016ryc,Shore:2017dqx}, this description of geodesic congruences in the plane wave spacetime
can be developed in many ways, notably in calculating the Van Vleck-Morette matrix which enters the 
loop-corrected propagators needed for QFT applications. Here, we focus on the plane waves which 
arise as Penrose limits of various gravitational shockwave backgrounds and discuss in detail
how they determine gravitational memory.

\subsection{Weak gravitational waves}\label{sect 2.5}

A very natural class of gravitational waves from an observational point of view are of course the
weak gravitational waves, viewed as a small perturbation around flat spacetime.
Here, we briefly review geodesic deviation and memory for weak gravitational waves from the viewpoint 
of the general formalism developed in this section.

A weak gravitational wave is described by the metric,
\begin{align}
ds^2 ~&=~ 2du dV + \bigl(\d_{ab} + \hh_{ab}(u)\bigr) dX^a dX^b  \ , ~~~~~~~~~~~~a,b = 1,2  \nonumber \\
{}&{} \nonumber  \\
&=~2du dV + \begin{pmatrix}\,&dX^1~&dX^2\,\end{pmatrix}\,
\begin{pmatrix} \,&1 + \hh_+(u) ~~&\hh_\times(u) \, \\ \,&\hh_\times ~~&1 - \hh_+(u) \, \end{pmatrix}\,
\begin{pmatrix}\,&dX^1\, \\ \,&dX^2\,\end{pmatrix} \ ,
\label{b30}
\end{align}
that is, the perturbation $\hh_{ab}(u)$ is transverse and traceless. The `weak' condition means that we may work 
to $O(\hh)$ only.

We immediately recognise the metric (\ref{b30}) as a plane wave in Rosen coordinates (\ref{b29aa}),
with
\begin{equation}
{\bf C}_{ab} ~=~ \begin{pmatrix} \,&1 + \hh_+ ~~&\hh_\times \, \\
\,&\hh_\times ~~&1 - \hh_+ \, \end{pmatrix} \ .
\label{b31}
\end{equation}
Writing ${\bf C}_{ab} = \left(E^T E\right)_{ab}$ in terms of the zweibein $E^i{}_a$, we find 
\begin{equation}
E^i{}_a(u) ~=~ \begin{pmatrix} &1 + \half \hh_+ ~~&\half \hh_\times \, \\
&\half\hh_\times ~~&1 - \half\hh_+ \, \end{pmatrix} ~~~~ +~ O(\hh^2) \ .
\label{b32}
\end{equation}
This allows us to re-express the metric (\ref{b30}) in Brinkmann form. 
Defining $\Omega_{ij} = \left(\dot{E} E^{-1}\right)_{ij}$ and $h_{ij} =  \left(\ddot{E} E^{-1}\right)_{ij}$, we clearly
have, to $O(\hh^2)$, 
\begin{equation}
\Omega_{ij} ~=~ \half \begin{pmatrix} &\dot{\hh}_+ ~~&\dot{\hh}_\times \, \\
&\dot{\hh}_\times ~~&- \dot{\hh}_+ \, \end{pmatrix} \ , ~~~~~~~~~~~~
h_{ij}~=~ \half \begin{pmatrix} &\ddot{\hh}_+ ~~&\ddot{\hh}_\times \, \\
&\ddot{\hh}_\times ~~&- \ddot{\hh}_+ \, \end{pmatrix} \ .
\label{b33}
\end{equation}
The Brinkmann plane wave metric is therefore 
\begin{equation}
ds^2 ~=~ 2 du dv + \ddot{\hh}_{ij}(u)\, x^i x^j \,du^2 + \d_{ij} dx^i dx^j \ .
\label{b34}
\end{equation}

With the metric in this form, we can simply transcribe everything we have described for plane waves in general,
substituting the specific forms (\ref{b32}), (\ref{b33}) for $E^i{}_a$, $\Omega_{ij}$ and $h_{ij}$. 
In particular, we can read off the optical tensors from $\Omega_{ij}$. This gives (see (\ref{b13})),
\begin{equation}
\hat{\theta} = 0\, ~~~~~~~~~~\hat{\s}_+ = \half \dot{\hh}_+ \ , ~~~~~~~~~~
\hat{\s}_\times = \half \dot{\hh}_\times \ , ~~~~~~~~~~ \hat{\omega } = 0 \ ,
\label{b35}
\end{equation}
and we see immediately that for these weak gravitational waves, the geodesic congruences exhibit shear,
but no expansion. This is a direct consequence of the perturbation ${\hh}_{ij}$ having zero trace.

Gravitational memory is usually discussed in this context by integrating the Jacobi equation (\ref{b3a}) in the
weak-field limit. That is, starting from
\begin{equation}
\ddot{z}^i(u) = - R^i{}_{uju}(u)\, z^j(u) \ ,
\label{b36}
\end{equation}
where $z^i(u)$ is the transverse connecting vector,\footnote{For definiteness, we write the equations 
for a null congruence. The timelike case is essentially the same, with the curvature replaced by $R^i{}_{0j0}(t)$
corresponding to the time coordinate along the geodesic $\c$.} and recognising that $R^i{}_{uju}$ and $\dot{z}^i$
are of $O(\hh)$, we can set $z^i(u) = \langle z^j\rangle + O(\hh)$ in (\ref{b36}) , with $\langle z^i\rangle$ constant,
and integrate to give
\begin{equation}
\dot{z}^i(u) = - \int_{-\infty}^u du'\,R^i{}_{uju}(u')\,  \langle z^j\rangle ~~~+~ O(\hh^2) \ .
\label{b37}
\end{equation}
If we now consider the relevant case of an initially parallel congruence with $\dot{z}^i(-\infty) = 0$
and spacetime with $\dot{\hh}(-\infty) = 0$, we find
\begin{align}
\Delta z^i(u) &\equiv z^i(u) - z^i(-\infty)  \nonumber \\
&= - \int_{-\infty}^u du'\, \int_{-\infty}^{u'} du''\, R^i{}_{uju}(u'')\,  \langle z^j\rangle ~~~+~ O(\hh^2) \ .
\label{b38}
\end{align}
Writing $R^i{}_{uju} = -\half \ddot{\hh}_{ij}(u)$, these simplify to give
\begin{align}
\Delta \dot{z}^i(u) &=  \half \dot{\hh}_{ij}(u) \, \langle z^j\rangle \ ,  
\label{b39} \\
\Delta z^i(u) &= \half \bigl(\hh_{ij}(u) - \hh_{ij}(-\infty) \bigr)  \,\langle z^j\rangle \ .
\label{b40}
\end{align}
Considering this in the context of a gravitational wave burst confined to a finite region of $u$, 
say $u_i \le u \le u_f$, we see that in order to find a purely {\it position-encoded} memory, the
integral in (\ref{b37}) should vanish for $u> u_f$ while (\ref{b38}) is non-zero. 
From (\ref{b39}), (\ref{b40}) this requirement in terms of the metric amplitudes is that
$\dot{\hh}_{ij}(u_f) = 0$ but $\Delta\hh_{ij} \equiv \hh_{ij}(u_f) - \hh_{ij}(u_i) \neq 0$.
If $\dot{\hh}_{ij}(u_f) \neq 0$, we have in addition a {\it velocity-encoded} memory.

These moments of the curvature can be related to specific astrophysical sources of gravitational waves,
{\it e.g.}~flybys, core-collapse supernovae, black hole mergers, {\it etc}. 
These are discussed at length in the literature;
see {\it e.g.}~\cite{Gibbons:1972fy, Favata:2010zu, Lasky:2016knh, McNeill:2017uvq, Talbot:2018sgr}
for a selection.

\subsection{Gravitational memory}\label{sect 2.6}

We now generalise this conventional description of gravitational memory to the case of potentially strong
gravitational wave bursts, especially shockwaves. This finds a very natural realisation in the language
of Penrose limits developed here.

As in section \ref{sect 2.1}, we start with the most general form of the Jacobi equation for geodesic
deviation, and immediately adopt the description in terms of Fermi normal coordinates.
Again, we present results for a null congruence, the timelike case following in an exactly analogous way.
The connecting vector $z^i(u)$ from a reference null geodesic $\c$ therefore satisfies,
\begin{equation}
\frac{d^2 z^i}{du^2} ~=~ h^i{}_j(u)\,z^j  \ ,
\label{b41} 
\end{equation}
with $h_{ij} = - R_{iuju}\big|_\c$.  As we have seen in (\ref{b4}), it is {\it always} possible to express the
curvature in the form
\begin{equation}
h_{ij} ~=~ \frac{d}{du} \Omega_{ij} + \left(\Omega^2\right)_{ij}  \ ,
\label{b42}
\end{equation}
for some $\Omega_{ij}$.  This is satisfied by $\Omega_{ij} = D_j k_i$, where  $k^\m$ is the tangent vector
to the geodesic $\c$. Then (\ref{b41}) can be immediately integrated\footnote{Explicitly, we have 
the self-consistent solution,
\begin{align*}
\dot{z}^i(u) - \dot{z}^i(-\infty) &= \int_{-\infty}^u du' \, \left(\dot{\Omega} + \Omega^2\right)_{ij} z^j \\
&= \int_{-\infty}^u du'\,\left(\dot{\Omega} z + \Omega \dot{z}\right)^i ~=~ \left(\Omega z\right)\Big|^u_{-\infty} \ .
\end{align*}   }
with solution,
\begin{equation}
\frac{dz^i}{du} ~=~\Omega^i{}_j(u) z^j \ ,
\label{b43}
\end{equation}
where we impose the initial condition of an initially parallel congruence, $\dot{z}^i(u<u_i) = 0$.
This is simply the defining equation for the connecting vector following from the alternative characterisation
in terms of the vanishing of its Lie derivative along $\c$, {\it i.e.} ${\cal L}_k z^i = 0$.

Now, provided only that we can write $\Omega_{ij}$ in the form $\Omega_{ij} = \bigl(\dot{E} E^{-1}\bigr)_{ij}$,
we can integrate (\ref{b43}) directly, giving
\begin{equation}
z^i(u) ~=~ E^i{}_a(u) X^a \ ,
\label{b44}
\end{equation}
for some integration constants $X^a$.  Of course, expressing $\Omega_{ij}$ in this form necessarily
implies $h_{ij} = \bigl(\ddot{E} E^{-1}\bigr)_{ij}$, which is just the defining oscillator equation
$\ddot{E}^i{}_a - h^i{}_j(u) E^j{}_a = 0$ for $E^i{}_a(u)$.

As explained above, these expressions are precisely those following from the geodesic equations
for the Brinkmann transverse coordinate $x^i(u)$ in the Penrose limit of the original spacetime.
Memory for a general background spacetime is therefore entirely encoded in an appropriate
gravitational plane wave.

To describe gravitational memory, we need to compare the relative positions and velocities
of neighbouring geodesics before and after an encounter with a gravitational wave burst 
confined to $u_i\le u \le u_f$. The velocity-encoded memory is then,
\begin{equation}
\Delta \dot{z}^i(u) ~=~ \dot{E}^i{}_a(u>u_f) \,X^a \ ,
\label{b45}
\end{equation}
and the position-encoded memory is
\begin{equation}
\Delta z^i(u) ~=~ \left( E^i{}_a(u>u_f) - E^i{}_a(u<u_i) \right) \,X^a \ ,
\label{b46}
\end{equation}
determined entirely by the zweibein $E^i{}_a(u)$.

The subtle point here is that we are considering gravitational wave bursts and shockwaves
which in the initial $u<u_i$ and final $u>u_f$ regions are simply flat spacetime. 
Nevertheless, gravitational memory requires $E^i{}_a(u>u_f) \ne E^i{}_a(u<u_i)$.
These regions must therefore be described by two different, {\it non-equivalent}, descriptions
of flat spacetime. This is best seen in the Rosen metric (\ref{b29aa}), where
${\bf C}_{ab} = \bigl(E^T E\bigr)_{ab}$.  The corresponding curvature is 
$R_{aubu} = (E^T \ddot{E})_{ab}$ and vanishes for a zweibein which is at most linear in $u$.
Any metric of the form (\ref{b29aa}) with such a zweibein is therefore diffeomorphic to flat spacetime,
being related by the Brinkmann-Rosen coordinate transformation (\ref{b26}).
In effect, the original spacetime links inequivalent and distinguishable copies of flat spacetime, 
gravitational memory being the physical signature.
In the language adopted in discussions of memory in Bondi-Sachs gravitational wave backgrounds 
(for a review, see \cite{Strominger:2017zoo}),
these metrics describe inequivalent gravitational vacua.\footnote{This equivalence may be made more 
precise by comparing the formula (\ref{b46}) for position-encoded memory involving the
zweibein $E^i{}_a(u)$ for the gravitational shockwave with the corresponding result for the displacement
memory $s^{\bar{z}}$ in the Bondi-Sachs metric \cite{Strominger:2017zoo},
\begin{equation*}
\Delta^+ s^{\bar{z}} = \frac{\gamma^{z\bar{z}}}{2r}\, \Delta^+ C_{zz}\,s^{\bar{z}} \ .
\end{equation*}
The zweibein $E^i{}_a(u)$ and the associated flat-space preserving Rosen coordinate transformations 
play the r\^ole here of the Bondi-Sachs metric coefficients $C_{zz}$ 
(where $N_{zz}=\partial_u C_{zz}$ is the Bondi news function) and the BMS supertranslations 
which interpolate between inequivalent vacua.\label{Bondi-Sachs}}
All this is especially evident in the Aichelburg-Sexl shockwave considered below, where the full
spacetime may be described by a Penrose `cut and slide' construction, dividing flat spacetime
along the shockwave localised on $u=0$.

Now consider how these forms of the zweibein give rise to memory.
For a purely position-encoded memory, an idealised form which realises (\ref{b46}) 
would be
\begin{equation}
E^i{}_a(u) ~\sim~ \d^i{}_a  + a^i{}_a \,\theta(u) \ .
\label{b48}
\end{equation}
The corresponding velocity change proportional to $\d(u)$ would be localised at $u=0$
and so would not affect the future memory region. However, taken literally, the form (\ref{b48}) would 
require a too-singular dependence $R_{iuju} \sim \d'(u)$ for the curvature, so a physical
realisation would have to be smoothed. Even so, $R_{iuju}$ would necessarily be negative for 
some values of $u$, and to be compatible with the null energy condition for the source
we would still need to impose the non-trivial constraint $R_{uu} = \tr R_{iuju} \ge 0$.

Next, consider zweibeins of the form 
\begin{equation}
E^i{}_a(u) ~\sim~ \d^i{}_a + a^i{}_a \,u\,\theta(u) \ .
\label{b49}
\end{equation}
In this case, $\dot{E}^i{}_a(u) \sim \theta(u)$, so we have velocity-encoded memory 
\cite{Hollowood:2009qz, Hollowood:2015elj, Zhang:2017jma}.
This solution corresponds to a localised source with curvature $R_{iuju} \sim \d(u)$, which is characteristic
of an impulsive gravitational shockwave. Clearly, the same comments apply to a smoothed
or extended wave burst, where we would find both position and velocity memory.

In the following sections, we will see how all this is realised in two important examples of gravitational
wave bursts -- the Aichelburg-Sexl shockwave (including non-impulsive extensions) and its
spinning generalisation, the gyraton.

\section{Gravitational Shockwaves}\label{sect 3}

This brings us to the main topic of this paper, the characterisation of memory, via the optical tensors
and Penrose limits, for important classes of gravitational wave bursts. In this section, we consider in detail
the Aichelburg-Sexl shockwave \cite{Aichelburg:1970dh}, 
together with a smoothed (sandwich wave) extension in which the impulsive limit is relaxed.

We therefore consider the metric,
\begin{equation}
ds^2 = 2 du dv + f(r) \chi_F(u) du^2 + dr^2 + r^2 d\phi^2 \ ,
\label{c1}
\end{equation}
This has a manifest symmetry with Killing vector $\partial_V$ and is a pp wave. The Christoffel symbols are
\begin{align}
\Gamma^v_{uu} &= \frac{1}{2}f(r) \,\chi'_F(u)  \ ,~~~~~~~~
\Gamma^v_{ur} = \frac{1}{2} f'(r)\, \chi_F(u)  \ ,  \nonumber \\
\Gamma^r_{uu} &= -\frac{1}{2} f'(r) \, \chi_F(u) \ , ~~~~~~~~
\Gamma^r_{\phi\phi} = -r \ , ~~~~~~~~
\Gamma^\phi_{r\phi} = 1/r \ ,
\label{c2}
\end{align}
while the non-vanishing curvature components are
\begin{equation}
R_{ruru} = - \frac{1}{2} f''(r) \, \chi_F(u) \ , ~~~~~~~~~~~~ R_{\phi u \phi u} = - \frac{1}{2} r f'(r) \, \chi_F(u)  \ ,
\label{c3}
\end{equation}
and
\begin{equation}
R_{uu} = -\frac{1}{2} \Delta f\, \chi_F(u) \ ,
\label{c4}
\end{equation}
with $\Delta$ denoting the two-dimensional Laplacian in polar coordinates $(r,\phi)$.

The original Aichelburg-Sexl shockwave describes an impulsive gravitational wave localised on the
lightcone $u=0$, corresponding to the profile $\chi_F(u) = \d(u)$. The most important case is the shockwave 
formed by an infinitely boosted particle\footnote{It is also interesting to consider a homogeneous beam source
\cite{Ferrari:1988cc, Shore:2002in, Hollowood:2015elj} 
for which $T_{uu} = \r \d(u)$ with $\r$ constant, for which $f(r) = -4\pi G \r \,r^2$. },
with energy-momentum tensor $T_{uu} = \r(r) \d(u)  = E \d^2(\underline{x}) \d(u)$. Solving the Einstein 
equations using (\ref{c4}) gives
\begin{equation}
f(r) = -4G E \, \log\left(\frac{r}{r_0}\right)^2 \ ,
\label{c5}
\end{equation}
for some short-distance cut-off scale $r_0$.

\subsection{Geodesics for gravitational bursts and shockwaves}\label{sect 3.1}

The geodesic equations for a general profile are
\begin{align}
&\ddot{v} + \half f(r) \chi_F'(u) + f'(r) \chi_F(u) \dot{r} = 0 \ ,  \nonumber \\
&\ddot{r} - \half f'(r) \chi_F(u) - r\dot{\phi}^2 = 0 \ ,  \nonumber \\
&\ddot{\phi} + \frac{2}{r} \dot{r} \dot{\phi} = 0 \ ,
\label{c6}
\end{align}
and we can immediately write the integrated expression for $\dot{v}$ directly from the metric
as
\begin{equation}
2 \dot{v} + f(r) \chi_F(u) + \dot{r}^2 + r^2 \dot{\phi}^2 = 2 \eta \ ,
\label{c7}
\end{equation}
with $\eta = 0$ for a null geodesic, $\eta < 0$ for a timelike geodesic.

The cylindrical symmetry of the original metric (\ref{c1}) allows the $\ddot{\phi}$ equation to be integrated,
implying
\begin{equation}
r^2 \dot{\phi} = \ell = {\rm const.} \ ,
\label{c8}
\end{equation}
where $\ell$ is the conserved angular momentum about the $r=0$ axis. In what follows, we choose the 
natural initial condition (before the incidence of the gravitational wave burst on the test particle described by (\ref{c6}))
$\dot{\phi}(u)=0$. Then $\ell=0$ and the geodesics are curves with $\phi(u)$ constant,
taking $\phi=0$ for the chosen geodesic $\c$.

Now focus on the Aichelburg-Sexl (AS) shockwave with profile $\chi_F(u) = \d(u)$. In this case we can solve the
geodesic equations exactly. For the first integrals of (\ref{c6}) we have
\begin{align}
&\dot{v} = \eta - \half f(b) \d(u) - \frac{1}{8} f'(b)^2 \theta(u) \ ,
&\dot{r} = \half f'(b) \theta(u) \ , 
\label{c9}
\end{align}
and so 
\begin{align}
&v = V + \eta u - \half f(b) \theta(u) - \frac{1}{8} f'(b)^2 u \theta(u) \ , \nonumber \\
&r = b + \half f'(b) u \theta(u) \ ,
\label{c10}
\end{align}
where $b$ is the impact parameter for the chosen geodesic $\c$.

For $f'(b) < 0$, which follows from (\ref{c4}), (\ref{c5}) provided the null energy condition $T_{uu} > 0$ is respected,
the geodesics converge towards the source at $r=0$ after encountering the shockwave at $u=0$. 
The focal point, at $u = -2b/f'(b)$, depends on the impact parameter. 
The most striking feature of the solution, however, is the `back in time' jump $\Delta v = - \tfrac{1}{2} f'(b)$ 
in $v$ at the instant of collision. This raises immediate questions concerning causality in these
shockwave spacetimes. This has been thoroughly explored in our previous papers 
\cite{Shore:2002in, Hollowood:2015elj, Hollowood:2016ryc} (see also \cite{Camanho:2014apa, Papallo:2015rna})
both classically and including the special issues arising from vacuum polarisation in QFT in these spacetimes.

It is also interesting to relax the impulsive limit and consider an extended pulse of duration $L$ in the lightcone coordinate
$u$, centred on $u=0$. For definiteness, 
we consider\footnote{For numerical results, we frequently use a smoothed version
\begin{equation*}
\Theta(u,L) \sim 
\frac{1}{2L}\Big( \tanh\left(\a\left(u+ \tfrac{L}{2}\right)\right)- \tanh\left(\a\left(u- \tfrac{L}{2}\right)\right) \Big)
\end{equation*}
where $L$ gives the duration of the burst and the parameter $\a$ can be adjusted to smooth
the profile.}
\begin{equation}
\chi_F(u) ~=~ \Theta(u,L) ~\equiv~ 
\frac{1}{L}\left(\theta\left(u+ \tfrac{L}{2}\right) - \theta\left(u-\tfrac{L}{2}\right)\right) \ ,
\label{c11}
\end{equation}
which, like $\d(u)$, is normalised so that $\int du\, \Theta(u,L) = 1$ and has the impulsive limit  
$\lim_{L\rta 0} \Theta(u,L) = \d(u)$.  In this case, analytic solutions may still be found\footnote{With the 
extended profile $\chi_F(u) = \Theta(u,L)$, we solve the geodesic equations piecewise in the three
regions before, during and after the wave burst and match at the boundaries. In the interaction
region $-L/2 < u < L/2$, we can solve the radial equation in (\ref{c6}) with $f(r)$ for the particle source 
in terms of the inverse error function to give
\begin{equation*}
r(u) = b\, \exp\left[ -\,\left( {\rm erf}^{-1}\left(\frac{1}{b}\sqrt{\frac{8GE}{\pi L}}\,\tilde{u}\right)\right)^2 \right] \ ,
\end{equation*}
with $\tilde{u} = u + L/2$. Substituting into (\ref{c6}) and integrating then gives $v(u)$ in the form
\begin{equation*}
v(u) = V + \frac{4GE}{L} \,\left( \log b - 1\right) \tilde{u} +
b\,\sqrt{\frac{8GE}{L}} \,\, {\rm erf}^{-1}\left( \frac{1}{b} \sqrt{\frac{8GE}{\pi L}}\,\tilde{u}\right) \,\,
\exp\left[-  \left({\rm erf}^{-1}\left(\frac{1}{b}\sqrt{\frac{8GE}{\pi L}}\,\tilde{u}\right)\right)^2 \right] \ .
\end{equation*}
These analytic forms reproduce the numerical plots shown in Figs.~\ref{fig3.1} and \ref{fig3.4} for the
extended profile.  The shift $\Delta v$ across the interaction region is easily found (see appendix \ref{Appendix A})
from this expression for $v(u)$, with the $\log b$ term reproducing the Aichelburg-Sexl shift. \label{footnote AS geodesics}}
but are at first sight less illuminating than in the impulsive limit, so below we show plots of the geodesics 
found from numerical solutions of (\ref{c6}).

We now present illustrations of these solutions for both null and timelike geodesics, and for both the
impulsive and extended profiles $\chi_F(u)$. Numerical values of the parameters ($GE$, $L$ and subsequently $J$)
are chosen purely to demonstrate the key general properties of the geodesics.

In Fig.~\ref{fig3.1} we show the radial coordinate $r(u)$ for three values of the impact parameter $b$
for both $\chi_F(u) = \d(u)$ and $\chi_F(u) = \Theta(u,L)$. Notice that outside the gravitational wave burst,
the metric simply describes flat spacetime so the trajectories are straight lines, and may continue 
unperturbed through $r=0$. The radial geodesics are also insensitive to the parameter $\eta$,
so are the same for null and timelike geodesics.  Note that the asymptotic slope of the geodesics
for equal $b$ but different $\chi_F(u)$ are different, as illustrated in Fig.~\ref{fig3.2}. The final geodesic
remembers the nature of the gravitational wave burst.
\begin{figure}[h]
\centering
\includegraphics[scale=0.8]{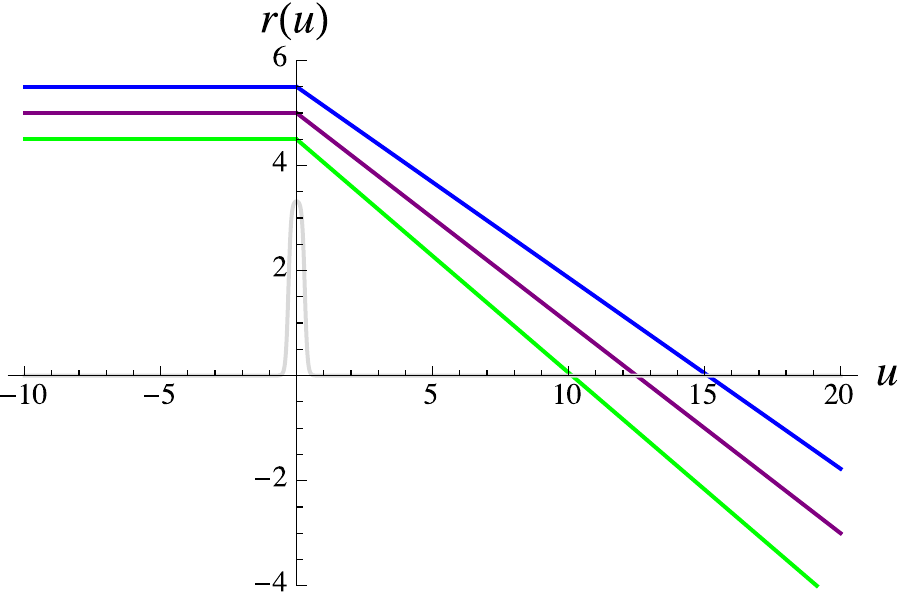} \hskip0.7cm
\includegraphics[scale=0.8]{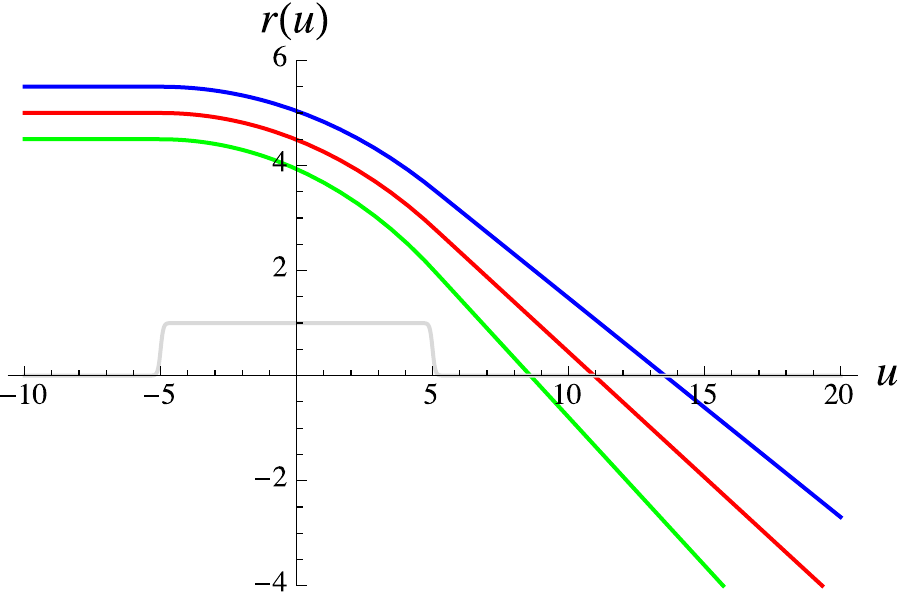}
\caption{The left-hand plot shows the behaviour of the radial coordinate $r(u)$ of geodesics with different
impact parameters $b$ as the corresponding test particle is struck by an impulsive gravitational shockwave
at $u=0$. The right-hand plot shows the same result for an extended gravitational wave burst with profile
$\chi_F(u)=\Theta(u,L)$ indicated (with a different vertical scale) by the grey curve. In this and 
all subsequent plots, we have taken $L=10$. }
\label{fig3.1}
\end{figure}

\begin{figure}[h]
\centering
\includegraphics[scale=0.8]{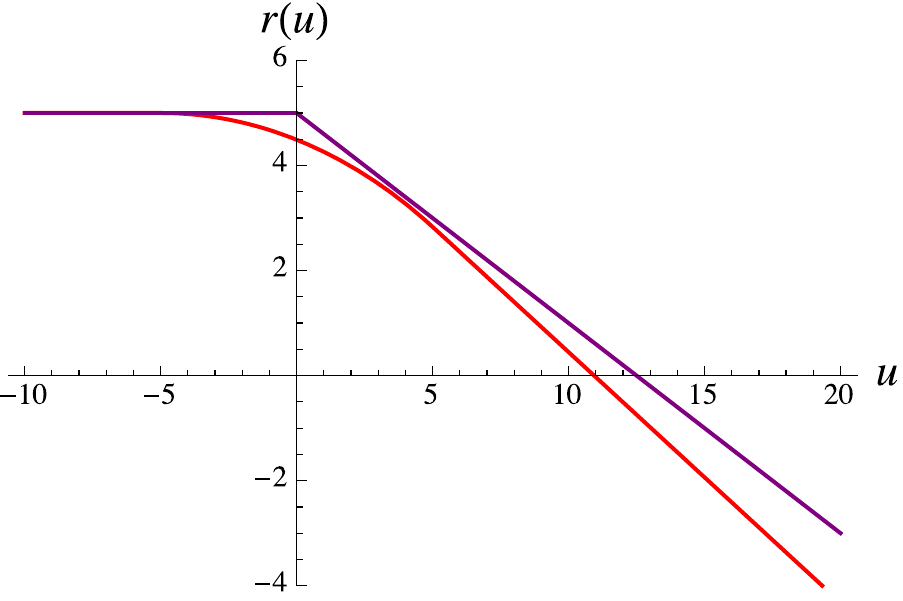} 
\caption{The final transverse velocity of the test particle has a memory of the gravitational wave
profile, and is greater for the extended wave burst.}
\label{fig3.2}
\end{figure}

Fig.~\ref{fig3.3} shows the 3-dim behaviour of a circle of geodesics with the same impact parameter but
different initial angles to the shockwave axis. The focusing of the geodesics is clearly seen.
\begin{figure}[h]
\centering
\includegraphics[scale=0.6]{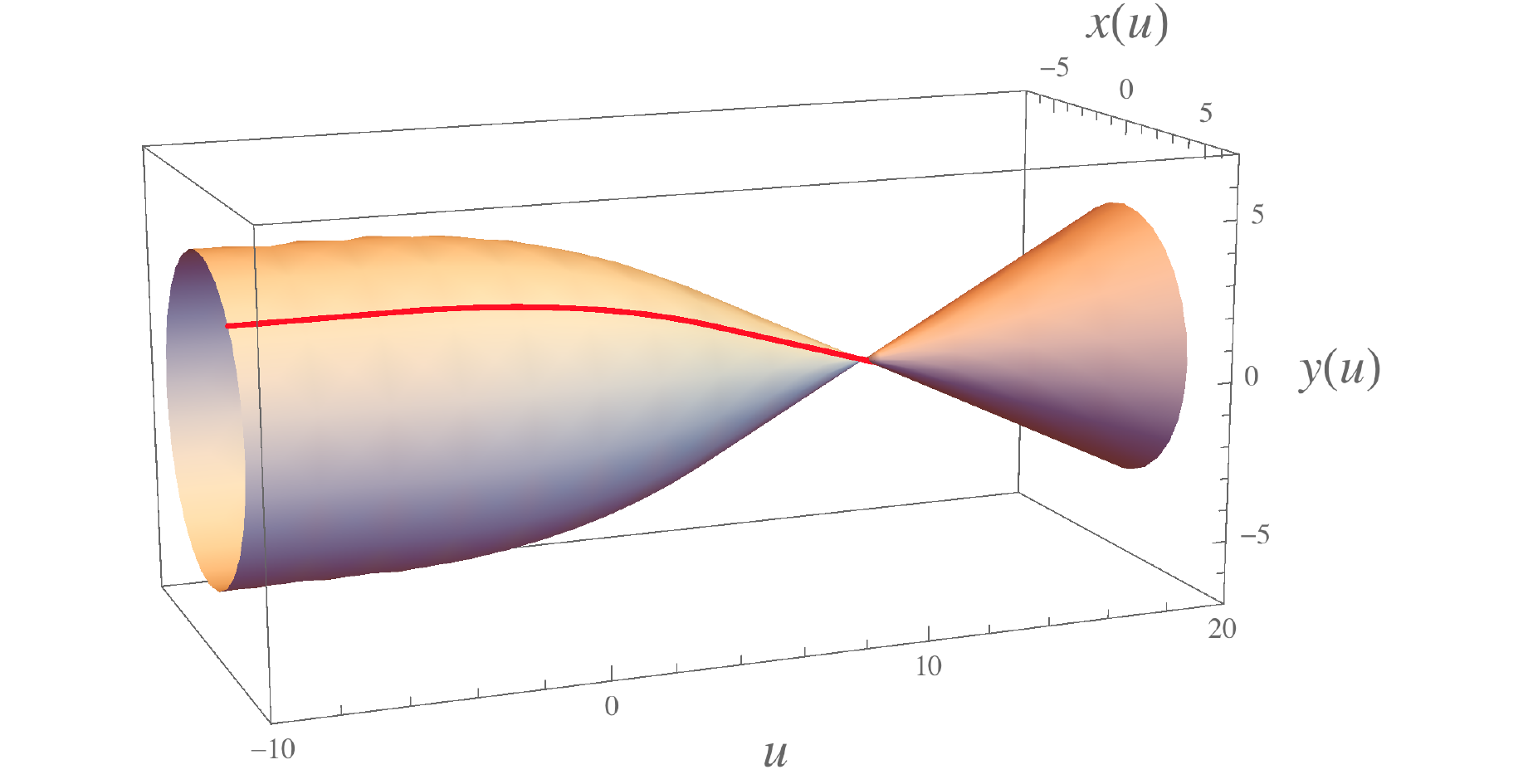} 
\caption{The red curve shows the trajectory of a geodesic in the background of a gravitational wave burst with
extended profile $\Theta(u,L)$. The parameters match those of the $r(u)$ plots in Fig.~\ref{fig3.1}. The shaded
surface is mapped out by geodesics with the same impact parameter $b$ but with different angles $\phi$ around
the gravitational wave source on the $r=0$ axis. The focusing effect for geodesics with the same impact
parameter is evident. }
\label{fig3.3}
\end{figure}

The behaviour of $v(u)$ is shown in Fig.~\ref{fig3.4} for null geodesics and with impulsive and
extended profiles.  A similar behaviour is seen for timelike geodesics, but of course with $v(u)$ not
constant before the arrival of the gravitational wave, as given in (\ref{c10}). 
The notable feature is the discontinuous jump in the lightcone coordinate $v(u)$ in the impulsive
shockwave case, not least since the jump is backwards in the corresponding time coordinate. 
This immediately raises issues of causality and observability.
Further discussion of these trajectories, and especially their implications for `time machines', may be found
in \cite{Shore:2002in, Hollowood:2015elj, Hollowood:2016ryc}. 

In Appendix \ref{Appendix A}, we discuss briefly how this shift $\Delta v$ across the shockwave allows us
to compute the scattering amplitude in the eikonal limit for ultra-high energy particle scattering, which is 
mediated by graviton exchange. In this picture, $\Delta v$ is directly related to the scattering phase
$\Theta(s,b)$, which depends on the CM energy and impact parameter $b$. For the Aichelburg-Sexl 
shockwave, (\ref{c10}) immediately gives $\Delta v = -\half f(b)$.
\begin{figure}[h!]
\centering
\includegraphics[scale=0.8]{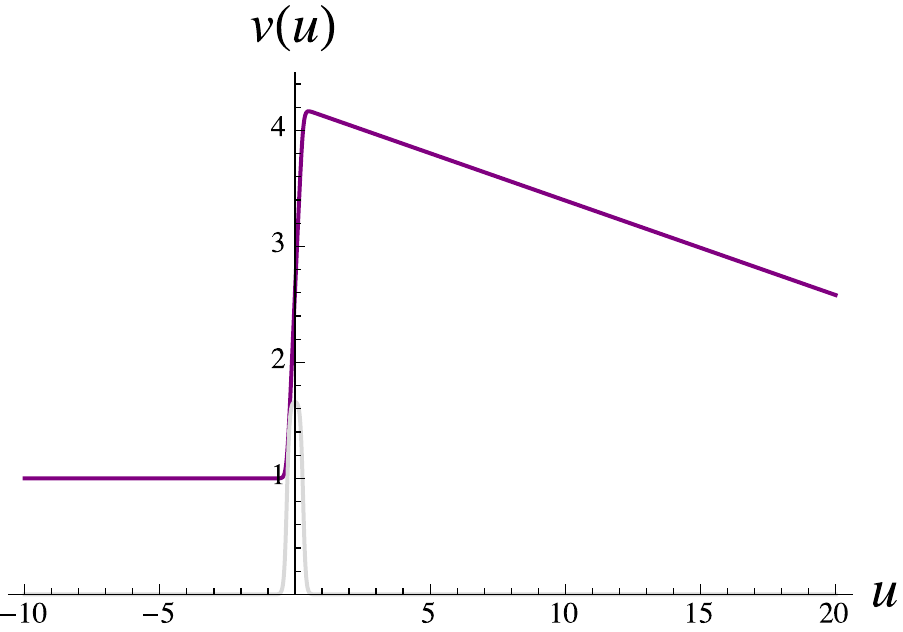} \hskip0.7cm
\includegraphics[scale=0.8]{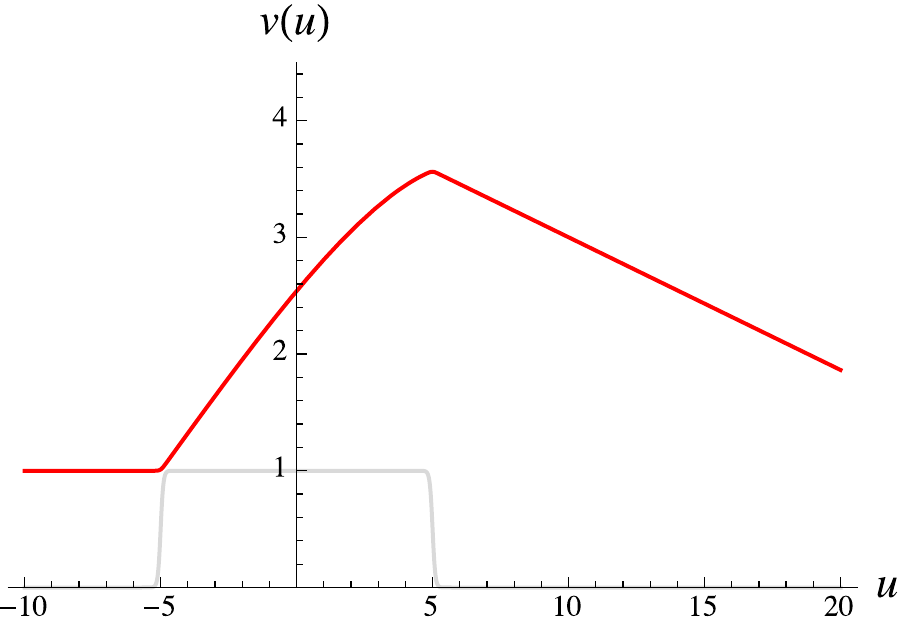}
\caption{The lightcone coordinate $v(u)$ of a test particle exhibits a jump as it encounters 
the gravitational wave burst, for both impulsive (left-hand figure) and extended (right-hand figure) profiles.
Note the discontinuous (and backwards in time) jump in the impulsive gravitational shockwave case. }
\label{fig3.4}
\end{figure}

This summarises the properties of a single geodesic in a gravitational shockwave background. 
To describe gravitational memory, however, we need to find the {\it relative motion} of neighbouring geodesics.
We therefore need to describe the full congruence centred on a chosen geodesic and determine the 
optical tensors.

\subsection{Optical tensors and memory}\label{sect 3.2}

To find the optical tensors, we start by calculating $\Omega_{\m\n}  = D_\n k_\m$ for the congruence centred 
on a chosen geodesic $\c$. Here, $k^\m$ is a vector field, the tangent vector to the individual geodesics 
in the congruence.  With the generalised shockwave metric (\ref{c1}), we have
\begin{equation}
k^\m = \begin{pmatrix}\,1\,\\\,\dot{v}\, \\ \,\dot{r}\, \\ \,\dot{\phi}\,\end{pmatrix} \ , ~~~~~~~~~~
k_\m = g_{\m\n} k^\n ~=~ \begin{pmatrix}\, \dot{v} + f(r) \chi_F(u) \, \\ \,1\, \\ \, \dot{r}\, \\
\, r^2 \dot{\phi}\, \end{pmatrix} \ ,
\label{c12a}
\end{equation}
and we can immediately use angular momentum conservation to set $\dot{\phi}=0$.

Taking the covariant derivatives (using the Christoffel symbols given in (\ref{c2}), a detailed calculation
using in particular the expression (\ref{c7}) for $\dot{v}$, now shows
\begin{equation}
\Omega_{\m\n} ~=~ \begin{pmatrix} \dot{r} \Omega_{rr} \dot{r} ~ &0  ~&-\Omega_{rr} \dot{r} ~& 0 \\
0 ~& 0 ~& 0 ~& 0 \\
-\dot{r} \Omega_{rr} ~&0 ~&\Omega_{rr} ~& 0 \\
0 ~& 0 ~& 0 ~&\Omega_{\phi\phi} \end{pmatrix} \ ,
\label{c12}
\end{equation}
where
\begin{align}
\Omega_{rr} &~=~ \partial_r k_r - \Gamma^\l_{rr}k_\l ~=~ \partial_r \dot{r}  \ , \nonumber \\
\Omega_{\phi \phi} &~=~\partial_\phi k_\phi - \Gamma^\l_{\phi\phi} k_\l ~=~ r \dot{r}  \ .
\label{c13}
\end{align}

\subsubsection{Null congruences}

Before evaluating these expressions more explicitly, we first introduce Fermi Normal Coordinates
and take the transverse projection $\hat{\Omega}_{ij} = \left(\hat{g} \,\Omega \,\hat{g}\right)_{ij}$ 
as described in (\ref{b9}) and (\ref{b15}).  While the final results for $\hat{\Omega}_{ij}$ will be the same for both 
null and timelike congruences, for definiteness we consider the null case first.

The null FNC basis vectors can be chosen in this case as ${\bf e}^A$, $A= u,v,x^i$, where by definition 
$e^u{}_\m = k_\m$.  Imposing $\dot\phi = 0$ immediately, we have
\begin{equation}
e^{u\m} = \begin{pmatrix} \,&1\,\\ \,&\dot{v}\,\\ \,&\dot{r}\,\\ \,&0\, \end{pmatrix} \ , ~~~~~~~~
e^{v\m} = \begin{pmatrix} \,&0\,\\ \,&-1\,\\ \,&0\,\\ \,&0\, \end{pmatrix} \ , ~~~~~~~~
e^{1\m} = \begin{pmatrix} \,&0\,\\ \,&-\dot{r}\,\\ \,&1\,\\ \,&0\, \end{pmatrix} \ , ~~~~~~~~
e^{2\m} = \begin{pmatrix} \,&0\,\\ \,&0\,\\ \,&0\,\\ \,&1/r\, \end{pmatrix} \ . 
\label{c14}
\end{equation}

It is readily checked that they satisfy the appropriate orthonormality condition 
$g_{\m\n} e^{A\m} e^{B\n} = \eta^{AB}$, with $\eta_{AB}$ as in (\ref{b5}). These all follow as purely algebraic
conditions except for the lightcone identity $g_{\m\n} e^{u\m} e^{u\n} = k^2 = 0$, which requires the null
geodesic condition (\ref{c7}) for $\dot{v}$. 

We also need to verify that this set of basis vectors is parallel transported along the null geodesic $\c$, 
that is, $k.D e^{A\m} = 0$.  The first identity, $k.D e^{u\m} = k.D k^\m = 0$ is simply the defining geodesic
equation itself so is satisfied by definition. The remaining identities for $e^{v\m}$ and $e^{i\m}$ are readily
verified using the Christoffel symbols (\ref{c2}).

With the FNC basis established, we can construct the projection matrix $\hat{g}_{\m\n}$ according to 
(\ref{b8}). The required transverse components $\hat{\Omega}_{ij}$ which determine the optical tensors
are then given directly from (\ref{b10}) as
\begin{equation}
\hat{\Omega}_{ij} = e^{i\m} \Omega_{\m\n} e^{j\n} \ ,
\label{c15}
\end{equation}
with $\Omega_{\m\n}$ in (\ref{c12}) and $e^{i\m}$ in (\ref{c14}).  It follows directly that
\begin{equation}
\hat{\Omega}_{ij} = 
\begin{pmatrix} ~&\Omega_{rr} ~~&0 \\ ~&0~~&\frac{1}{r^2}\Omega_{\phi\phi} \end{pmatrix} \ ,
\label{c16}
\end{equation}
and the optical tensors are read off from the decomposition (\ref{b11}). 

We see immediately that $\hat{\Omega}_{ij}$ is symmetric and therefore unsurprisingly the congruence has 
vanishing twist, $\hat{\omega}_{ij} = 0$. Explicitly, the expansion is
\begin{equation}
\hat{\theta} = \tr\,\hat{\Omega} ~=~ \Omega_{rr} + \frac{1}{r^2} \Omega_{\phi\phi} \ ,
\label{c17}
\end{equation}
leaving the shear as 
\begin{equation}
\hat{\sigma}_{ij} = \frac{1}{2} \left(\Omega_{rr} - \frac{1}{r^2} \Omega_{\phi\phi} \right) \,
\begin{pmatrix} \,&1 ~&0\, \\ \,&0 ~ &-1\, \end{pmatrix} \ .
\label{c18}
\end{equation}

\subsubsection{Timelike congruences}

For a timelike congruence, we need an FNC basis ${\bf e}^A$, $A = 0,3,i$, such that
$g_{\m\n} e^{A\m} e^{B\n} = \eta^{AB}$ with $\eta_{AB} = {\rm diag}(-1,1,1,1)$  and 
$e^{0\m} = k^\m/\sqrt{-2\eta}$, the normalisation being fixed by (\ref{c7}) such that $k^2 = 2\eta$.
In this case, a suitable choice which also satisfies the parallel transport condition $k.D e^{A\m} = 0$
is
\begin{equation}
e^{0\m} = \frac{1}{\sqrt{-2\eta}} \begin{pmatrix} \,&1\,\\ \,&\dot{v}\,\\ \,&\dot{r}\,\\ \,&0\, \end{pmatrix} \ , ~~~~
e^{3\m} = \sqrt{-2\eta} \begin{pmatrix} \,&0\,\\ \,&-1\,\\ \,&0\,\\ \,&0\, \end{pmatrix} - e^{0\m} \ , ~~~~
e^{1\m} = \begin{pmatrix} \,&0\,\\ \,&-\dot{r}\,\\ \,&1\,\\ \,&0\, \end{pmatrix} \ , ~~~~
e^{2\m} = \begin{pmatrix} \,&0\,\\ \,&0\,\\ \,&0\,\\ \,&1/r\, \end{pmatrix} \ . 
\label{c19}
\end{equation}

$\Omega_{\m\n}$ as given in (\ref{c12}), (\ref{c13}) is unchanged, so we can project the three-dimensional 
transverse components in the same way from
\begin{equation}
\hat{\Omega}_{rs} ~=~ e^{r\m}\,\Omega_{\m\n}\, e^{s\n} \ , ~~~~~~r,s = 3,1,2 \ .
\label{c20}
\end{equation}
A simple calculation with the basis vectors (\ref{c19}) now shows that the components $\hat{\Omega}_{33}$
and $\hat{\Omega}_{3i}$ vanish, leaving
\begin{equation}
\hat{\Omega}_{rs} ~=~ \begin{pmatrix} \,&0 ~ &0 ~&0\,\\
\,&0 ~&\Omega_{rr} ~&0\, \\ 
\,&0 ~&0 &\frac{1}{r^2} \Omega_{\phi\phi} \, \end{pmatrix}
\label{c21}
\end{equation}

For this metric, the three-dimensional transverse space in the timelike case becomes effectively two-dimensional.
The deeper reason for this, which is a property of pp waves, becomes clear below when we present the 
description of geodesic deviation from the perspective of the Penrose limits.
The optical tensors for the timelike congruence are therefore identical to those in the null case,
and can be visualised as before in terms of deformations of a Tissot ring.

\subsubsection{Aichelburg-Sexl shockwave}\label{sect 3.2.3}

We can now evaluate the optical tensors explicitly for the null congruence in the Aichelburg-Sexl metric
in the impulsive, shockwave limit. The only subtlety is that we have to use the geodesic solution (\ref{c10})
for $r(u;b)$ to define a vector field describing the whole congruence rather than a single geodesic specified
by the fixed impact parameter $b$. 

To achieve this, we invert the solution
\begin{equation}
r= b + \half f'(b) \,u\theta(u) \ ,
\label{c22}
\end{equation}
implicitly to define $b(u,r)$, {\it i.e.}~given a geodesic passing through the point $(u,r)$, this specifies
the corresponding impact parameter. Then, we may write
\begin{equation}
\dot{r} = \half f'(b) \theta(u) ~=~ \frac{1}{u} \left(r - b(u,r)\,\right) \ ,
\label{c23}
\end{equation}
and so,
\begin{equation}
\Omega_{rr} = \partial_r\,\dot{r} ~=~ \frac{1}{u} \left(1 - \frac{\partial b(u,r)}{\partial r} \right)  \ .
\label{c24}
\end{equation}
Taking the partial derivative of (\ref{c22}) now gives
\begin{equation}
1 = \frac{\partial b(u,r)}{\partial r} \, \left(1 + \half f''(b) \,  u\theta(u) \,\right)   \ ,
\label{c25}
\end{equation}
and so we find
\begin{equation}
\Omega_{rr} =  \half  f''(b) \theta(u) \left(1 + \half f''(b) u \theta(u) \right)^{-1} \ .
\label{c26}
\end{equation}
The $\Omega_{\phi\phi} = r \dot{r}$ component is found directly from (\ref{c22}), (\ref{c23}).

From (\ref{c16}), we therefore determine the projections $\hat{\Omega}_{ij}$ which specify the optical tensors 
for the null congruence centred on the geodesic $\c$ with impact parameter $b$ as
\begin{align}
\hat{\Omega}_{11} &= \half f''(b) \left(1 + \half f''(b) u \right)^{-1} \theta(u)  \ , \nonumber \\
\hat{\Omega}_{22} &= \half \frac{f'(b)}{b} \left( 1 + \half \frac{f'(b)}{b} u \right)^{-1} \theta(u) \ .
\label{c27}
\end{align}

Defining the optical scalars $\hat{\theta}$ (expansion) and $\hat{\s}_+$ (shear) as in section \ref{sect 2.2},
we find for a general profile $f(r)$ that
\begin{align}
\hat{\theta} &= \half \frac{1}{\left(1 + \half f''(b) u \right) \left( 1 + \half \frac{f'(b)}{b} u \right)}
\left(f''(b) + \half \frac{f'(b)}{b} + f''(b) \frac{f'(b)}{b} u \right) \theta(u) \ ,  \nonumber \\
{}  \nonumber \\
\hat{\s}_+ &= \frac{1}{4} \frac{1}{\left(1 + \half f''(b) u \right) \left( 1 + \half \frac{f'(b)}{b} u \right)}
\left(f''(b) - \frac{f'(b)}{b} \right) \theta(u)  \ .
\label{c28}
\end{align}
Evaluating for the particle shockwave, we find (the negative sign for $\hat{\theta}$ indicating focusing),
\begin{align}
\hat{\theta} &= -\,  \frac{2 (4 G E)^2 \,u \,\theta(u)}{b^4 - (4 G E)^2 u^2}  \ , \nonumber \\
{} \nonumber \\
\hat{\s}_+ &= \frac{4 G E b^2 \,\theta(u) }{b^4 - (4 G E)^2 u^2}  \ .
\label{c29}
\end{align}

We illustrate these properties of the congruence in Figs.~\ref{fig3.5}, \ref{fig3.6}  after we have derived these results 
in the Penrose limit formalism. Note immediately the singular feature at $u = b^2/4G E$ where the congruence
focuses in one direction while diverging in the orthogonal direction.

It is interesting to consider other shockwave sources, for example a uniform density beam 
\cite{Ferrari:1988cc, Shore:2002in, Hollowood:2009qz, Hollowood:2015elj, Hollowood:2016ryc}, 
for which $T_{uu} = \r \,\d(u)$ and $f(r) = -4\pi G\r\, r^2$. 
In this case the congruence has only expansion and no shear, and there is a single focal point for
all the geodesics independent of their impact parameter.\footnote{For the beam shockwave,
we have $f''(b) = f'(b)/b = -8\pi G\r$ and so $\hat{\s}_+ = 0$.  The focal point is at
$u = 1/4\pi G \r$ and the expansion is 
\begin{equation*}
\hat{\theta} = - \frac{8\pi G\r}{1-4\pi G \r u} \theta(u) \ .
\end{equation*}}

Interpreting in terms of gravitational memory, we see immediately that after the passage of the shockwave
the relative position of neighbouring geodesics, visualised by the Tissot ring, is $u$-dependent.
That is, the memory associated with an impulsive shockwave is of the purely velocity-encoded type.
After the encounter with the shockwave, neighbouring test particles (recall that the null and timelike
optical tensors are identical) fly apart with fixed velocities, in a pattern exhibiting both expansion (focusing)
and shear.

\subsection{Penrose limits and memory}\label{sect 3.3}

The Penrose limit of the generalised shockwave metric (\ref{c1}) is readily found using the FNC method
described in section \ref{sect 2.3}. This is already known from our previous work \cite{Hollowood:2009qz},
where it was derived using an alternative method involving the construction of the Rosen form
of the plane wave metric. Indeed  most of the results of this subsection are already known from our earlier papers 
on causality and quantum field theoretic effects in QED in shockwave backgrounds 
\cite{Hollowood:2009qz, Hollowood:2015elj, Hollowood:2016ryc}.

\subsubsection{Null congruences and plane waves}

We consider first the FNC construction of the Penrose limit corresponding to null geodesics. With the
identification of Brinkmann coordinates for the plane wave metric with FNCs along the geodesic $\c$ in the
original spacetime, the plane wave profile function $h_{ij}$ is simply the projection onto the FNC basis 
of the relevant components of the shockwave curvature tensor $R_{\r\m\s\n}$. That is, the Penrose limit
metric is
\begin{equation}
ds^2 ~=~ 2 du dv + h_{ij}(u) x^i x^j du^2 + \d_{ij} dx^i dx^j \ ,
\label{c30}
\end{equation}
with
\begin{equation}
h_{ij} = - R_{\r u \s u}\big|_\c \, e^{i\r} e^{j\s} \ ,
\label{c31}
\end{equation}
where the $e^{i\m}$ are the basis vectors in (\ref{c14}). The only non-vanishing components of the curvature 
are $R_{ruru}$ and $R_{\phi u \phi u}$ given in (\ref{c3}) and we immediately find
\begin{equation}
h_{ij} ~=~ \begin{pmatrix} \, &h_{11} ~ &0\, \\ \, &0 ~&h_{22}\,\end{pmatrix} \ ,
\label{c32}
\end{equation}
with
\begin{align}
h_{11} ~&=~ - R_{ruru}\big|_\c  ~=~ \half f''(r)\big|_\c \,\chi_F(u)  \nonumber \\
h_{22} ~&=~ - \frac{1}{r^2} \,R_{\phi u \phi u}\big|_\c  ~=~ \half \frac{f'(r)}{r}\Big|_\c \, \chi_F(u) \ .
\label{c33}
\end{align}
In these expressions, the function $r(u)$ is the solution of the geodesic equation (\ref{c6}) defining $\c$.
For the Aichelburg-Sexl shockwave only, where $\chi_F(u) \rta \d(u)$, we can replace $f(r) \rta f(b)$
for constant impact parameter $b$.

The geodesics for the transverse coordinates are then,
\begin{equation}
\ddot{x}^i - h^i{}_j(u) x^j = 0 \ ,
\label{c34}
\end{equation}
as described in section \ref{sect 2.4}, and solutions are plotted in Fig.~\ref{fig3.5} below.

Our key assertion is that the geodesics in this plane wave metric are the same as those of the congruence
in the tubular neighbourhood of the null geodesic with impact parameter $b$ in the original shockwave 
spacetime (\ref{c1}).

To see this explicitly in the impulsive limit, recall from section \ref{sect 2.4} that the solutions of the
 geodesic equations for the $x^i(u)$ in the plane wave are written in terms of a zweibein $E^i{}_a(u)$ 
which solves the oscillator equation
\begin{equation}
\ddot{E}^i{}_a - h^i{}_j E^j{}_a = 0 \ .
\label{c35}
\end{equation}
The solutions with $h_{ij}(u)$ given by (\ref{c33}) with $\chi_F(u) = \d(u)$ are easily found and, with boundary 
conditions appropriate for a congruence of initially parallel geodesics, we have $E^i{}_a(u)$ diagonal
with \cite{Hollowood:2009qz}:
\begin{align}
E^1{}_1(u) ~&=~ 1 + \half f''(b) \,u\, \theta(u) \ , \nonumber \\
E^2{}_2(u) ~&=~ 1 + \half \frac{f'(b)}{b} \,u\,\theta(u) \ .
\label{c36}
\end{align}
The optical tensors in this plane wave are found from the tensor $\Omega_{ij} \equiv (\dot{E} E^{-1})_{ij}$.
Evaluating this, we find
\begin{align}
\Omega_{11} ~&=~ \half f''(b) \left(1 + \half f''(b) u\right)^{-1}\,\theta(u) \ , \nonumber \\
\Omega_{22} ~&=~ \half \frac{f'(b)}{b} \left(1 + \half \frac{f'(b)}{b} u\right)^{-1} \, \theta(u) \ .
\label{c37}
\end{align}

This confirms the identification of $\Omega_{ij}$ in the Penrose limit plane wave and the $\hat{\Omega}_{ij}$
of (\ref{c27}) established directly in the full shockwave spacetime.

The optical tensors are therefore identical, and we can directly verify the Raychoudhuri equations,
written here as
\begin{equation}
\dot{\Omega}_{ij} + \left(\Omega\right)^2_{ij} ~=~ h_{ij} \ .
\label{c38}
\end{equation}
This discussion therefore confirms how for null geodesics, geodesic deviation and gravitational memory 
is entirely encoded in the corresponding Penrose limit plane wave.

\subsubsection{Timelike congruences}

For timelike geodesics, corresponding to massive test particles/detectors, we use the analogous formalism 
from section \ref{sect 2.3}. What we have called the `timelike Penrose limit' has the metric (\ref{b23}),
\begin{equation}
ds^2 = - \left(1 - h_{rs}(t) x^r x^s\right) dt^2 + \d_{rs} dx^r dx^s \ ,
\label{c39}
\end{equation}
where $h_{rs}(t)$ is the projection of the curvature tensor of the shockwave onto the timelike FNC basis vectors 
(\ref{c19}), {\it i.e.}
\begin{equation}
h_{rs}(t) ~=~ - R_{r0s0}\big|_\c ~=~ - R_{\r\m\s\n}\big|_\c e^{r\r} e^{0\m} e^{s\s} e^{0\n}  \ .
\label{c40}
\end{equation}
We now see immediately that $h_{rs} = 0$ if either $r$ or $s$ is 3. This follows from the symmetries of the curvature 
tensor together with the fact that there is no non-vanishing $v$ component in $R_{\r \m \s \n}$. In turn, this can
be traced to the $v$-translation symmetry of the shockwave metric, the existence of the corresponding Killing vector 
$\partial_v$ being a defining property of pp waves. It is therefore a general feature of pp waves, including the 
shockwave, that $h_{rs}$ is effectively two-dimensional. 

Then, evaluating as before, we find
\begin{equation}
h_{rs} ~=~ \begin{pmatrix} \,&0~ &0~ &0 \, \\ \,&0 ~ &h_{11} ~&0\, \\ \,&0 ~&0 ~&h_{22}\, \end{pmatrix} \ ,
\label{c41}
\end{equation}
with 
\begin{equation}
h_{11} = - R_{r u r u}\big|_\c \ , ~~~~~~~~~~  h_{22} = - \frac{1}{r^2} R_{\phi u \phi u}\big|_\c \ .
\label{c42}
\end{equation}

Following section \ref{sect 2.3}, we now see that the geodesics for the transverse coordinates
$x^i(u)$ in the metric (\ref{c39})  are identical (for small $x^i$, see footnote \ref{timelike}) 
to those in the null Penrose limit. This confirms that the optical tensors are the same for the
null and timelike congruences in the shockwave metric, as is already implicit in section \ref{sect 3.2}.

\subsubsection{Congruences and memory for the gravitational shockwave}

As we have seen, the behaviour of nearby geodesics in the congruence, and therefore the gravitational memory,
is described by the geodesic equations (\ref{c34}) in the appropriate Penrose limit metric.

We illustrate this here by solving (\ref{c34}) explicitly for the impulsive Aichelburg-Sexl shockwave,
with $\chi_F(u) = \d(u)$, and the generalisation with an extended profile, $\chi_F(u) = \Theta(u,L)$. 
Analytic solutions have been given above for the impulsive limit, whereas in the extended (sandwich wave) case 
a numerical solution is used. This is necessary since the functions $f(r)$ in the Penrose limit geodesic
equation (\ref{c34}) involve the solutions $r(u)$ characterising the geodesic $\c$ with 
impact parameter $b$ in the original metric.

In Fig.~\ref{fig3.5}, we show the behaviour of $x^i(u)$ (specifying the null congruence for definiteness) in the
impulsive and extended cases. This demonstrates the initial convergence in the $x^2$ direction and divergence
in $x^1$ implied by (\ref{c34}) and governed by the optical scalars given in (\ref{c29}).
\begin{figure}[h]
\centering
\includegraphics[scale=0.7]{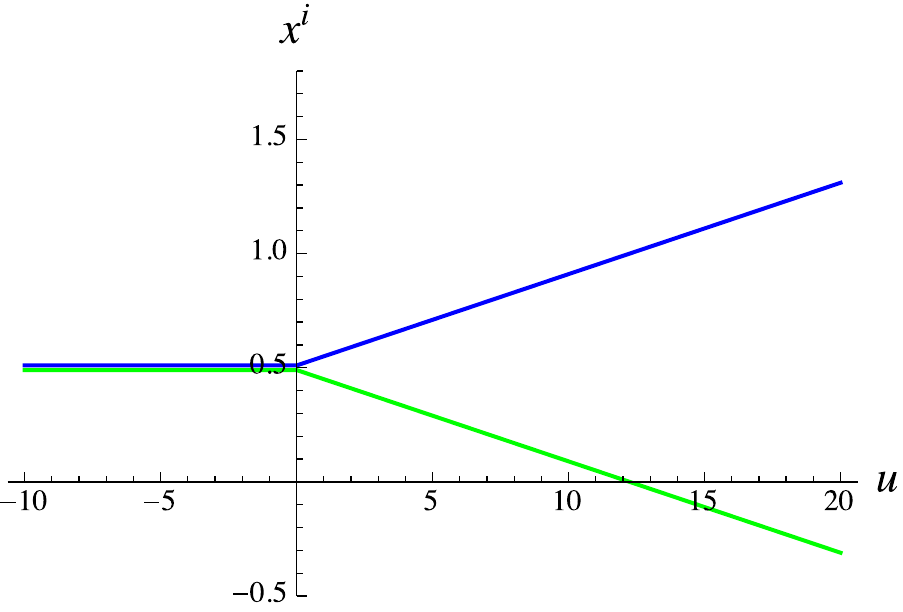} \hskip1cm
\includegraphics[scale=0.7]{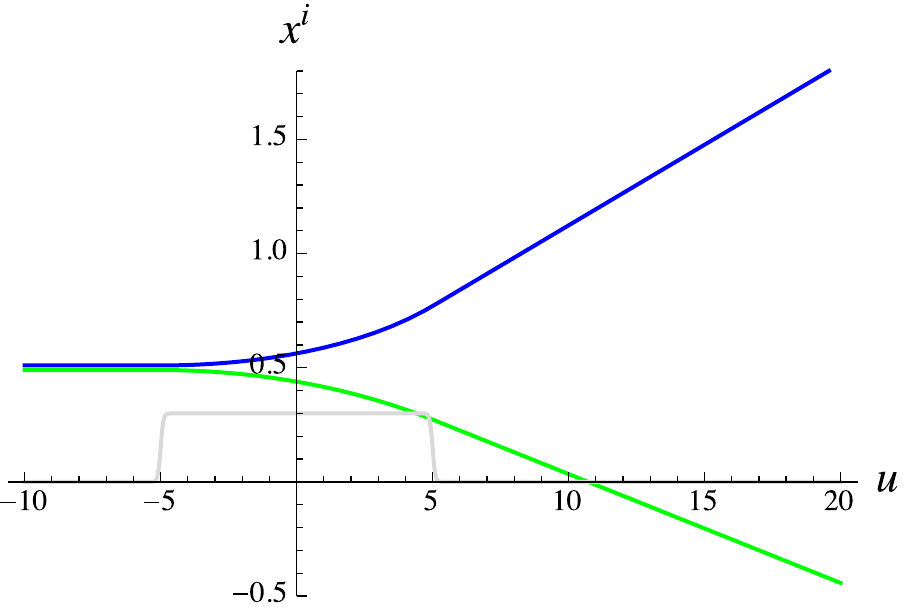}
\caption{These plots show the behaviour of the transverse coordinates $x^1(u)$ and $x^2(u)$ of a geodesic in the 
neighbourhood of $\c$ (with $x^i = 0$) through the encounter with a  gravitational wave burst. The green curves
denote $x^1(u)$ and exhibit an initial focusing followed by divergence, while the blue curves denoting
$x^2(u)$ show a divergence.
The left-hand figure refers to the Penrose limit plane wave with impulsive profile $\chi_F(u) = \d(u)$, 
while the right-hand figure describes the case of an extended profile $\chi_F(u) = \Theta(u,L)$.}
\label{fig3.5}
\end{figure}

The behaviour of the Tissot ring is illustrated in Fig.~\ref{fig3.6}.
\begin{figure}[h]
\centering
\includegraphics[scale=0.6]{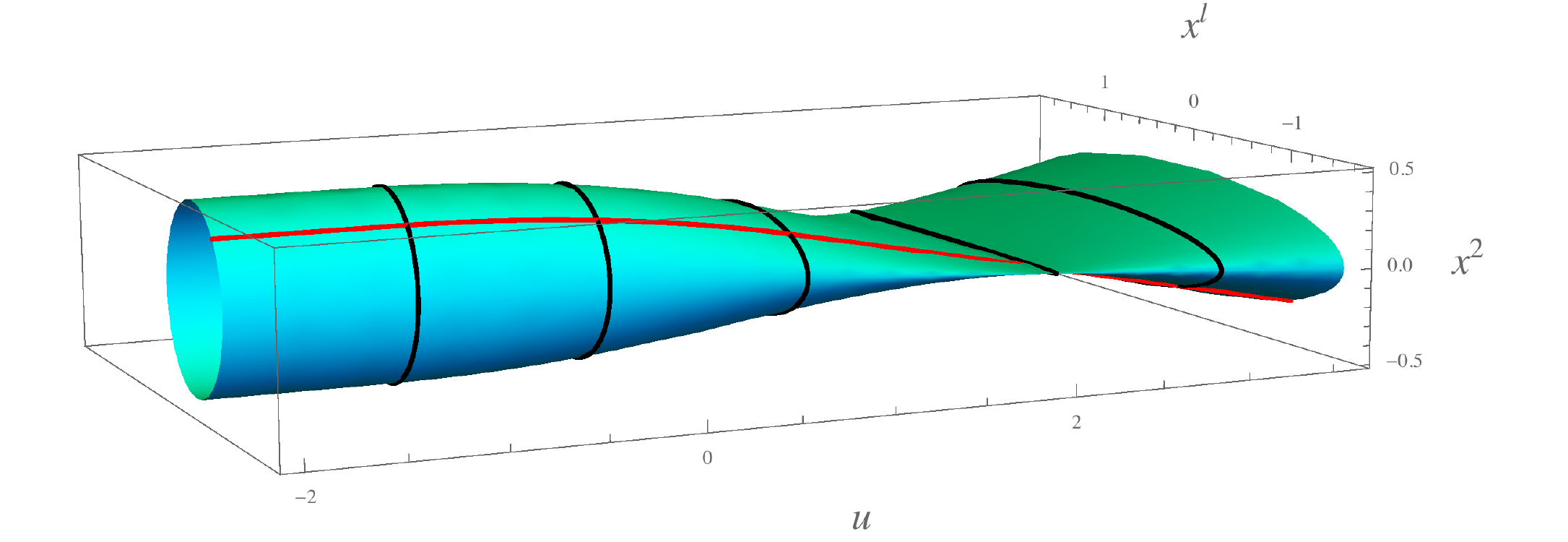} \hskip0.3cm
\caption{Illustration of the evolution of the Tissot ring of geodesics through the passage of an extended
plane-wave burst. The red curve shows a single geodesic described by $(x^1,x^2)$ as in Fig.~\ref{fig3.5}.
The combination of expansion $\hat{\theta}$ and shear $\hat{\s}_+$ causes the Tissot circle 
to deform to an ellipse, degenerate to a line, then form an expanding ellipse in the far memory region
following the gravitational wave burst.
(Note that the $u$ axis has been rescaled by a factor 5 compared
to the previous figures for clarity.) }
\label{fig3.6}
\end{figure}
This clearly shows the combination of shear and (initially negative) expansion described above.
The initial circle squashes due to the + oriented shear, and contracts up to the point in $u$ where 
the geodesics in the $x^1$ direction focus on to the original geodesic $\c$ (with $x^i(u) = 0$ by definition)
before diverging as the denominator in (\ref{c28}) changes sign. For the impulsive shockwave, this degenerate
line is reached at $u = b^2/4G\m$. 

In terms of memory, it is evident from Figs.~\ref{fig3.5} and \ref{fig3.6} that after the passage of the gravitational
wave burst, there is a velocity-encoded memory with neighbouring geodesics eventually diverging with
straight-line trajectories. In the case of the extended-profile wave, there is in addition a shift in position
in the geodesics compared immediately before and after the interaction with the gravitational wave burst. 
This is in accord with the general expectations discussed in section \ref{sect 2.6}.

\section{Gyratons}\label{sect 4}

\subsection{Gyraton metric}

The special form of the gyraton metric we consider here is the simplest extension of the Aichelburg-Sexl
shockwave to accommodate a spinning source \cite{Frolov:2005zq, Frolov:2005in, Podolsky:2014lpa}. 
The motivations for this choice are described briefly in 
Appendix \ref{Appendix C} together with a discussion of more general gyraton metrics.

The metric in the vacuum region outside the spinning source centred at $r=0$ is
\begin{equation}
ds^2 ~=~ 2du dv + f(r) \chi_F(u) du^2 - 2J \chi_J(u) du d\phi + dr^2 + r^2 d\phi^2 \ ,
\label{d1}
\end{equation}
where $\chi_F(u)$ and $\chi_J(u)$ are profile functions, which in this case
may be chosen independently. The angular momentum of the source is proportional to $J$.

The non-vanishing Christoffel symbols of this gyraton metric are
\begin{align}
\Gamma^v_{uu} &= \half f(r) \chi_F'(u) - \frac{J^2}{r^2} \chi_J(u) \chi_J'(u) \ , ~~~~~~
\Gamma^v{}_{ur} = \half f'(r) \chi_F(u) \ , ~~~~~~~ \Gamma^v{}_{r\phi} = \frac{J}{r} \chi_J(u) \ , \nonumber \\
\Gamma^r_{uu} &= -\half f'(r) \chi_F(u) \ , ~~~~~~~~~ \Gamma^r_{\phi \phi} = -r \ , ~~~~~~~~
\Gamma^\phi_{uu} = - \frac{J}{r^2} \chi_J'(u) \ , ~~~~~~~~ \Gamma^\phi_{r\phi} = \frac{1}{r} \ ,
\label{d2}
\end{align}
while the curvature components  are
\begin{equation}
R_{ruru} = -\half f''(r) \chi_F(u) \ , ~~~~~~~~  R_{\phi u \phi u} = - \half r f'(r) \chi_F(u) \ , ~~~~~~~~
R_{r u \phi u} = \frac{J}{r} \chi'_J(u) \ , ~~~~~~~~
\label{d3}
\end{equation}
and
\begin{equation}
R_{uu} = -\half \Delta f(r) \chi_F(u) = 0 \ ,
\label{d4}
\end{equation}
in the vacuum region where $\Delta f(r) = 0$ as in the Aichelburg-Sexl case.

The fact that the profiles $\chi_F(u)$ and $\chi_J(u)$ may be chosen independently is actually a consequence 
of the cylindrical symmetry we have assumed for the metric. In the more general case, the Einstein equations
link $\chi_F(u)$ and $\chi_J(u)$ (see Appendix \ref{Appendix A}) and there is a constraint $\chi_F(u) \sim \chi_J'(u)$,
which raises issues with the null energy condition.

Now as we show below, the impulsive choice $\chi_J(u) = \d(u)$ is of relatively little interest as the effect
on a test particle is merely to give it a sideways kick. On the other hand, an extended $\chi_J(u)$ typical
of a sandwich wave produces an orbital motion in the geodesics.  It is of particular interest to verify
explicitly how, despite this orbital motion for a single geodesic, the corresponding congruence 
does not acquire a non-vanishing twist, in accordance with the general theory of section \ref{sect 2.4}.
We therefore choose
\begin{equation}
\chi_J(u) = \Theta(u,L)  \ ,
\label{d5}
\end{equation}
with $\Theta(u,L)$ as in (\ref{c11}), together with the smoothed form for some of the numerical plots. 
For the most part, we also use the same form for the profile $\chi_F(u)$, rather than the impulsive limit.

\subsection{Geodesics and orbits}

We now study the geodesics, null and timelike, for the gyraton metric (\ref{d1}), extending our earlier 
analysis for the spinless gravitational shockwave. 

The geodesics are
\begin{align}
&\ddot{v} + \half f(r) \,\chi_F'(u) - \frac{J^2}{r^2} \,\chi_J(u) \chi_J'(u) + f'(r) \,\chi_F(u) \dot{r} 
+ \frac{2J}{r} \,\chi_J(u) \dot{r} \dot{\phi} = 0 \ , \nonumber \\
&\ddot{r} - \half f'(r) \,\chi_F(u) - r \dot{\phi}^2 = 0 \ ,  \nonumber \\
&\ddot{\phi} - \frac{J}{r^2} \,\chi_J'(u) + \frac{2}{r} \dot{r} \dot{\phi} = 0 \ , 
\label{d6}
\end{align}
where we have immediately exploited the $v$-translation symmetry of the metric, which implies the
geodesic equation $\ddot{u} = 0$, to choose $u$ as the affine parameter.
It is usually simpler to use the integrated form of the $v$ equation directly from the metric, {\it viz.}
\begin{equation}
2\dot{v} +  f(r) \,\chi_F(u) +  \dot{r}^2  -  \frac{J^2}{r^2} \,\chi_J^2(u) = 2\eta    \ ,
\label{d7}
\end{equation}
where $\eta = 0 ~ (<0)$ for a null (timelike) geodesic and we have already used (\ref{d8}).

The azimuthal equation is also immediately integrable, giving the angular momentum $\ell(u)$ as
\begin{equation}
\ell(u) = r^2 \dot{\phi}  = J \chi_J(u) \ .
\label{d8}
\end{equation}
Substituting back into the radial geodesic gives a characteristic orbit equation
\begin{equation}
\ddot{r} + \frac{\partial}{\partial r} V_{\rm eff}(u,r) = 0 \ ,
\label{d9}
\end{equation}
with
\begin{equation}
V_{\rm eff} (u,r) = -\half f(r)\, \chi_F(u) + \half \frac{J^2}{r^2} \,\chi_J(u)^2 \ ,
\label{d10}
\end{equation}
independently of whether the geodesic is null or timelike.

Now, we can readily see that in order to find solutions for which the test particle exhibits orbital motion
rather than simply receiving a kick at first encounter with the gyraton and a second kick as it passes\footnote{
For example, if we take $f(r) = 0$ and just keep the angular momentum term with $\chi_J(u) = \Theta(u,L)$,
we can solve the geodesic equations exactly in the region $-L/2 < u < L/2$, giving
\begin{align*}
v(u) &= v_0 + \left( \eta - \half \frac{J^2}{b^2} \right) \tilde{u} + J \arctan\left(\frac{J\tilde{u}}{b^2}\right) \ , \nonumber \\
r(u) &= b\left(1 + \frac{J^2 \tilde{u}^2}{b^4}\right)^{1/2} \ , \nonumber \\
\phi(u) &= \arctan \left(\frac{J\tilde{u}}{b^2}\right) \ ,
\end{align*}
for impact parameter $b$ and initial $v=v_0$. Here, $\tilde{u} = u + L/2$. This describes a straight line
trajectory.}, 
we need {\it both} profiles to be extended. Choosing both $\chi_F(u)$ and $\chi_J(u)$ to be $\Theta(u,L)$,
and considering a particle source for the gyraton shockwave, we then have
\begin{equation}
V_{\rm eff}(u,r) ~=~ \left( 4GE \log\frac{r}{r_0} + \half \frac{J^2}{r^2} \right) \Theta(u,L) ~~\equiv~~ 
\tilde V_{\rm eff}(r) \,\Theta(u,L) \ .
\label{d11}
\end{equation}
This becomes a typical central force problem with a logarithmic attractive potential provided by $f(r)$
and gives a bound orbit in the region of $u$ where $\Theta(u,L) = 1$.
For a central potential, Bertrand's theorem states that every bound orbit is periodic for potentials proportional
to $r^2$ or $1/r$ only. So the orbit corresponding to (\ref{d11}) will precess (in contrast to that with a 
homogeneous beam source for the gyraton, where $f(r) \sim r^2$ and we find a stable, closed orbit).

To be more explicit, integrating (\ref{d9}) gives
\begin{align}
\frac{d}{du} \left( \half \dot{r}^2 + V_{\rm eff}(u,r) \right) &~=~ \frac{\partial}{\partial u} V_{\rm eff}(u,r) \nonumber \\
&~\equiv~ \tilde V_{\rm eff}(r) \left( \d\left(u + \tfrac{L}{2}\right) - \d\left(u - \tfrac{L}{2} \right) \right) \ .
\label{d12}
\end{align}
So away from the initial and final kicks from the straight line trajectories for the initial region $u < - L/2$
and into the `memory' region $u>L/2$ after the passage of the gyraton, we have
\begin{equation}
\half\dot{r}^2 + \tilde V_{\rm eff}(r) = E\ , ~~~~~~~~~~~~\left(- L/2 < u < L/2 \right)
\label{d13}
\end{equation}
with $E = {\rm const.}$ 

For the logarithmic potential characterising the particle-source gyraton, we do not have 
analytic expressions for the geodesic orbits, so we illustrate the key features with numerical solutions.
A typical orbiting solution is shown in Fig.~\ref{fig4.1} (for a slightly smoothed approximation
to $\Theta(u,L)$), clearly showing the precessing orbit and the final kick at $u=L/2$ into the memory region.
Fig.~\ref{fig4.2} shows the form of the geodesic as it evolves with the lightcone coordinate $u$.
\begin{figure}[h]
\centering
\includegraphics[scale=0.8]{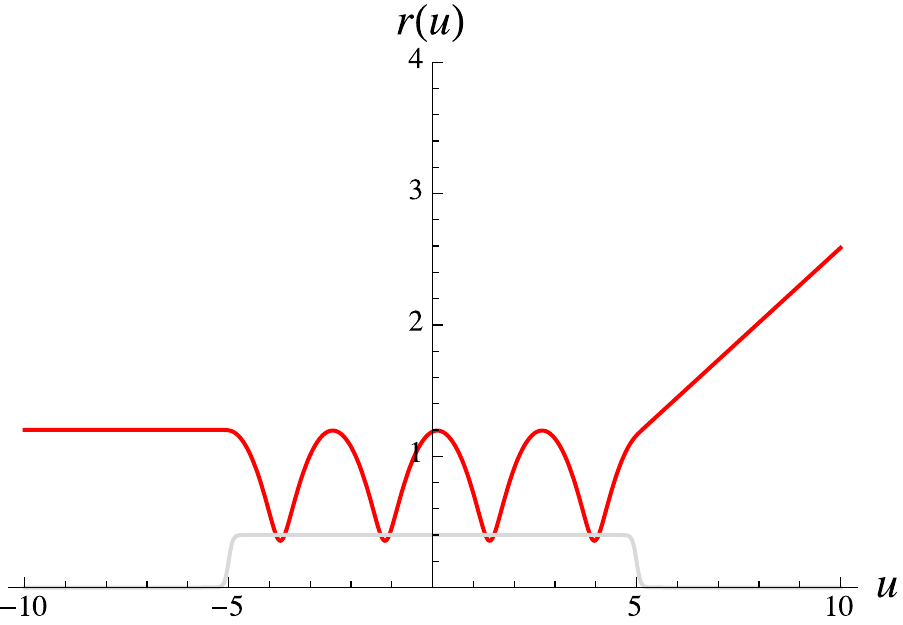} \hskip0.7cm 
\includegraphics[scale=0.8]{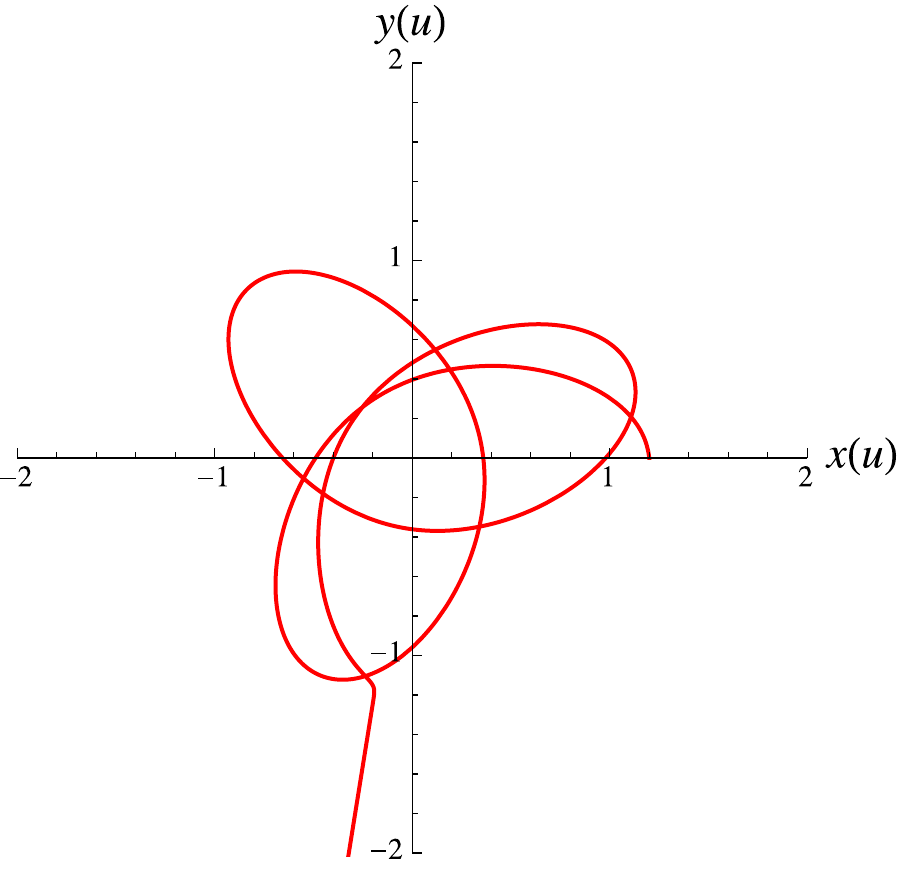} 
\caption{The left-hand plot shows the behaviour of $r(u)$ through the encounter with a gyraton with 
profile $\chi_F(u) = \chi_J(u) = \Theta(u,L)$, shown here with $L=10$. The right-hand plot shows the 
test particle following a precessing orbit around the gyraton axis before emerging as a straight line in the
memory region after the passage of the gyraton. }
\label{fig4.1}
\end{figure}

\begin{figure}[h]
\centering
\includegraphics[scale=0.65]{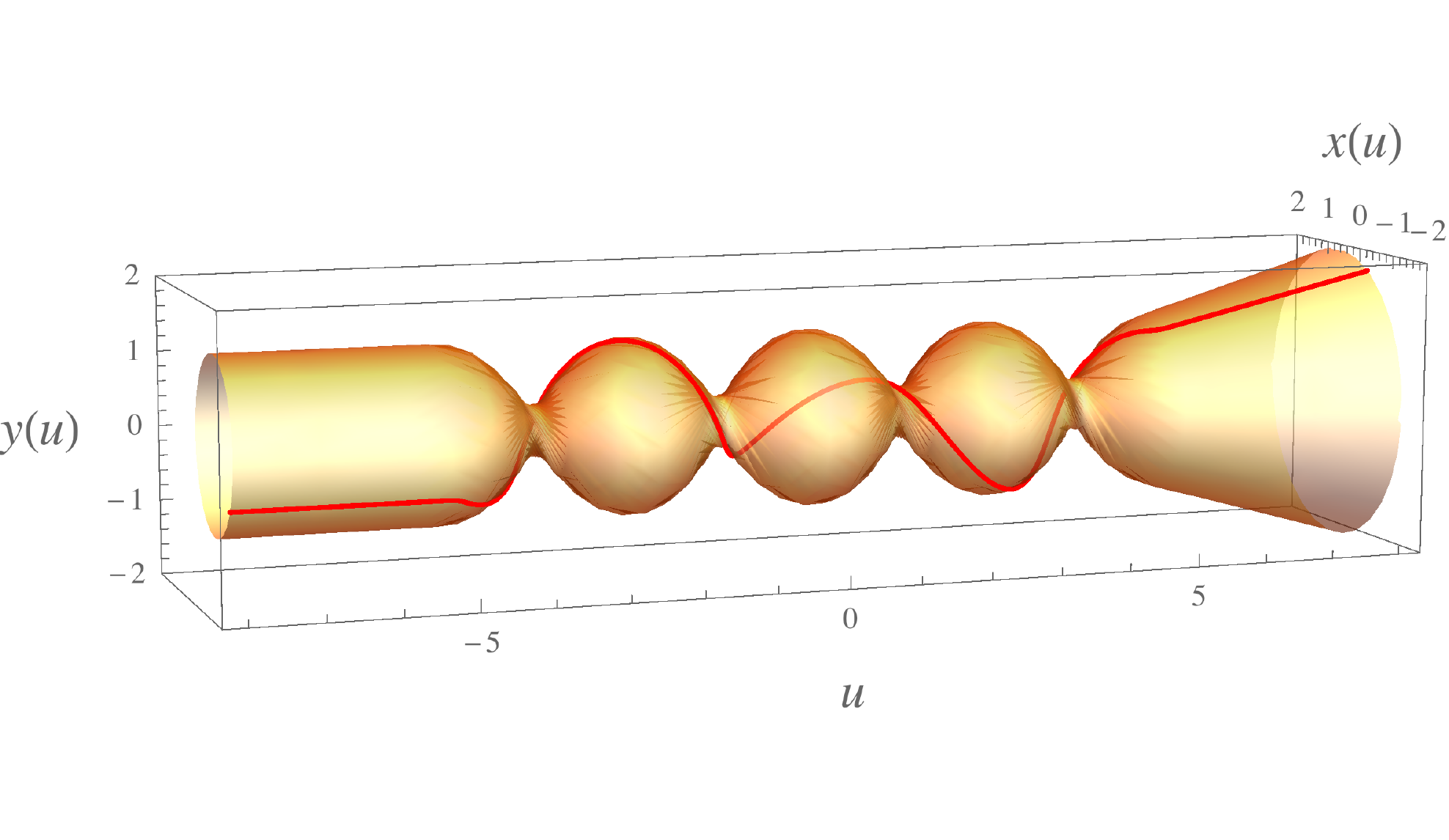} 
\caption{The red curve shows a single geodesic orbiting around the gyraton centred on $r=0$. The shaded
envelope is the set of geodesics with the same impact parameter $b$ but different initial angles $\phi$
to the gyraton axis.}
\label{fig4.2}
\end{figure}

All this clearly illustrates the difference between the geodesics in the Aichelburg-Sexl shockwave and the
gyraton. While the initial and final trajectories are of course straight lines with the test particle being
deflected by its encounter with the gyraton, the angular momentum of the gyraton metric induces
an orbital motion for the test particle geodesics in the region where the gyraton profiles $\chi_F(u)$
and $\chi_J(u)$ are non-vanishing. 

In Fig.~\ref{fig4.v}, we show the analogue of the jump in the lightcone coordinate $v(u)$ we found for the
Aichelburg-Sexl or extended shockwave in Fig.~\ref{fig3.4}, for different values of the angular momentum
parameter $J$ in the gyraton metric. Naturally, for a trajectory covering several orbits, $v(u)$ reflects the
oscillations in $r(u)$. Note also that depending on the metric parameters, the jump in $v(u)$ as the gyraton
passes may have either sign. See Appendix \ref{Appendix A} for a brief discussion of the relevance of the 
jump $\Delta v$ in ultra-high energy gravitational scattering.

\begin{figure}[h]
\centering
\includegraphics[scale=0.8]{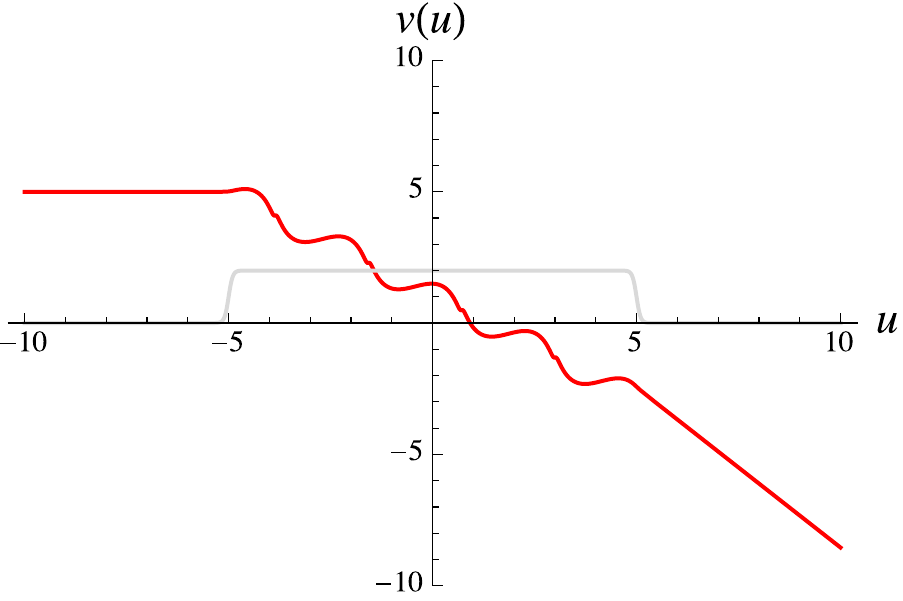} \hskip0.7cm 
\includegraphics[scale=0.8]{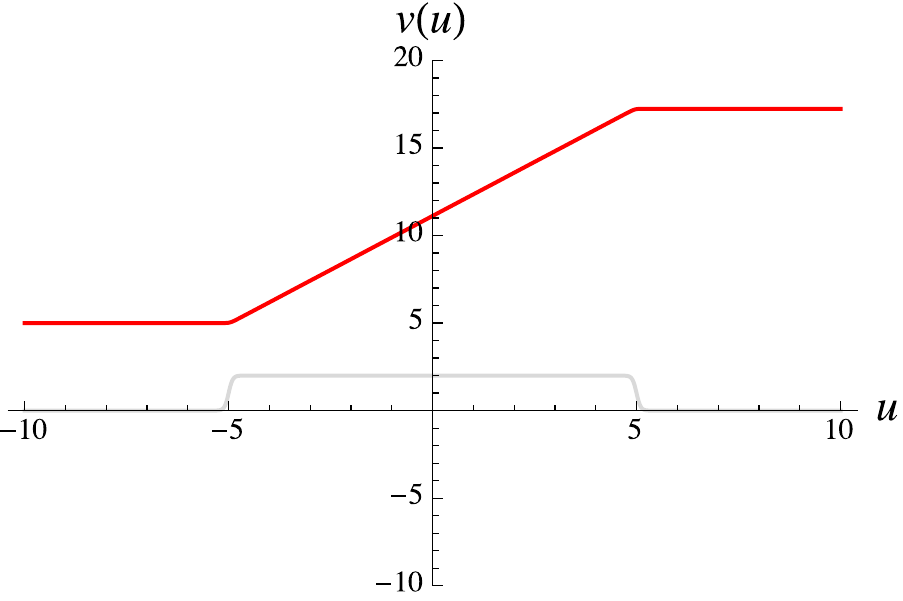} 
\caption{Plots of the lightcone coordinate $v(u)$ through the encounter with a gyraton. The left-hand plot 
is for relatively small angular momentum $J$, with the bumpy change in $v(u)$ reflecting the number of orbits.
The right-hand plot is for bigger $J$ and covers a single orbit. }
\label{fig4.v}
\end{figure}

With this description of the behaviour of an individual geodesic in the gyraton background,  we now move on
to analyse the congruence in the neighbourhood of such a geodesic, in particular to see
whether this rotation is inherited in the optical tensors in the form of memory with twist.

\subsection{Optical tensors and memory} 

The first step in calculating the optical tensors for the gyraton background is to evaluate 
$\Omega_{\m\n} = D_\n k_\m$, where $k^\m$ is the tangent vector field corresponding to the geodesics
in a congruence based on the solutions described above.

With the gyraton metric, we have 
\begin{equation}
k^\m = \begin{pmatrix} \,&1\, \\ \,&\dot{v}\, \\ \,&\dot{r}\, \\ \,&\dot{\phi}\, \end{pmatrix} \ , ~~~~~~~~~~~~
k_\m = g_{\m\n} k^\n = \begin{pmatrix} \, &\dot{v} - J\chi_J(u) \dot{\phi} + f(r) \chi_F(u)\, \\
\, &1\, \\ \,&\dot{r}\, \\ \,&r^2\dot{\phi} - J \chi_J(u) \, \end{pmatrix} \ ,
\label{d14}
\end{equation}
where $k^\m$ is given by the first integrals of the geodesic equations (\ref{d6}). From (\ref{d7}) and (\ref{d8}) we
can immediately express $\dot{v}$ and $\dot{\phi}$ in terms of $\dot{r}$, since
\begin{align}
\dot{v} &= -\half \dot{r}^2 - \half f(r)\,\chi_F(u) + \half \frac{J^2}{r^2}\,\chi_J(u)^2 + \eta \ , \nonumber \\
\dot{\phi} &= \frac{J}{r^2} \,\chi_J(u) \ .
\label{d15}
\end{align}
In particular, this gives
\begin{align}
k_u &= -\half \dot{r}^2 + \half f(r)\,\chi_F(u) - \half \frac{J^2}{r^2} \,\chi_J(u)^2 + \eta \ , \nonumber \\
k_\phi &= 0 \ ,
\label{d16}
\end{align}
while $k_v = 1$, $k_r = \dot{r}$ as in (\ref{d14}).

Given the Christoffel symbols for the gyraton metric in (\ref{d2}), we may now evaluate $\Omega_{\m\n}$.
After some calculation, we find that $\Omega_{\m\n}$ may be expressed in the form
\begin{equation}
\Omega_{\m\n} = \begin{pmatrix} \,&\Omega_{uu} ~&0 ~&\Omega_{ur} ~&0\, \\
\,&0 ~&0 ~&0 ~&0\, \\
\,&\Omega_{ru} ~&0 ~&\Omega_{rr} ~&\Omega_{r\phi}\, \\
\,&0 ~&0 ~&\Omega_{\phi r} ~&\Omega_{\phi\phi}\, \end{pmatrix} \ ,
\label{d17}
\end{equation}
with 
\begin{align}
\Omega_{uu} &= \dot{r} \,\Omega_{rr} \,\dot{r} + \dot{r}\,\Omega_{r\phi}\,\dot{\phi} + \dot{\phi}\,\Omega_{\phi r}\,\dot{r} 
+ \dot{\phi}\, \Omega_{\phi \phi}\, \dot{\phi}  \ ,  \nonumber \\
\Omega_{ur} &= \Omega_{ru} = - \Omega_{rr}\,\dot{r} - \Omega_{r\phi} \,\dot{\phi} \ , \nonumber \\
\Omega_{rr} &= \partial_r \dot{r} \ , ~~~~~~~~ \Omega_{r\phi} = \Omega_{\phi r} = - r\dot{\phi} \ , ~~~~~~~~
\Omega_{\phi\phi} = r \dot{r} \ ,
\label{d18}
\end{align}
where $\dot{\phi} = J\chi_J(u)/r^2$. At this point, we have not yet had to specify $\dot{r}$, and the result
holds for both null and timelike congruences.

Next we need to find a basis for Fermi normal coordinates. We show this explicitly for the null congruence, 
with FNCs for the timelike congruence being constructed similarly as described in section~\ref{sect 3.2}.
These give the same result for the optical tensors in the effectively two-dimensional transverse space.

It is relatively straightforward to see that an appropriate basis which satisfies the required orthonormality conditions
(\ref{b5}) {\it at a point} is (compare (\ref{c19}) for the spinless shockwave),
\begin{equation}
\tilde{e}^{u\m} = \begin{pmatrix} \,&1\,\\ \,&\dot{v}\,\\ \,&\dot{r}\,\\ \,&\dot{\phi}\, \end{pmatrix} \ , ~~~~~~~~
\tilde{e}^{v\m} = \begin{pmatrix} \,&0\,\\ \,&-1\,\\ \,&0\,\\ \,&0\, \end{pmatrix} \ , ~~~~~~~~
\tilde{e}^{1\m} = \begin{pmatrix} \,&0\,\\ \,&-\dot{r}\,\\ \,&1\,\\ \,&0\, \end{pmatrix} \ , ~~~~~~~~
\tilde{e}^{2\m} = \begin{pmatrix} \,&0\,\\ \,&J\chi_J(u)/r - r\dot{\phi}\,\\ \,&0\,\\ \,&1/r\, \end{pmatrix} \ ,
\label{d19}
\end{equation}
to be compared with (\ref{c19}) for the spinless shockwave. 
However, this basis is {\it not} parallel transported along the chosen geodesic $\c$ with tangent vector $k^\m$. 
While $k.D \tilde{e}^{u\m} =0$ and $k.D \tilde{e}^{v\m} = 0$, a short calculation shows that in fact
\begin{equation}
k.D \tilde{e}^{1\m} = \dot{\phi}\,\tilde{e}^{2\m} \ , ~~~~~~~~~~~~ 
k.D \tilde{e}^{2\m} = -\dot{\phi}\,\tilde{e}^{1\m} \ .
\label{d20}
\end{equation}
It follows that the correct choice of FNC basis with $k.D e^{1\m} =0$ and $k.D e^{2\u}=0$ is a rotated set
defined by
\begin{equation}
\begin{pmatrix} \,&e^{1\m}\, \\ \,&e^{2\m}\, \end{pmatrix} ~=~
\begin{pmatrix}  \,&\cos\phi ~~&-\sin\phi \, \\ \,&\sin\phi ~~&\cos\phi\, \end{pmatrix} \,
\begin{pmatrix} \,&\tilde{e}^{1\m}\, \\ \,&\tilde{e}^{2\m}\, \end{pmatrix} \ , 
\label{d21}
\end{equation}
that is,
\begin{equation}
e^{i\m} ~=~ O^i{}_j \, \tilde{e}^{j\m} \ ,
\label{d22}
\end{equation}
where $O_{ij},~i,j=1,2$ is the orthogonal matrix in (\ref{d21}). Note that in $O_{ij}$, the angle $\phi(u)$ is a solution
of the geodesic equation, $\dot{\phi}(u) = J \chi_j(u) /r(u)$.

Now, following the construction described in section \ref{sect 2.2}, we define the optical tensors from the
projection
\begin{equation}
\hat{\Omega}^{ij} ~=~ e^{i\m}\,\Omega_{\m\n}\,e^{j\n} \ ,
\label{d23}
\end{equation}
with the basis vectors defined in (\ref{d22}). We find,
\begin{equation}
\hat{\Omega}_{ij} ~=~ O\, \begin{pmatrix} \,&\Omega_{rr} ~~&\frac{1}{r} \Omega_{r\phi} \, \\
\,&\frac{1}{r} \Omega_{\phi r} ~~&\frac{1}{r^2} \Omega_{\phi\phi} \, \end{pmatrix} \, O^T \ .
\label{d24}
\end{equation}

This is a very natural generalisation of $\hat{\Omega}_{ij}$ for the ordinary shockwave to incorporate the
spin inherent in the gyraton spacetime. This is evident first in the appearance of the off-diagonal terms
$\hat{\Omega}_{r\phi} = \hat{\Omega}_{\phi r}$, and in the $\phi(u)$-dependent rotation of the FNC basis.
Writing (\ref{d24}) in full we therefore have
\begin{equation}
\hat{\Omega}_{ij} ~=~ \begin{pmatrix}  \,&\cos\phi ~~&-\sin\phi \, \\ \,&\sin\phi ~~&\cos\phi\, \end{pmatrix}\,
\begin{pmatrix} \,&\partial_r \dot{r} ~~&- \frac{J}{r} \chi_J(u) \, \\ 
\,&-\frac{J}{r} \chi_J(u) ~~&\frac{1}{r}\dot{r} \, \end{pmatrix}
\,\begin{pmatrix}  \,&\cos\phi ~~&\sin\phi \, \\ \,&-\sin\phi ~~&\cos\phi\, \end{pmatrix}\ .
\label{d25}
\end{equation}
To interpret this, recall that $r(u)$, $\dot{r}(u)$ and $\phi(u)$ are the solutions of the geodesic equations
{\it for the chosen geodesic} $\c$, which we take as the null geodesic with initial conditions $r=b$,
$\phi = 0$.
The optical tensors -- expansion, shear and twist -- are then read off from (\ref{d25}) with the usual definitions,
\begin{equation}
\hat{\Omega}_{ij} = \half \hat{\theta}\, \d_{ij} + \hat{\s}_{ij} + \hat{\omega}_{ij} \ .
\label{d26}
\end{equation}

We see immediately that $\hat{\Omega}_{ij}$ is symmetric, so the twist $\hat{\omega}_{ij}$ vanishes. Even in the
gyraton background, the fact that an individual geodesic orbits around the source does {\it not} imply a 
relative rotation of neighbouring geodesics in the congruence.

The expansion is given by the trace of $\hat{\Omega}_{ij}$, so we simply find
\begin{equation}
\hat{\theta} = \tr\,\hat{\Omega}_{ij} = \partial_r \dot{r} + \dot{r}/r \ ,
\label{d27}
\end{equation}
since the rotation of the FNC basis plays no r\^ole.
The presence here of the off-diagonal terms proportional to $\Omega_{r\phi} = - J\chi_J(u)/r$ however means 
that in this case we have non-vanishing shear in {\it both} $+$ and $\times$ orientations. Of course, since 
$\hat{\Omega}_{ij}$ is symmetric, it can be diagonalised to find a rotating basis in which the shear is 
non-vanishing in a single orientation only -- however, this does {\it not} coincide with the basis defining the FNC
coordinates. Explicitly,
\begin{align}
\hat{\s}_+ &= \half\left(\partial_r\,\dot{r} - \frac{1}{r} \dot{r}\right) \cos 2\phi  + \frac{J}{r} \chi_J(u) \sin 2\phi \ ,
\nonumber \\
\hat{\s}_{\times} &= \half\left(\partial_r\,\dot{r} - \frac{1}{r} \dot{r}\right) \sin 2\phi - \frac{J}{r} \chi_J(u) \cos 2\phi \ .
\label{d28}
\end{align}

To evaluate further we would need to find explicit solutions for $r(u)$ and $\phi(u)$ along the geodesic $\c$ 
and carry through an analysis analogous to section \ref{sect 3.2.3}.
These are not known in analytic form for a logarithmic central potential. Instead, we first re-express these 
results in terms of the Penrose limit, then study the behaviour of the congruences numerically.

\subsection{Penrose limit and memory}

The Penrose limit is now readily found given the gyraton curvature tensors (\ref{d3}) and the
FNC basis (\ref{d19}), (\ref{d22}).  Recall that for the null geodesic $\c$,\footnote{The timelike case follows 
in the same way as in section \ref{sect 3.3}. The fact that the gyraton is also a pp wave again means that the 
$v$-components of the curvature tensor vanish, so the three-dimensional $h_{rs}$ in section \ref{sect 2.3}
degenerates to a two-dimensional $h_{ij}$ identical to that considered here for the null congruence.}
the Penrose limit metric is the plane wave,
\begin{equation}
ds^2 = 2 du dv + h_{ij}(u)\, x^i\,x^j\,du^2 + \d_{ij}\,dx^i\,dx^j \ ,
\label{d30}
\end{equation}
with profile function,
\begin{align}
h_{ij} ~&=~ - R_{\r u \s u}\,e^{i\r}\, e^{j\s}  \nonumber \\
&= - O~\begin{pmatrix} \, &R_{ruru} ~~ &\frac{1}{r} R_{r u \phi u} \, \\
\,&\frac{1}{r} R_{\phi u r u} ~~&\frac{1}{r^2} R_{\phi u \phi u} \, \end{pmatrix} \, O^T \ ,
\label{d31}
\end{align}
with $O$ defined in (\ref{d21}), (\ref{d22}). Explicitly,
\begin{equation}
h_{ij} ~=~ O(\phi) \, \begin{pmatrix}
\,&\half f''(r)\,\chi_F(u) ~~&-\frac{J}{r^2}\,\chi_J'(u) \, \\
&{}&{} \\
\,&-\frac{J}{r^2}\,\chi_J'(u) ~~&\half \frac{f'(r)}{r}\,\chi_F(u) \,\end{pmatrix} \, O^T(\phi) \ .
\label{d32}
\end{equation}

Now according to the general theory in section \ref{sect 2}, we should have
\begin{equation}
h_{ij} = \frac{d}{du}\,\hat{\Omega}_{ij} + \hat{\Omega}^2_{ij} \ ,\label{d33}
\end{equation}
with $\hat{\Omega}_{ij}$ as in (\ref{d25}). To verify this, note first that 
\begin{equation}
\frac{d}{du}\,\hat{\Omega}_{ij} = O \left( \frac{d}{du}\,\tilde{\Omega} - \left[\e,\tilde{\Omega}\right]\,\dot{\phi}
\right)\,O^T \ ,
\label{d34}
\end{equation}
where $\e_{ij}$ is the antisymmetric symbol and we use the temporary notation 
$\hat{\Omega} = O\,\tilde{\Omega} \,O^T$.
We can then verify (\ref{d33}) component by component.  Equation (\ref{d34}) implies 
$h = O\,\tilde{h}\, O^T$ with, for example,
\begin{align}
\tilde{h}_{12} &= \frac{d}{du}\,\left(\frac{1}{r}\Omega_{r\phi} \right) + 
\left(\Omega_{rr} - \frac{1}{r^2} \Omega_{\phi\phi}\right)\,\dot{\phi} + \frac{1}{r} \Omega_{rr} \Omega_{r\phi} 
+ \frac{1}{r^3} \Omega_{r\phi} \Omega_{\phi\phi} \nonumber \\
&= \ddot{\phi} + \left(\partial_r \dot{r} - \frac{1}{r} \dot{r} \right) \,\dot{\phi} - \dot{\phi} \,\partial_r \dot{r} 
- \frac{1}{r} \,\dot{r}\,\dot{\phi} \nonumber \\
&= - \ddot{\phi} - \frac{2}{r}\,\dot{r}\,\dot{\phi}  \nonumber \\
&= - \frac{J}{r^2}\,\chi_J'(u) \ ,
\label{d35}
\end{align}
using the geodesic equation (\ref{d6}) in the final step.
The other components follow similarly and we confirm the link between the derivatives of the optical tensors
found directly from $\hat{\Omega}_{ij}$ and the geodesic congruences in the Penrose plane wave limit.
These are found by solving the plane wave geodesic equations,
\begin{equation}
\ddot{x}^i - h^i{}_j(u) \,\dot{x}^j = 0 \ ,
\label{d36}
\end{equation}
wuth $h_{ij}$ defined in (\ref{d32}). We have solved these equations numerically for the extended profiles
$\chi_F(u) = \chi_J(u) = \Theta(u,L)$, and the particle source $f(r) = -4G E \,\log r^2/r_0^2$. 

The results are illustrated in the following figures. Fig.~\ref{fig4.3} shows the behaviour of the transverse
coordinates for a member of the geodesic congruence as the gyraton passes through. We have chosen parameters
so that the evolution of $(x^1,x^2)$ shown covers a single orbit of the original geodesic $\c$ around the
gyraton axis. The right-hand plot shows the how the transverse position of the geodesic, {\it i.e.} the
connecting vector, evolves. Clearly, there is a {\it position} shift from before to after the encounter with
the gyraton. Subsequently the geodesic follows a straight line, exhibiting {\it velocity-encoded} memory.

\begin{figure}[h]
\centering
\includegraphics[scale=0.75]{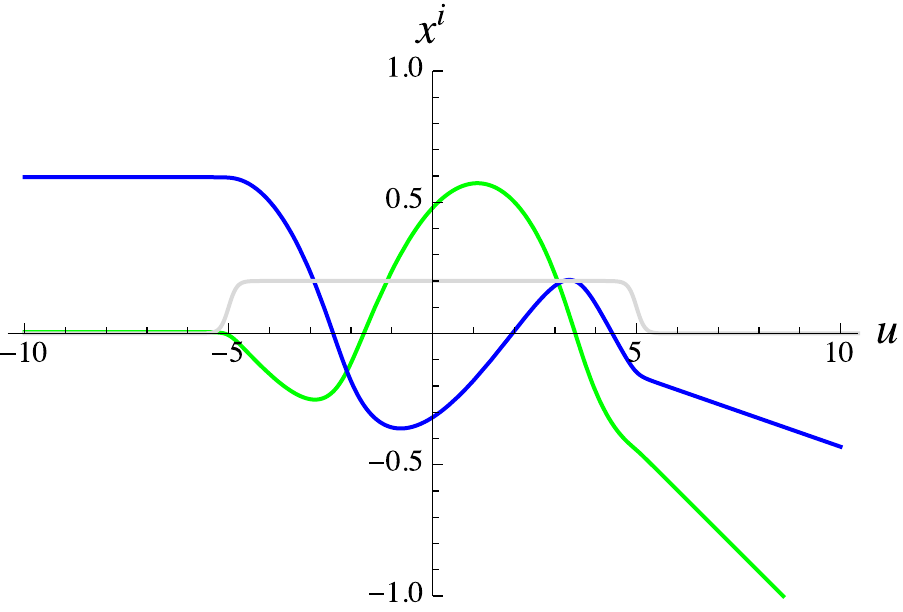} \hskip1cm
\includegraphics[scale=0.65]{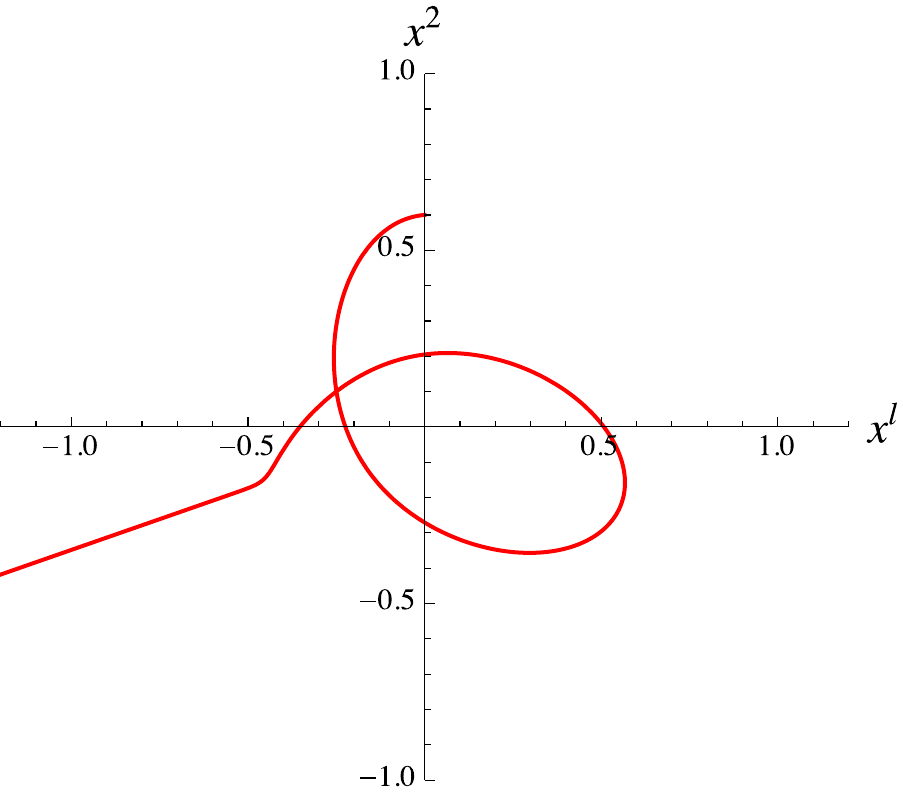}
\caption{The left-hand plot shows the behaviour of the transverse coordinates $x^1(u)$ (in green) and 
$x^2(u)$ (in blue) of a geodesic in the neighbourhood of $\c$ through the passage of the gyraton,
shown here with profile $\chi_F(u) = \chi_J(u) = \Theta(u,L)$ with $L=10$. Parameters are chosen such that
the reference geodesic $\c$ makes one orbit of the gyraton axis between $u=-5$ and $u=5$.
The right-hand plot shows this motion in the transverse $(x^1,x^2)$ plane.  }
\label{fig4.3}
\end{figure}

\begin{figure}[h]
\centering
\includegraphics[scale=0.6]{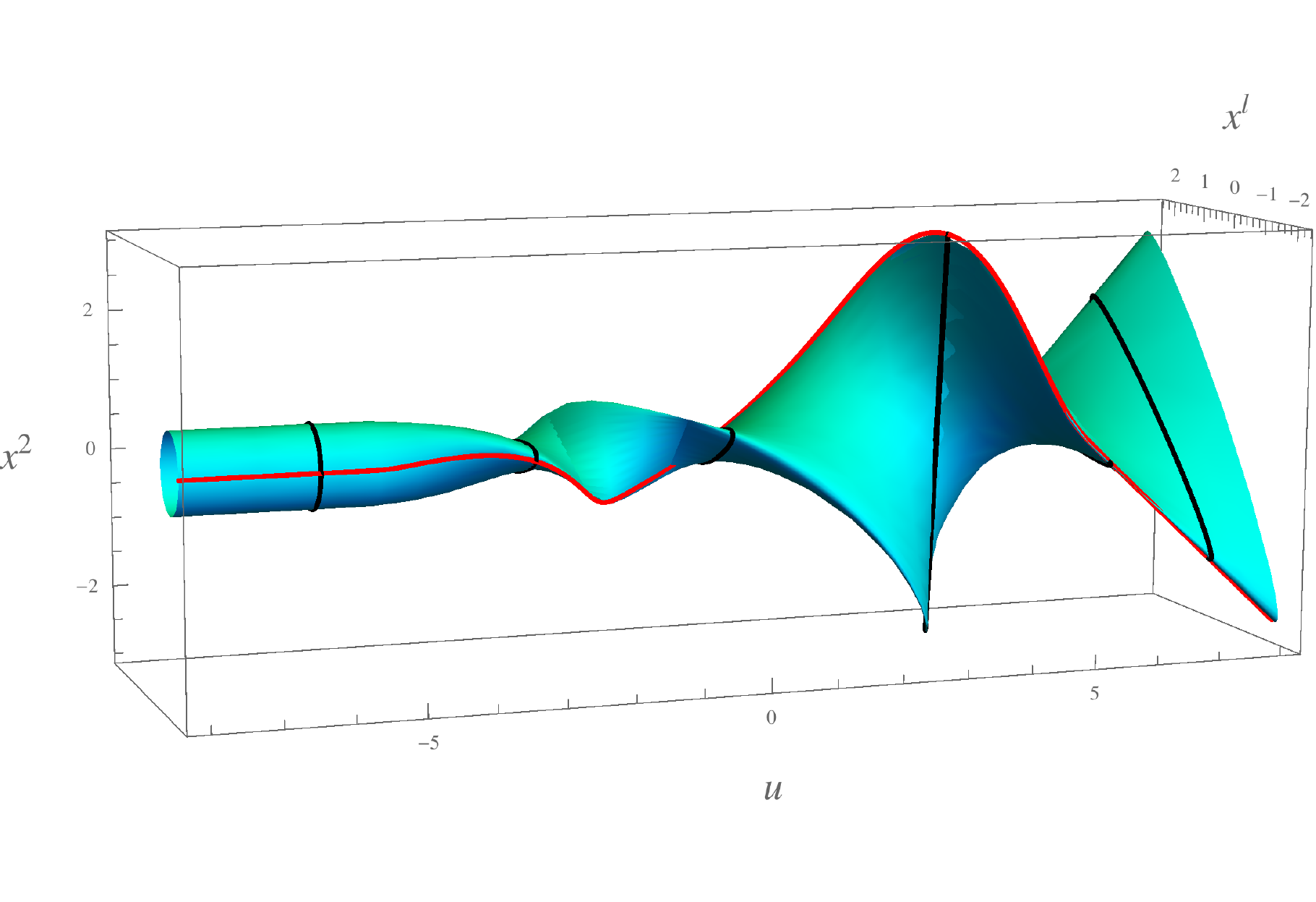} 
\caption{The evolution of the Tissot circle through the passage of the gyraton. The red curve shows a single geodesic
in the congruence as described in Fig.~\ref{fig4.3}. In this case, the Tissot circle evolves to an expanding 
ellipse in the far memory region, with orientation determined by the interplay of the two shear scalars 
$\hat{\s}_+$ and $\hat{\s}_{\times}$.}
\label{fig4.4}
\end{figure}

The evolution of the Tissot circle is shown in Fig.~\ref{fig4.4}.  Here, under the influence of non-vanishing
and $u$-dependent expansion $\hat{\theta}$ and both orientations $\hat{\s}_+$ and $\hat{\s}_{\times}$
of shear, the Tissot circle is deformed in a complicated way during the passage of the gyraton. 
Eventually, in the far memory region, the Tissot ring settles to become an expanding ellipse, whose
orientation is governed by diagonalising the shear matrix.  Despite superficial appearances, this
change in orientation is not due to any twist of the congruence, simply to the interplay
of the two directions of shear, confirming the general analysis in section \ref{sect 2.4}.

\section{Discussion}\label{sect 6}

In this paper, we have developed the geometric description of gravitational memory in a formalism which 
encompasses strong gravitational waves, and have applied our results to shockwave spacetimes.

A key observation is that memory is encoded in the Penrose limit of the original gravitational wave spacetime.
For null congruences, the Penrose limit is a plane wave so our analysis enhances the range of applications
of existing studies involving geodesic deviation and memory in plane wave spacetimes, which include
the weak-field approximations relevant for gravitational wave observations in astronomy.
For timelike congruences, we defined a new `timelike Penrose limit' spacetime, which is less well-studied.
However, we showed that if the original spacetime is in the wide class of pp waves, then the
transverse geodesic equations determining memory are the same as those for the plane waves in the
null Penrose limit.

The geometric formalism was applied to two examples of strong gravitational waves of particular interest --
gravitational shockwaves of the Aichelburg-Sexl type and their spinning generalisations, gyratons.
Analytic and numerical methods were used to illustrate the evolution of null and timelike geodesic congruences 
through their encounter with the gravitational wave burst, and the optical tensors -- expansion, shear and twist --
were used to characterise the eventual gravitational memory.

Gravitational wave astronomy has been revolutionised with the recent LIGO and Virgo observations of 
gravitational waves from black hole mergers \cite{Abbott:2016blz, Abbott:2016nmj}
and neutron star inspirals \cite{TheLIGOScientific:2017qsa}. As well as the observed oscillatory signal, 
these and other astrophysical sources may also produce a gravitational memory effect, potentially 
observable at LIGO/Virgo \cite{Lasky:2016knh, McNeill:2017uvq, Talbot:2018sgr}
and more certainly with satellite detectors such as eLISA \cite{AmaroSeoane:2012je}
(see also \cite{Loeb:2015ffa, Kolkowitz:2016wyg}).
Of course, these observed signals are weak-field gravitational waves, but it may be hoped that our analysis 
of gravitational shockwaves may also eventually find applications in astrophysics. As discussed earlier,
these shockwaves would be produced, for example, by fly-bys of extremely highly-boosted black holes.

One theoretical area of intense current interest is the relation of gravitational memory and soft graviton 
theorems, and more generally with the infra-red physics of quantum gravity 
(for a review, see \cite{Strominger:2017zoo}). 
Much of the research in this area has focused on the asymptotic symmetries of radially propagating
gravitational waves, described by the Bondi-Sachs spacetime. Here, we have established the geometric 
foundations to apply similar ideas to gravitational memory in shockwave spacetimes. In particular, the
Aichelburg-Sexl spacetime may be viewed as Minkowski spacetime cut along the $u=0$ plane and with the
past and future halves glued back with a coordinate displacement $\Delta v$. These two flat spacetime
regions are described in Rosen coordinates by different metrics, distinguished by the metric coefficient
${\bf C}_{ab} = (E^T E)_{ab}$ 
in which the zweibein $E^i{}_a(u)$ is at most linear in $u$. In the language of \cite{Strominger:2017zoo},
we may say that the shockwave localised at $u=0$ represents a domain wall separating diffeomorphic but
physically inequivalent copies of flat spacetime, {\it i.e.}~gravitational vaciua.
The shockwave scattering phase reflects this map between `gauge inequivalent' flat regions. 
The full web of connections between symmetries, vacua and gravitational memory on one hand
and scattering amplitudes and soft graviton theorems on the other is, however, left for future work.

Finally, we have shown in previous work 
\cite{Hollowood:2007ku, Hollowood:2008kq, Hollowood:2011yh, Hollowood:2015elj, Hollowood:2016ryc}
how quantum loop contributions to photon propagation,
and to Planck energy scattering, are governed by the same geometry of geodesic deviation that
determines gravitational memory. Here, we have extended the analysis of gravitational memory
in the Aichelburg-Sexl shockwaves relevant for ultra-high energy scattering to include 
spin effects in the form of gyratons. The nature of gravitational memory in the gyraton background
was clearly illustrated through the evolution of the Tissot circle in Fig.~\ref{fig4.4} and displays
both position and velocity-encoded memory. This establishes the essential geometric framework
for future investigations of gravitational spin effects in quantum field theory.

\vskip0.7cm
\noindent{\bf Acknowledgements}

\noindent I am grateful to Tim Hollowood for our earlier collaboration on quantum field theory and gravitational
shockwaves, from which the current work has developed.
This research was supported in part by STFC grant ST/P00055X/1.


\newpage

\appendix

\section{Planck energy scattering}\label{Appendix A}

One of the most interesting applications of the gravitational shockwave geometry is in ultra-high energy
scattering. At CM energies of order the Planck mass, particle scattering is dominated by the gravitational interactions. 
As shown in \cite{tHooft:1987vrq, Muzinich:1987in, Amati:1987wq} 
(see also \cite{Hollowood:2015elj, Hollowood:2016ryc} for QFT loop effects), 
in the eikonal limit where the interaction may be approximated by a sum
of ladder graviton-exchange diagrams, the phase shift determining the scattering amplitude may be calculated
from the shift $\Delta v$ in the lightcone coordinate for a null geodesic in the Aichelburg-Sexl shockwave 
background.\footnote{In Brinkmann coordinates, $\Delta v$ represents the shift by which the future and
past Minkowski spacetimes are displaced when they are glued back together along $u=0$ to form the 
global AS metric in the Penrose cut and paste construction. It is in this sense that the scattering phase reflects 
the map between these two inequivalent copies of flat spacetime.} 
One of our principal motivations in studying geodesics in the gyraton metric is to develop some
insight into how the gravitational effects of particle spin would influence Planck energy scattering amplitudes.

To see what is involved, recall the formula for the scattering amplitude ${\cal A}(s,t)$ in terms of the phase
$\Theta(s,b)$, which depends on the CM energy through $s=4E E'$, where $E,E'$ are the energies of the scattering 
particles and ${\bf b}$ is the (vector) impact parameter:
\begin{equation}
{\cal A}(s,t) = -2is \,\int d^2{\bf b}\, e^{i{\bf q}.{\bf b}}\, \left( e^{i\Theta(s,b)} - 1\right) \ .
\label{cc1}
\end{equation}
Here, $t=-q^2$, where ${\bf q}$ is the exchanged transverse momentum. 

In the shockwave picture, the phase is identified (with our metric conventions) 
as $\Theta(s,b) = -2E' \Delta v(E,b)$. 
Evaluating the integral over the angular dependence of ${\bf b}$ then gives,
\begin{equation}
{\cal A}(s,t) = -4\pi i s \, \int_0^\infty db\,b\,\left( e^{-2i E' \Delta v(E,b)} - 1 \right)\,J_0(qb) \ .
\label{cc2}
\end{equation}Given the shift $\Delta v$ as a function of the impact parameter, we can therefore determine the 
scattering amplitude by performing the Hankel transform in (\ref{cc2}).

For the Aichelburg-Sexl shockwave, the discontinuous shift in the Brinkmann coordinate $\Delta v$ is given by
\begin{equation}
\Delta v(E,b) = -\half f(b) =  2G E \log \frac{b^2}{r_0^2} \ ,
\label{cc3}
\end{equation}
implying (since $G= 1/M_p^2$),
\begin{equation}
\Theta(s,b) = - \frac{s}{M_p^2} \,\log\frac{b^2}{r_0^2} \ .
\label{cc4}
\end{equation}
We therefore have
\begin{equation}
{\cal A}(s,t) = -4\pi i s\, \int_0^\infty db\,b\,\left[ \left(\frac{b^2}{r_0^2}\right)^{-is/M_p^2} -~ 1~\right] \, J_0(qb) \ .
\label{cc5}
\end{equation}
The integral is standard,\footnote{The required Hankel tansform is
\begin{equation*}
\int_0^\infty dz\, z^p J_\n(a z) = 
2^p a^{-1-p} \, \Gamma\left(\frac{1+p+\n}{2} \right) / \, \Gamma\left(\frac{1-p+\n}{2}\right) \ .
\end{equation*} }
and setting $\L = 1/r_0$ as the momentum cut-off, we find
\begin{equation}
{\cal A}(s,t) = 8\pi i \,\frac{s}{t} \,\left(\frac{-t}{4\L^2}\right)^{is/M_p^2} \, 
\frac{\Gamma\left(1 - is/M_p^2\right)}{\Gamma\left(is/M_p^2\right)} \ .
\label{cc6}
\end{equation}
It follows directly that
\begin{equation}
\big|{\cal A}(s,t)\big|^2 = (8\pi)^2\, \frac{1}{M_p^4} \,\frac{s^4}{t^2} \ .
\label{cc7}
\end{equation}

It is remarkable that the complex pole structure, with poles at
$i s/M_p^2 = n$ with $n =1,2,\ldots$ implied by the gamma functions in the amplitude ${\cal A}(s,t)$,
as well as the extremely simple final result for $\big|{\cal A}(s,t)\big|^2$, is reproduced so elegantly by the
classical calculation of $\Delta v(E,b)$ in the Aichelburg-Sexl spacetime.

Now of course our ability to perform the Hankel transform to find ${\cal A}(s,t)$ in analytic form depends 
on knowing the functional dependence of $\Delta v(E,b)$ on the impact parameter $b$. 
For the impulsive shockwave profile, we have the simple solution (\ref{cc3}) for $\Delta v$, while in 
section \ref{sect 3.1} we have also found an analytic solution for the extended shockwave profile $\Theta(u,L)$.
In the case of the gyraton, however, the shift $\Delta v$ across the extended shockwave is determined by
solving (\ref{d7}) for $\dot{v}(u)$ after substituting the solution $r(u)$ for the precessing geodesic orbit.
Evidently, this is not so straightforward and a range of behaviours for $v(u)$ can arise as the impact 
parameter $b$ and metric parameters $E$ and $J$ are varied, as illustrated in the numerical plots in 
Fig.~\ref{fig4.v}. Naturally, we can still obtain numerical results for ${\cal A}(s,t)$, though it is not clear what
insight this would bring, in contrast to the analytic solution (\ref{cc6}) for the Aichelburg-Sexl shockwave. 
It is therefore not obvious at present how to make  progress in this direction, and we leave further investigation 
of scattering using the gyraton metric to future work.

As a first look at the effect of an extended profile on the scattering amplitude, however, we can calculate 
${\cal A}(s,t)$ for the Aichelburg-Sexl shockwave with profile $\chi_F(u) = \Theta(u,L)$. 
From the geodesic solution $v(u)$ in footnote \ref{footnote AS geodesics}, section \ref{sect 3.1}, 
we easily find the shift $\Delta v$ across the range $-L/2 < u < L/2$ where the test geodesic interacts
with the shockwave. This is shown in Fig.~\ref{fig3.4}. We find,
\begin{align}
&\Delta v = 4GE (\log b - 1) + 
b\,\sqrt{\frac{8GE}{L}}\,\, {\rm erf}^{-1}\left(\frac{1}{b} \sqrt{\frac{8GEL}{\pi}}\right)\,
\exp\left[- \left( {\rm erf}^{-1} \left( \frac{1}{b} \sqrt{\frac{8GEL}{\pi}}\right) \right)^2 \right]  \nonumber \\
&{}
\label{cc8}
\end{align}
giving the exact dependence on the impact parameter $b$.

While we do not have an analytic form for the Hankel transform of (\ref{cc8}), we can make progress 
by expanding in the parameter $L$ describing the duration of the extended shockwave interaction.
As this is equivalent to an expansion in large $b$, this will also give an approximation to the scattering
amplitude for small momentum exchange $t$. After some reparametrisation, we find
\begin{equation}
E'\,\Delta v ~=~ \frac{s}{M_p^2}\,\left[\, \log b ~-~ \frac{1}{3}\, \frac{s}{M_p^2}\, \frac{L}{E'}\, \frac{1}{b^2}
~-~ \frac{1}{15}\, \frac{s^2}{M_p^4}\, \frac{L^2}{E^{\prime 2}}\, \frac{1}{b^4} \, ~+~\,  O(L^3/b^6) \,\,\right] \ .
\label{cc9}
\end{equation}
Substituting into (\ref{cc2}) for ${\cal A}(s,t)$ and performing the Hankel transform, we find an expansion 
of the form,
\begin{align}
{\cal A}(s,t) ~&=~ 8\pi i \, \frac{s}{t} \left(\frac{-t}{4\Lambda^2}\right)^{is/M_p^2} ~ \nonumber \\
&~~~\times~ \left[\, \frac{\Gamma(1-is/M_p^2)}{\Gamma(is/M_p^2)} ~+~ 2i\,\frac{s}{M_p^2}\, 
\sum_{r=1}^\infty\, a_r\, \left(\frac{s\, t \,L}{M_p^2\, E'}\right)^r \, 
\frac{\Gamma(1-r - is/M_p^2)}{\Gamma(r + is/M_p^2)}\,\right] \ ,
\label{cc10}
\end{align}
where $a_r$ are numerical coefficients.

This has an interesting effect on the pole structure, arising from the new gamma functions in (\ref{cc10}).
As each new term in the series is included, an extra pole is added on the imaginary $s$-axis. That is, the
$r^{\rm th}$ term in the series has poles at $is/M_p^2 = -r + n$, with $n=1,2,\ldots$, with the exception that
there is never a pole at $s=0$, where the pre-factors impose a zero.
Eq.~(\ref{cc10}) also shows that, for fixed $s/M_p^2$, the expansion parameter is the Lorentz invariant
combination $\left(t\,L/E'\right)$. 
This makes clear how the corrections due to the extension $L$ of the profile
depend on the momentum transfer $t$ and test particle energy $E'$.

To complete the calculation keeping only the leading correction, we now find explicitly,
\begin{align}
{\cal A}(s,t) ~&=~8\pi i \, \frac{s}{t} \left(\frac{-t}{4\Lambda^2}\right)^{is/M_p^2} ~
\left[\, \frac{\Gamma(1-is/M_p^2)}{\Gamma(is/M_p^2)}  ~-~ \frac{i}{6} \,\frac{s^2}{M_p^4} \,
\left(\frac{t\,L}{E'}\right)\, \frac{\Gamma(-is/M_p^2)}{\Gamma(1 + is/M_p^2)} ~+~\ldots \,\right] \nonumber \\
&{} \nonumber \\
&=~ 8\pi i \, \frac{s}{t} \left(\frac{-t}{4\Lambda^2}\right)^{is/M_p^2} \, \frac{\Gamma(1-is/M_p^2)}{\Gamma(is/M_p^2)}  \,
\left[ 1 ~-~ \frac{i}{6} \,\left(\frac{t\,L}{E'}\right)\, ~+~\ldots \, \right] \ , \nonumber \\
&{} 
\label{cc11}
\end{align}
and so,
\begin{equation}
\big|{\cal A}(s,t)\big|^2 ~=~ (8\pi)^2\, \frac{1}{M_p^4} \,\frac{s^4}{t^2} \,
\left[ 1 ~+~ \frac{1}{36} \, \left(\frac{t\,L}{E'}\right)^2 \, ~+~\ldots \,\right] \ ,
\label{cc12}
\end{equation}
showing clearly the parametrisation of the correction due to the extended profile.

\section{Symmetries of gravitational shockwaves}\label{Appendix B}

A gravitational shockwave with an impulsive profile exhibits an enhanced symmetry compared to generic pp waves.
In this appendix, we describe these symmetries and discuss similar issues for the corresponding plane waves
arising as their Penrose limits.

We focus on the Aichelburg-Sexl shockwave with metric,
\begin{equation}
ds^2 = 2 du dv + f(r) \d(u) du^2 + dx^2 + dy^2 \ .
\label{bb1}
\end{equation}
Evidently, this has the symmetry
\begin{equation}
v\rta v+\a\  ~~~~~~~~   \Rightarrow ~~~~~~ K_Z = \partial_V \ ,
\label{bb2}
\end{equation}
with Killing vector $K_Z = \partial_V$ characteristic of pp waves. Cylindrical symmetry of $f(r)$ immediately implies 
the rotational symmetry,
\begin{equation}
x \rta x - \a y \ , ~~~~~~ y \rta y + \a x ~~~~~~~~ \Rightarrow ~~~~~~  K_J = x \partial_y - y \partial_x \ .
\label{bb3}
\end{equation}
However, for the impulsive profile proportional to $\d(u)$, there are two further $u$-dependent translation
symmetries \cite{Aichelburg:1995fi}. Inspection of (\ref{bb1}) shows these are,
\begin{align}
v \rta v - \a x \ , ~~~~~~ x \rta x + \a u ~~~~~~~~ &\Rightarrow ~~~~~~ K_{P_1} = u \partial_x - x \partial_v 
\nonumber \\
v \rta v - \a y \ , ~~~~~~ y \rta y + \a u ~~~~~~~~ &\Rightarrow ~~~~~~ K_{P_2} = u \partial_y - y \partial_v \ ,
\label{bb4}
\end{align}
where the $x,y$ translations, which must be linear in $u$,  must also be accompanied by a compensating 
transformation of $v$.

The corresponding generators satisfy the commutation relations,
\begin{align}
&\left[P_1, J\right] = P_2 \ , ~~~~~~~~ \left[P_2, J\right] = - P_1 \ , ~~~~~~~~ \left[P_1, P_2\right] = 0 \ , 
\nonumber \\
&\left[Z, P_1\right] = 0 \ , ~~~~~~~~~\, \left[Z, P_2\right] = 0 \ , ~~~~~~~~~~~~ \left[Z, J\right] = 0 \ .
\label{bb5}
\end{align}
This determines the 4-parameter isometry group as $ISO(2) \times \mathbb{R}$.
Recall that the Euclidean group $ISO(2)$ is the semi-direct product $ISO(2) = SO(2) \ltimes \mathbb{R}^2$.

Now consider the Penrose limit. This is the plane wave with metric,
\begin{equation}
ds^2 = 2du dv + h_{ij}(u) x^i x^j du^2 + (dx^i)^2 \ ,
\label{bb6}
\end{equation}
where for the particle shockwave,
\begin{equation}
h_{ij}(u) = h \begin{pmatrix} &\,1\,&\,\,0\, \\ &\,0\,&\,-1\,\end{pmatrix} \d(u)  \ ,
\label{bb7}
\end{equation}
defining $h = \half f''(b) = - \half f'(b)/b$ for ease of notation.

The symmetries of general plane waves have been widely studied (see especially \cite{Blau:2002js, 
Duval:2017els, Zhang:2017rno, Zhang:2017geq, Shore:2017dqx}
for some particularly relevant recent discussions) and we follow here the approach and notation
of \cite{Shore:2017dqx}. The generic isometry group\footnote{Plane wave metrics with specific forms for $h_{ij}(u)$ 
may possess a further symmetry. A notable case is the extra symmetry comprising $u$-translations
with a compensating rotation of the transverse coordinates which arises in one of the two classes of
homogeneous plane waves \cite{Blau:2002js, Shore:2017dqx}, including the Ozsv\'ath-Sch\"ucking plane wave 
\cite{Ozsvath} analysed in \cite{Shore:2017dqx}. The same symmetry also occurs in oscillatory polarised 
plane waves \cite{Zhang:2018srn, Ilderton:2018lsf}. \label{HPW}}
for a plane wave with arbitrary profile $h_{ij}(u)$ is the 
5-parameter Heisenberg group with generators $Q_r$, $P_r$ and $Z$ $(r,s=1,2)$ satisfying the
commutation relations,
\begin{align}
&\left[Q_r, Q_s\right] = 0 \ , ~~~~~~~~~~\left[P_r,P_s\right] = 0 \ , ~~~~~~~~~~ \left[Q_r,P_s\right] = - \d_{rs} Z \ ,
\nonumber \\
&\left[Z, Q_r\right] = 0 \ , ~~~~~~~~~~~\, \left[Z, P_r\right] = 0 \ .
\label{bb8}
\end{align}

The corresponding symmetry transformations and Killing vectors are known to be \cite{Blau:2002js, Shore:2017dqx},
\begin{align}
&x^i \rta x^i + \a^{(r)} f^i_{(r)} \ , ~~~~~~v \rta v - \a^{(r)} \dot{f}^i_{(r)} x^i  ~~~~~~\Rightarrow ~~~~
K_{Q_r} = - \dot{f}^i_{(r)} x_i \partial_v + f^i_{(r)} \partial_i  \nonumber \\
&x^i \rta x^i + \a^{(r)} g^i_{(r)} \ , ~~~~~~v \rta v - \a^{(r)} \dot{g}^i_{(r)} x^i  ~~~~~~\Rightarrow ~~~~
K_{P_r} = - \dot{g}^i_{(r)} x_i \partial_v + g^i_{(r)} \partial_i   \nonumber \\
&v \rta v+ \a ~~~~~~\Rightarrow ~~~~ K_Z = \partial_v \ ,
\label{bb9}
\end{align}
where $f^i_{(r)}$ and $g^i_{(r)}$ are independent solutions of the key oscillator equation, 
\begin{align}
&\ddot{f}^i_{(r)} - h^i{}_j(u) f^j_{(r)} = 0  \nonumber \\
&\ddot{g}^i_{(r)} - h^i{}_j(u) g^j_{(r)} = 0 \ ,
\label{bb10}
\end{align}
which are conveniently chosen to satisfy the canonical boundary conditions at some $u=u_0 < 0$,
\begin{align}
&f^i_{(r)} (u_0) = \d^i_r \ , ~~~~~~~~~~  \dot{f}^i_{(r)} (u_0) = 0 \ , \nonumber \\
&g^i_{(r)} (u_0) = 0 \ , ~~~~~~~~~~~  \dot{f}^i_{(r)} (u_0) = \d^i_r \ . 
\label{bb11}
\end{align}
The boundary conditions for $f^i_{(r)}$ correspond to those for a parallel congruence and we can therefore
identify the $f^i_{(r)}(u)$ with the zweibein $E^i{}_a(u)$ from (\ref{c35}), (\ref{c36}). The solutions $g^i_{(r)}$ 
are satisfied by `spray' boundary conditions, corresponding to geodesics emanating from a fixed point
at $u_0 < 0$.

We therefore already have the solutions $f^i_{(r)}(u)$, given by\footnote{For a general source for the shockwave, 
we simply replace the $\pm h$ factors in the Killing vectors shown here by $\half f''(b)$ and $\half f'(b)/b$
respectively, as in (\ref{c36}).}
\begin{equation}
f^i_{(1)} = \begin{pmatrix} &\,1 + h\, u \,\theta(u) \,\\ &\,0\,\end{pmatrix} \ , ~~~~~~~~~~
f^i_{(2)} = \begin{pmatrix} &0\, \\ &\,1 - h\, u \,\theta(u) \, \end{pmatrix} \ .
\label{bb12}
\end{equation}
To determine the solutions $g^i_{(r)}(u)$ systematically, we use the Wronskian condition,
\begin{equation}
\sum_i \left( f^i_{(r)} \,\dot{g}^i_{(s)} \,-\, \dot{f}^i_{(r)} \,g^i_{(s)} \right) = \d_{rs} \ .
\label{bb13}
\end{equation}
A short calculation now shows that the required solutions are
\begin{equation}
g^i_{(1)} = \begin{pmatrix} &\, u - u_0\left(1 + h\, u \,\theta(u)\right)  \,\\ &\,0\,\end{pmatrix} \ , ~~~~~~~~~~
g^i_{(2)} = \begin{pmatrix} &0\, \\ &\,u - u_0\left(1-h\,u\, \theta(u)\right) \, \end{pmatrix} \ .
\label{bb14}
\end{equation}

The explicit form for the Killing vectors is then,
\begin{align}
K_{Q_1} &= - h\, \theta(u) x^1 \partial_v \,+\, \left(1 + h\, u\, \theta(u) \right) \partial_{x^1}  \nonumber \\
K_{Q_2} &= h\, \theta(u) x^2 \partial_v \,+\, \left(1 - h\, u\, \theta(u) \right) \partial_{x^2} \ ,
\label{bb15}
\end{align}
and 
\begin{align}
K_{P_1} &= - \left( 1 - h\, u_0 \,\theta(u) \right) x^1 \partial_v \,
+\, \bigl(u - u_0\left(1 + h\, u\, \theta(u)\right) \bigr) \partial_{x^1}  \nonumber \\
K_{P_2} &= - \left( 1 + h\, u_0\, \theta(u) \right) x^2 \partial_v \,
+\, \bigl(u - u_0\left(1 - h\, u\, \theta(u)\right) \bigr) \partial_{x^2}  \ .
\label{bb16}
\end{align}
The commutation relations are readily checked, {\it e.g.}
\begin{align}
\left[K_{Q_1}, K_{P_1}\right] &= - \left(1 + h\, u\, \theta(u)\right) \left(1 - h\,u_0 \,\theta(u) \right) \,\partial_v
\,+\, \bigl(u - u_0\left(1 + h\, u \,\theta(u) \right) \bigr) h\, \theta(u) \,\partial_v \nonumber \\
&= - \partial_v \nonumber \\
&= - K_Z \ .
\label{bb17}
\end{align}

These expressions for the generators and Killing vectors have already made use of the fact that the 
metric coefficient $h_{ij}(u)$ is impulsive. Nevertheless, we can ask whether there are still more symmetries
for this special profile compared to the Heisenberg algebra for a generic plane wave. For example, the 
particular form of $h_{ij}(u)$ characterising a homogeneous plane wave is known to give rise to a further symmetry
related to $u$-transformations \cite{Blau:2002js, Shore:2017dqx} (see footnote \ref{HPW}).

The obvious approach is to look for analogues of the $u$-dependent translations of the transverse coordinates 
shown for the original Aichelburg-Sexl shockwave in (\ref{bb4}), that is
\begin{equation}
x^i \rta x^i + \a^{(r)} u\, \d^i_r \ , ~~~~~~~~~~  v \rta v - \a^{(r)} x^i\, \d_{ir} \ .
\label{bb18}
\end{equation}
This is indeed a symmetry of the metric (\ref{bb6}), (\ref{bb7}).  However, we see immediately from (\ref{bb16}) 
that these are simply the $u_0\rta 0$ limit of the general transformations defining the generators $P_r$.
No other extended symmetries are apparent. We therefore conclude that even with an impulsive profile,
the plane wave metric exhibits only the generic 5-parameter isometry group with Heisenberg algebra (\ref{bb8}).

\newpage

\section{Gyraton metrics}\label{Appendix C}

In this appendix, we review briefly more general gyraton metrics and discuss issues arising 
with the choice of profile functions and coordinate redefinitions.\footnote{A very clear presentation 
of these results for gyratons may be found in the paper \cite{Podolsky:2014lpa}.}

To motivate the choice of metric (\ref{d1}), we start with a more general gyraton metric, {\it viz.}~the
pp wave with metric
\begin{equation}
ds^2 = 2 du dv + F(u,r,\phi) \,du^2 - 2J(u,r,\phi)\,du\,d\phi + dr^2 + r^2\,d\phi^2 \ .
\label{aa1}
\end{equation}
The corresponding Ricci tensor components are (with subscript commas denoting partial derivatives),
\begin{align}
R_{uu} &= -\half \Delta F + \frac{1}{2r^2}\, \left(J_{,r}\right)^2 - \frac{1}{r^2} \, J_{,u\phi}  \nonumber \\
R_{ur} &= - \frac{1}{2r^2} \, J_{,r\phi}  \nonumber \\
R_{u\phi} &= \half\left(J_{,rr} - \frac{1}{r}\, J_{,r}\right) \ ,
\label{aa2}
\end{align}
and $R_{uu} = 0$. In the vacuum region outside a source localised at $r=0$, the metric coefficient $J(u,r,\phi)$
is therefore constrained by $J_{,r\phi} = 0$ and $J_{,rr} - \frac{1}{r}\, J_{,r}=0$, which implies
\begin{equation}
J(u,r,\phi) = \omega(u) r^2 + \tilde{J}(u,\phi) \ .
\label{aa3}
\end{equation}

Now consider the effect of coordinate redefinitions on the metric (\ref{aa1}).
First, 
\begin{equation}
\phi ~\rta~  \phi + \a(u) \ , 
\label{aa4}
\end{equation}
changes the metric coefficients by 
\begin{align}
F(u,r,\phi)~&\rta~ F(u,r,\phi) + r^2 \a'(u)^2 - 2 J(u,r,\phi) \a'(u)  \ ,\nonumber \\
J(u,r,\phi) ~&\rta~ J(u,r,\phi) - r^2 \a'(u) \ .
\label{aa5}
\end{align}
It follows that we can eliminate the $\omega(u)$ term in (\ref{aa3}) and with no loss of generality take 
$J(u,r,\phi) \rta \tilde{J}(u,\phi)$, {\it i.e.}~with no $r$-dependence in the coeffcient of $du d\phi$
in the metric. This considerably simplifies the curvatures in (\ref{aa2}), leaving only
\begin{equation}
R_{uu} = -\half \Delta F(u,r,\phi) - \frac{1}{r^2} \tilde{J}_{,u\phi}(u,\phi) 
\label{aa6}
\end{equation}
non-vanishing.

Next, consider the redefinition
\begin{equation}
v ~\rta~ v + \b(u,\phi) \ ,
\label{aa7}
\end{equation}
under which
\begin{align}
F(u,r,\phi) ~&\rta~ F(u,r,\phi) + 2 \b_{,u}(u,\phi)  \ ,\nonumber \\
\tilde{J}(u,\phi) ~&\rta~ \tilde{J}(u,\phi) - \b_{,\phi}(u,\phi) \ .
\label{aa8}
\end{align}
This means that the whole $du d\phi$ term in the metric can be removed by a coordinate redefinition {\it if and only if} 
$\tilde{J}(u,\phi)$ is expressible as a partial derivative $\partial \b(u,\phi)/\partial\phi$.
Locally, this is always true but not not necessarily globally. This is the case here since the vacuum region where (\ref{aa1})
applies (which excludes the source at $r=0$) is topologically non-trivial and admits non-contractible loops ${\cal C}$
encircling the axis $r=0$.\footnote{This is clearest \cite{Podolsky:2014lpa} if we consider the more general pp wave metric
\begin{equation*}
ds^2 = 2 du dv + F(u,x^i) du^2 - 2 H_i(u,x^i) du dx^i + \d_{ij} dx^i dx^j \ .
\end{equation*}
A coordinate redefinition $v \rta v + \b(u,x^i)$ then sends $H_i(u,x^i) \rta H_i(u,x^i) - \b_{,i}(u,x^i)$, so $H_i(u,x^i)$
can be eliminated if and only if it satisfies the integrability condition
\begin{equation*}
H_{i,j} - H_{j,i} = \b_{,ij} - \b_{,ji} = 0 \ .
\end{equation*}
In the language of differential forms, we may define ${\bf H} = H_i {\bf d}x^i$ so the integrability condition corresponds
to ${\bf d} {\bf H} = 0$. Now, with the simpler metric (\ref{aa1}) considered here, ${\bf H} = J {\bf d} \phi$ and
so ${\bf d} {\bf H} = J_{,r} {\bf d}r \wedge  {\bf d}\phi$.  Since we have established above that with no loss of generality 
we can take $J_{,r} = 0$, it follows that ${\bf d}{\bf H} = 0$, {\it i.e.}~that ${\bf H}$ is a closed form. The Poincar\'e lemma 
now implies it is {\it locally} exact, {\it i.e.} $H_i = \partial_i \b$ for some $\b(u,\phi)$ and can be removed locally by the 
coordinate redefinition (\ref{aa7}). However, the Poincar\'e lemma does {\it not} imply global exactness in the presence
of non-contractible loops as we have here in the topologically non-trivial vacuum region around the gyraton.\label{forms}}

Now consider the special cases where we can factorise the $u$-dependence of the metric coefficients in terms of 
the profile functions introduced in section \ref{sect 4}.  Without imposing cylindrical symmetry, the metric is then of the
form (\ref{aa1}) with
\begin{equation}
F(u,r,\phi) = f(r,\phi)\,\chi_F(u) \ , ~~~~~~~~~~~~ J(u,r,\phi) = J(\phi)\,\chi_J(u) \ .
\label{aa9}
\end{equation}
However, these profile functions are not independent, since the vacuum curvature equations (\ref{aa6}) now imply
\begin{equation}
\Delta f(r,\phi) \, \chi_F(u) = - \frac{2}{r^2} \frac{\partial J(\phi)}{\partial \phi}\, \chi_J'(u) \ .
\label{aa10}
\end{equation}
The profiles are then related by $\chi_F(u) \sim \chi_J'(u)$.

In fact, this is problematic for a physical interpretation. If we take $\chi_J(u) \sim \d(u)$ to be impulsive, this
requires $\chi_F(u)$ and the Ricci tensor $R_{uu}$ to be proportional to $\d'(u)$, which is too singular for a 
physical source.  On the other hand, if $\chi_J(u) \sim \Theta(u,L)$, then 
$\chi_F(u) \sim \d\left(u + \tfrac{L}{2}\right) - \d\left(u - \tfrac{L}{2} \right)$, which necessarily gives a negative contribution 
to $R_{uu}$ at some values of $u$ where it would violate the null energy condition $R_{uu} = 8\pi G T_{uu} > 0$.

This difficulty, which would require the metric (\ref{aa1}) to be embedded in a modified spacetime allowing a positive definite 
$R_{uu}$, is entirely avoided in the case of cylindrical symmetry. This seems in any case to be the most natural physical
situation. Then, in the metric (\ref{aa1}), we set
\begin{equation}
F(u,r,\phi) = f(r) \, \chi_F(u) \ , ~~~~~~~~~~~~  J(u,r,\phi) = J \, \chi_J(u) \ ,
\label{aa11}
\end{equation}
with $J$ constant and the profiles $\chi_F(u)$ and $\chi_J(u)$ uncorrelated. This is the metric (\ref{d1}) studied 
in detail in the main text.

\section{Gravitational spin memory and gyratons}\label{Appendix D}

In this paper, we have been concerned with the displacement memory effect, whether of position-encoded or
velocity-encoded type. A different type of gravitational memory was introduced in \cite{Pasterski:2015tva} in the context
of Bondi-Sachs gravitational waves --  {\it spin memory}. It was shown that whereas displacement memory is associated 
with BMS supertranslations, spin memory is related to superrotations, and an observational signature was proposed
relating spin memory to the angular momentum flux.

Here, we show how gravitational spin memory is realised in the case of the gyraton. Following \cite{Pasterski:2015tva},
we consider the non-geodesic scenario of two light beams constrained to follow circular paths, one rotating clockwise
and the other anticlockwise. A time difference between the two paths, which is manifested as an interference pattern, reveals
the presence of an angular momentum flux through the circles. This is the spin memory effect.

First, we review \cite{Podolsky:2014lpa} how the angular momentum of the gyraton source is related to the metric 
coefficient $J(u,r,\phi)$ in (\ref{aa1}). The longitudinal component of the angular momentum 
is given in terms of moments of the energy-momentum tensor as
\begin{equation}
{\bf J}^z = \int_{u_i}^{u_f} du\,{\cal J}(u)  \ ,
\label{dd1}
\end{equation}
with
\begin{equation}
{\cal J}(u) = \int d^2 x\,\left(x^1 T^{t2} - x^2 T^{t1} \right) \ ,
\label{dd2}
\end{equation}
where in our conventions $u=t-z,\, v=-\tfrac{1}{2}(t+z)$ and, evaluating on an equal-time hypersurface,
we have exchanged the integration over $z$ for an integration over $u$ in writing (\ref{dd1}).
Then, 
\begin{align}
{\cal J}(u) ~&=~-\int d^2 x \,\left(x^1 T^{v2} - x^2 T^{v1}\right) \nonumber \\
&=~ - \frac{1}{8\pi G}\,\int d^2 x\, r^2 R^{v\phi} \nonumber \\
&=~ - \frac{1}{8\pi G}\, \int d^2 x\, R_{u\phi}  \nonumber \\
&=~- \frac{1}{8\pi G}\, \iint  dr\,  d\phi\,r^2 \frac{\partial}{\partial r} \left(\frac{J_{,r}}{2r}\right) \ ,
\label{dd3}
\end{align}
from (\ref{aa2}). Now, integrating by parts, noting from Appendix \ref{Appendix C} that $J_{,r}=0$ in the vacuum region
and assuming $J(u,r,\phi)$ is non-singular at $r=0$, we find\footnote{In the differential form notation of 
footnote \ref{forms}, with ${\bf H} = J {\bf d}\phi$, we have ${\bf d H} = J_{,r} {\bf d}r \wedge {\bf d}\phi$ and (\ref{dd4})
is written as
\begin{align*}
{\cal J}(u) ~&=~\frac{1}{8\pi G} \, \iint J_{,r} {\bf d}r \wedge {\bf d}\phi  ~~~=~\frac{1}{8\pi G} \, \iint {\bf dH}  \nonumber \\
&=~\frac{1}{8\pi G} \,\oint \, {\bf H} ~~~=~\frac{1}{8\pi G} \, \oint\, J\,{\bf d}\phi \ ,
\end{align*}
by Stokes' theorem, reproducing (\ref{dd5}).  }
\begin{equation}
{\cal J}(u) ~=~ \frac{1}{8\pi G}\, \iint dr\,d\phi\, J_{,r} \ .
\label{dd4}
\end{equation}
Evaluating, with the notation (\ref{d1}) for the gyraton metric in the vacuum region, then gives,
\begin{equation}
{\cal J}(u) ~=~ \frac{1}{8\pi G} \,\chi_J(u) \, \oint d\phi \,J \ ,
\label{dd5}
\end{equation}
and so
\begin{equation}
{\bf J}^z ~=~ \frac{1}{8\pi G} \, \oint d\phi \,J ~=~ \frac{1}{4G} \,J  \ .
\label{dd6}
\end{equation}

\begin{figure}[h!]
\centering
\includegraphics[scale=0.6]{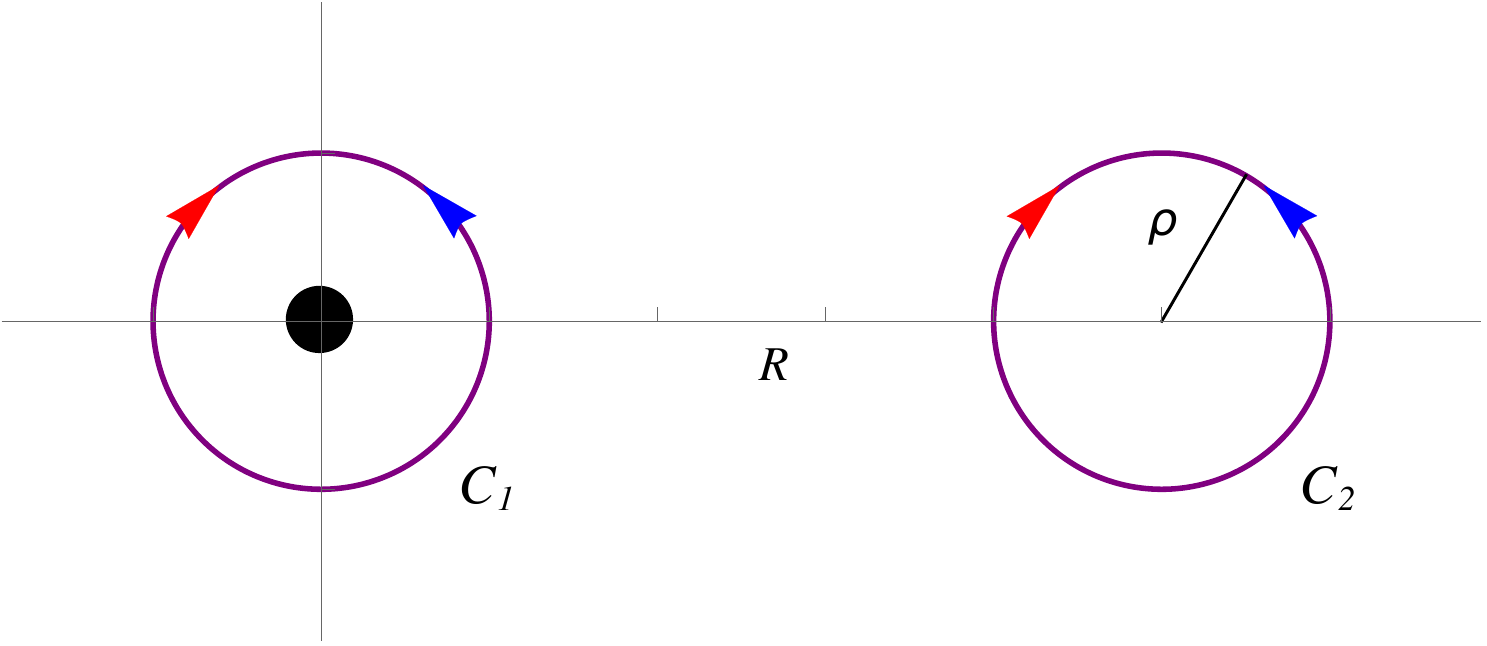} 
\caption{This illustrates the case of a light path ${\cal C}_1$ encircling the gyraton core, 
and one lying entirely outside at a radius $R$, with contour ${\cal C}_2$.  
Only the path encircling the gyraton exhibits non-vanishing spin memory.}
\label{figD.1}
\end{figure}

Now consider a photon constrained to follow a circle of radius $\r$ and polar angle $\theta$ in the 
plane $z=$ constant centred at $r=R$  (see Fig.~\ref{figD.1}). 
From the metric (\ref{d1}), this null path satisfies
\begin{equation}
0 ~=~ \big(-1 + f(r) \chi_F(u)\big) du^2 + \r^2 d\theta^2 
- 2 J \chi_J(u) \frac{1}{r^2} \left(\r^2 + R\r \cos\theta\right) \, du\,d\theta  \ ,
\label{dd7}
\end{equation}
where here $r^2 \equiv R^2 + 2R\r \cos\theta + \r^2$.

Following \cite{Pasterski:2015tva}, we assume that the metric coefficients are slowly varying over the 
timescale for an orbit of the circle. This is automatically satisfied in the case studied in section \ref{sect 4}
where we take $\chi_F(u) = \chi_J(u) = \Theta(u,L)$. Integrating over a clockwise orbit, 
and setting $L=1$ for clarity, we find
\begin{multline}
\int_0^{u_+} du ~=~ \int_0^{2\pi} d\theta\, \,\frac{1}{1- f(r)}  \\
\times  ~\left( - \frac{J}{r^2}(\r^2 + R\r\cos\theta)  
~+~ 
\sqrt{\r^2\left(1 -f(r) \right) +  \frac{J^2}{r^4} (\r^2 + R\r\cos\theta)^2 } ~\right) \ .
\label{dd8}
\end{multline}
For the anticlockwise path, integrating from $0$ to $-2\pi$, we must choose the opposite sign for the
square root, which recovers $u_- = 2\pi \r$ in flat spacetime. This gives the time difference (equal to the
difference in $u$) for the two paths as
\begin{equation}
\Delta u ~\equiv~ u_+ - u_- ~=~
-2 \,\int_0^{2\pi} d\theta\, \frac{1}{1 - f(r)} \,\, \frac{J}{r^2}\, (\r^2 + R \r \cos\theta)  , 
\label{dd9}
\end{equation}
with $r^2$ as above. 

\begin{figure}[h!]
\centering
\includegraphics[scale=0.6]{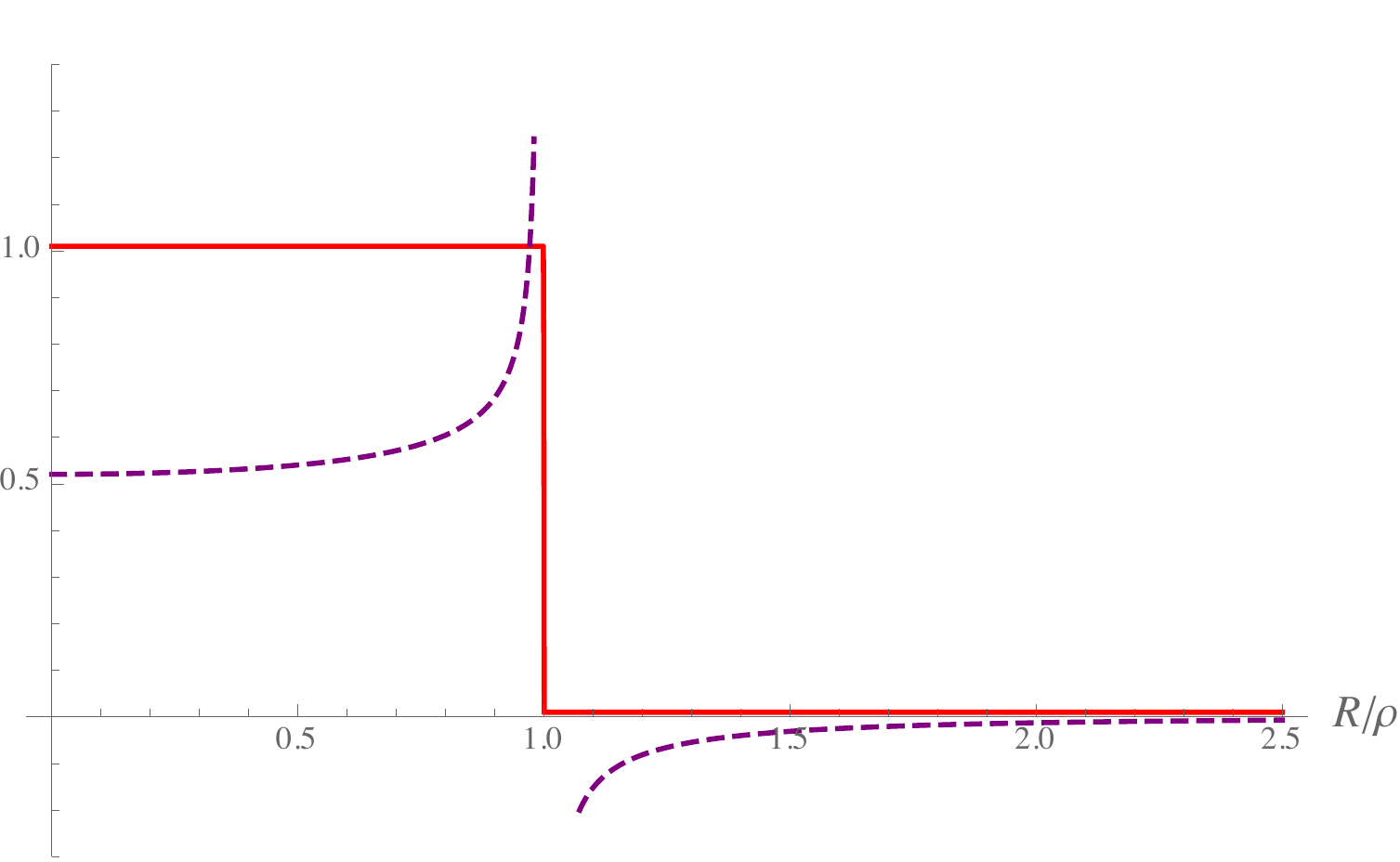}
\caption{
The right-hand figure shows the behaviour of the integral
(\ref{dd9}) for $\Delta u$ in the case when $f(r)$ is neglected (red curve) and included (purple dashed curve).
The parameters for $f(r)/L$ are chosen for illustration only.
When $R <\r$, the integration contour encircles the gyraton core and the corresponding time difference
$\Delta u$ is a measure of spin memory.}
\label{figD.2}
\end{figure}

To understand this integral, it is simplest first to neglect the $f(r)$ contribution.
Then we have an exact solution in terms of the step function,
\begin{align}
\Delta u \big|_{f=0} ~&=~ - 2 J \, \int_0^{2\pi} d\theta\, \frac{1 + \frac{R}{\r} \cos\theta}
{1 + 2\frac{R}{\r} \cos\theta + \bigl(\frac{R}{\r}\bigr)^2 } \nonumber \\
&=~ - 4\pi \,J\, \theta\left(1 - R/\r\right) \ .
\label{dd10}
\end{align}
Including $f(r) = -4 G E \log r^2/r_0^2$, we have the numerical solution also
shown in Fig.~\ref{figD.2}. For $R\gg\r$ the integral again vanishes, while 
$\Delta u \rightarrow -4\pi J /\left(1 - f(\r)\right)$ for $R\ll \r$.
The divergence occurs due to the behaviour of the integrand in the region $\theta \simeq \pi$
for $R\simeq \r$, where the circular path intersects the gyraton core and the vacuum metric 
is not valid.

This illustrates clearly the phenomenon of spin memory for the gyraton. 
Neglecting $f(r)$, the two light paths exhibit a time difference $\Delta u$, measurable as an interference 
fringe, in the case where the paths enclose the gyraton core, while there is no time difference for
paths outside the core. Including $f(r)$, the sharp distinction is smoothed as shown in 
Fig.~\ref{figD.2}.
In either case, the limiting value for the encircling paths,
\begin{equation}
\Delta u ~\simeq~  \frac{-2}{1 - f(\r)} \,\, \oint_{{\cal C}_1} d\theta \,J  \ ,
\label{dd11}
\end{equation}
provides a measure of the angular momentum flux through the loop.\footnote{The equivalent result 
for the Bondi-Sachs gravitational wave can be expressed in terms of the coefficients 
$D^z C_{zz}$ and $D^{\bar{z}}C_{\bar{z} \bar{z}}$ of the
$du dz$ and $du d\bar{z}$ terms in the metric as
\begin{equation*}
\Delta u = \oint \, \left(D^z C_{zz} \,dz \,+\, D^{\bar{z}} C_{\bar{z}\bar{z}} \,d\bar{z}\right) \ ,
\label{dd12}
\end{equation*}
and the relation of the r.h.s~to the angular momentum flux is quoted in eq.(5.9) of \cite{Pasterski:2015tva}.
Just as in footnote \ref{Bondi-Sachs} for the displacement memory, there is a clear correspondence
between the realisations of gravitational spin memory in the Bondi-Sachs and gyraton metrics. }


\newpage

\begin{thebibliography}{99}

\enlargethispage{\baselineskip}

\bibitem{Zeldovich}
  Ya.~B.~Zel'dovich and A.~G.~Polnarev,
  {\it ``Radiation of gravitational waves by a cluster of super-dense stars,''}
  Sov.~Astron., {\bf 18} (1974) 17 ~
   [Astron.~Zh. {\bf 51} (1974) 30].

\bibitem{Braginsky:1986ia}
  V.~B.~Braginsky and L.~P.~Grishchuk,
  {\it ``Kinematic Resonance and Memory Effect in Free Mass Gravitational Antennas,''}
  Sov.\ Phys.\ JETP {\bf 62} (1985) 427 ~
   [Zh.\ Eksp.\ Teor.\ Fiz.\  {\bf 89} (1985) 744].

\bibitem{Bondi:1957dt}
  H.~Bondi,
  {\it ``Plane gravitational waves in general relativity,''}
  Nature {\bf 179} (1957) 1072.

\bibitem{Grishchuk:1989qa}
  L.~P.~Grishchuk and A.~G.~Polnarev,
  {\it ``Gravitational wave pulses with 'velocity coded memory',''}
  Sov.\ Phys.\ JETP {\bf 69} (1989) 653 ~
   [Zh.\ Eksp.\ Teor.\ Fiz.\  {\bf 96} (1989) 1153].

\bibitem{Penrose:1965rx}
  R.~Penrose,
  {\it ``A remarkable property of plane waves in general relativity,''}
  Rev.\ Mod.\ Phys.\  {\bf 37} (1965) 215.

\bibitem{Penrose}
  R.~Penrose,
  {\it ``Any Space-Time has a Plane Wave as a Limit,''}
  in `Differential Geometry and Relativity: A Volume in Honour of
  Andr\'e Lichnerowicz on his 60th Birthday', 271, Springer Netherlands, Dordrecht (1976).

\bibitem{Blau:2006ar}
  M.~Blau, D.~Frank and S.~Weiss,
  {\it ``Fermi coordinates and Penrose limits,''}
  Class.\ Quant.\ Grav.\  {\bf 23} (2006) 3993,
  [hep-th/0603109]

\bibitem{Hollowood:2009qz}
  T.~J.~Hollowood, G.~M.~Shore and R.~J.~Stanley,
  {\it ``The Refractive Index of Curved Spacetime II: QED, Penrose Limits and Black Holes,''}
  JHEP {\bf 0908} (2009) 089,
  [arXiv:0905.0771 [hep-th]].

\bibitem{Stephani:2003tm}
  H.~Stephani, D.~Kramer, M.~A.~H.~MacCallum, C.~Hoenselaers and E.~Herlt,
  {\it ``Exact solutions of Einstein's field equations,''}
  Cambridge University Press, 2nd edition,  2003.

\bibitem{Griffiths:2009dfa}
  J.~B.~Griffiths and J.~Podolsky,
  {\it ``Exact Space-Times in Einstein's General Relativity,''},
  Cambridge University Press, 1st edition, 2009.

\bibitem{Gibbons:1972fy}
  G.~W.~Gibbons and S.~W.~Hawking,
  {\it ``Theory of the detection of short bursts of gravitational radiation,''}
  Phys.\ Rev.\ D {\bf 4} (1971) 2191.

\bibitem{Favata:2010zu}
  M.~Favata,
  {\it ``The gravitational-wave memory effect,''}
  Class.\ Quant.\ Grav.\  {\bf 27} (2010) 084036,
  [arXiv:1003.3486 [gr-qc]].

\bibitem{Lasky:2016knh}
  P.~D.~Lasky, E.~Thrane, Y.~Levin, J.~Blackman and Y.~Chen,
  {\it ``Detecting gravitational-wave memory with LIGO: implications of GW150914,''}
  Phys.\ Rev.\ Lett.\  {\bf 117} (2016) no.6,  061102,
  [arXiv:1605.01415 [astro-ph.HE]].

\bibitem{McNeill:2017uvq}
  L.~O.~McNeill, E.~Thrane and P.~D.~Lasky,
  {\it ``Detecting Gravitational Wave Memory without Parent Signals,''}
  Phys.\ Rev.\ Lett.\  {\bf 118} (2017) no.18,  181103,
  [arXiv:1702.01759 [astro-ph.IM]].

\bibitem{Talbot:2018sgr}
  C.~Talbot, E.~Thrane, P.~D.~Lasky and F.~Lin,
  {\it ``Gravitational-wave memory: waveforms and phenomenology,''}
  Phys.\ Rev.\ D {\bf 98} (2018) no.6,  064031,
  [arXiv:1807.00990 [astro-ph.HE]].

\enlargethispage{\baselineskip}

\bibitem{Harte:2012jg}
  A.~I.~Harte,
  {\it ``Strong lensing, plane gravitational waves and transient flashes,''}
  Class.\ Quant.\ Grav.\  {\bf 30} (2013) 075011,
  [arXiv:1210.1449 [gr-qc]].

\bibitem{Harte:2015ila}
  A.~I.~Harte,
  {\it ``Optics in a nonlinear gravitational plane wave,''}
  Class.\ Quant.\ Grav.\  {\bf 32} (2015) no.17,  175017,
  [arXiv:1502.03658 [gr-qc]].

\bibitem{Hollowood:2015elj}
  T.~J.~Hollowood and G.~M.~Shore,
  {\it ``Causality Violation, Gravitational Shockwaves and UV Completion,''}
  JHEP {\bf 1603} (2016) 129
  [arXiv:1512.04952 [hep-th]].

\bibitem{Duval:2017els}
  C.~Duval, G.~W.~Gibbons, P.~A.~Horvathy and P.-M.~Zhang,
  {\it ``Carroll symmetry of plane gravitational waves,''}
  Class.\ Quant.\ Grav.\  {\bf 34} (2017) no.17,  175003,
  [arXiv:1702.08284 [gr-qc]].

\bibitem{Zhang:2017rno}
  P.-M.~Zhang, C.~Duval, G.~W.~Gibbons and P.~A.~Horvathy,
  {\it ``The Memory Effect for Plane Gravitational Waves,''}
  Phys.\ Lett.\ B {\bf 772} (2017) 743,
  [arXiv:1704.05997 [gr-qc]].

\bibitem{Zhang:2017geq}
  P.-M.~Zhang, C.~Duval, G.~W.~Gibbons and P.~A.~Horvathy,
  {\it ``Soft gravitons and the memory effect for plane gravitational waves,''}
  Phys.\ Rev.\ D {\bf 96} (2017) no.6,  064013,
  [arXiv:1705.01378 [gr-qc]].

\bibitem{Shore:2017dqx}
  G.~M.~Shore,
  {\it ``A New Twist on the Geometry of Gravitational Plane Waves,''}
  JHEP {\bf 1709} (2017) 039,
  [arXiv:1705.09533 [gr-qc]].

\bibitem{Zhang:2017jma}
  P.-M.~Zhang, C.~Duval and P.~A.~Horvathy,
  {\it ``Memory Effect for Impulsive Gravitational Waves,''}
  Class.\ Quant.\ Grav.\  {\bf 35} (2018) no.6,  065011,
  [arXiv:1709.02299 [gr-qc]].

\bibitem{Zhang:2018srn}
  P.~M.~Zhang, C.~Duval, G.~W.~Gibbons and P.~A.~Horvathy,
  {\it ``Velocity Memory Effect for Polarized Gravitational Waves,''}
  JCAP {\bf 1805} (2018) no.05,  030,
  [arXiv:1802.09061 [gr-qc]].

\bibitem{Zhang:2018gzn}
  P.~M.~Zhang, M.~Elbistan, G.~W.~Gibbons and P.~A.~Horvathy,
  {\it ``Sturm–Liouville and Carroll: at the heart of the memory effect,''}
  Gen.\ Rel.\ Grav.\  {\bf 50} (2018) no.9,  107,
  [arXiv:1803.09640 [gr-qc]].

\bibitem{Zhang:2018upz}
  P.-M.~Zhang, M.~Cariglia, C.~Duval, M.~Elbistan, G.~W.~Gibbons and P.~A.~Horvathy,
  {\it ``Ion Traps and the Memory Effect for Periodic Gravitational Waves,''}
  Phys.\ Rev.\ D {\bf 98} (2018) no.4,  044037,
  [arXiv:1807.00765 [gr-qc]].

\bibitem{Aichelburg:1970dh}
  P.~C.~Aichelburg and R.~U.~Sexl,
  {\it ``On the Gravitational field of a massless particle,''}
  Gen.\ Rel.\ Grav.\  {\bf 2} (1971) 303.

\bibitem{Dray:1984ha}
  T.~Dray and G.~'t Hooft,
  {\it ``The Gravitational Shock Wave of a Massless Particle,''}
  Nucl.\ Phys.\ B {\bf 253} (1985) 173.

\bibitem{tHooft:1987vrq}
  G.~'t Hooft,
  {\it ``Graviton Dominance in Ultrahigh-Energy Scattering,''}
  Phys.\ Lett.\ B {\bf 198} (1987) 61.

\bibitem{Muzinich:1987in}
  I.~J.~Muzinich and M.~Soldate,
  {\it ``High-Energy Unitarity of Gravitation and Strings,''}
  Phys.\ Rev.\ D {\bf 37} (1988) 359.

\bibitem{Amati:1987wq}
  D.~Amati, M.~Ciafaloni and G.~Veneziano,
  {\it ``Superstring Collisions at Planckian Energies,''}
  Phys.\ Lett.\ B {\bf 197} (1987) 81.

\bibitem{Hollowood:2007ku}
  T.~J.~Hollowood and G.~M.~Shore,
  {\it ``The Refractive index of curved spacetime: The Fate of causality in QED,''}
  Nucl.\ Phys.\ B {\bf 795} (2008) 138,
  [arXiv:0707.2303 [hep-th]].

\bibitem{Hollowood:2008kq}
  T.~J.~Hollowood and G.~M.~Shore,
  {\it ``The Causal Structure of QED in Curved Spacetime: Analyticity and the Refractive Index,''}
  JHEP {\bf 0812} (2008) 091,
  [arXiv:0806.1019 [hep-th]].

\bibitem{Hollowood:2011yh}
  T.~J.~Hollowood and G.~M.~Shore,
  {\it ``The Effect of Gravitational Tidal Forces on Renormalized Quantum Fields,''}
  JHEP {\bf 1202} (2012) 120,
  [arXiv:1111.3174 [hep-th]].

\bibitem{Hollowood:2016ryc}
  T.~J.~Hollowood and G.~M.~Shore,
  {\it ``Causality, Renormalizability and Ultra-High Energy Gravitational Scattering,''}
  J.\ Phys.\ A {\bf 49} (2016) no.21,  215401,
  [arXiv:1601.06989 [hep-th]].

\bibitem{Lousto:1988ej}
  C.~O.~Lousto and N.~G.~Sanchez,
  {\it ``Gravitational Shock Waves of Ultrahigh Energetic Particles on Curved Space-times,''}
  Phys.\ Lett.\ B {\bf 220} (1989) 55.

\bibitem{Ferrari:1990}
 V.~Ferrari and P.~Pendenza, 
 {\it ``Boosting the Kerr metric,''}
 Gen.~Relat.~Gravit. (1990) 22: 1105. 

\bibitem{Balasin:1994tb}
  H.~Balasin and H.~Nachbagauer,
  {\it ``The Ultrarelativistic Kerr geometry and its energy momentum tensor,''}
  Class.\ Quant.\ Grav.\  {\bf 12} (1995) 707,
  [gr-qc/9405053].

\bibitem{Balasin:1995tj}
  H.~Balasin and H.~Nachbagauer,
  {\it ``Boosting the Kerr geometry into an arbitrary direction,''}
  Class.\ Quant.\ Grav.\  {\bf 13} (1996) 731,
  [gr-qc/9508044].

\bibitem{Hayashi:1994rf}
  K.~Hayashi and T.~Samura,
  {\it ``Gravitational shock waves for Schwarzschild and Kerr black holes,''}
  Phys.\ Rev.\ D {\bf 50} (1994) 3666,
  [gr-qc/9404027].

\bibitem{Lousto:1989ha}
  C.~O.~Lousto and N.~G.~Sanchez,
  {\it ``The Ultrarelativistic Limit of the {Kerr-Newman} Geometry and Particle Scattering at the Planck Scale,''}
  Phys.\ Lett.\ B {\bf 232} (1989) 462.

\bibitem{Lousto:1992th}
  C.~O.~Lousto and N.~G.~Sanchez,
  {\it ``The Ultrarelativistic limit of the boosted Kerr-Newman geometry and the scattering of spin 1/2 particles,''}
  Nucl.\ Phys.\ B {\bf 383} (1992) 377.

\bibitem{Yoshino:2004ft}
  H.~Yoshino,
  {\it ``Lightlike limit of the boosted Kerr black holes in higher-dimensional spacetimes,''}
  Phys.\ Rev.\ D {\bf 71} (2005) 044032,
  [gr-qc/0412071].

\bibitem{Cai:1998ii}
  R.~G.~Cai, J.~Y.~Ji and K.~S.~Soh,
  {\it ``Ultrarelativistic limits of boosted dilaton black holes,''}
  Nucl.\ Phys.\ B {\bf 528} (1998) 265,
  [gr-qc/9801097].

\bibitem{Eardley:2002re}
  D.~M.~Eardley and S.~B.~Giddings,
  {\it ``Classical black hole production in high-energy collisions,''}
  Phys.\ Rev.\ D {\bf 66} (2002) 044011,
  [gr-qc/0201034].

\bibitem{Yoshino:2005hi}
  H.~Yoshino and V.~S.~Rychkov,
  {\it ``Improved analysis of black hole formation in high-energy particle collisions,''}
  Phys.\ Rev.\ D {\bf 71} (2005) 104028
   Erratum: [Phys.\ Rev.\ D {\bf 77} (2008) 089905],
  [hep-th/0503171].

\enlargethispage{\baselineskip}

\textheight=612pt

\bibitem{Yoshino:2006dp}
  H.~Yoshino and R.~B.~Mann,
  {\it ``Black hole formation in the head-on collision of ultrarelativistic charges,''}
  Phys.\ Rev.\ D {\bf 74} (2006) 044003,
  [gr-qc/0605131].

\bibitem{Yoshino:2007ph}
  H.~Yoshino, A.~Zelnikov and V.~P.~Frolov,
  {\it ``Apparent horizon formation in the head-on collision of gyratons,''}
  Phys.\ Rev.\ D {\bf 75} (2007) 124005,
  [gr-qc/0703127].

\bibitem{Frolov:2005zq}
  V.~P.~Frolov, W.~Israel and A.~Zelnikov,
  {\it ``Gravitational field of relativistic gyratons,''}
  Phys.\ Rev.\ D {\bf 72} (2005) 084031,
  [hep-th/0506001].

\bibitem{Frolov:2005in}
  V.~P.~Frolov and D.~V.~Fursaev,
  {\it ``Gravitational field of a spinning radiation beam-pulse in higher dimensions,''}
  Phys.\ Rev.\ D {\bf 71} (2005) 104034,
  [hep-th/0504027].

\bibitem{Podolsky:2014lpa}
  J.~Podolsky, R.~Steinbauer and R.~Svarc,
  {\it ``Gyratonic pp-waves and their impulsive limit,''}
  Phys.\ Rev.\ D {\bf 90} (2014) no.4,  044050,
  [arXiv:1406.3227 [gr-qc]].

\bibitem{Pasterski:2015tva}
  S.~Pasterski, A.~Strominger and A.~Zhiboedov,
  {\it ``New Gravitational Memories,''}
  JHEP {\bf 1612} (2016) 053,
  [arXiv:1502.06120 [hep-th]].

\bibitem{Poisson}
 E.~Poisson, {\it ``A Relativist's Toolkit -- The Mathematics of Black-Hole Mechanics,''} 
 Cambridge University Press (2004). 

\bibitem{Blau:2002js}
  M.~Blau and M.~O'Loughlin,
  {\it ``Homogeneous plane waves,''}
  Nucl.\ Phys.\ B {\bf 654} (2003) 135,
  [hep-th/0212135].

\bibitem{Strominger:2017zoo}
  A.~Strominger,
  {\it ``Lectures on the Infrared Structure of Gravity and Gauge Theory,''}
  Princeton University Press, 2018; 
  [arXiv:1703.05448 [hep-th]].

\bibitem{Ferrari:1988cc}
  V.~Ferrari, P.~Pendenza and G.~Veneziano,
  {\it ``Beamlike Gravitational Waves and Their Geodesics,''}
  Gen.\ Rel.\ Grav.\  {\bf 20} (1988) 1185.

\bibitem{Shore:2002in}
  G.~M.~Shore,
  {\it ``Constructing time machines,''}
  Int.\ J.\ Mod.\ Phys.\ A {\bf 18} (2003) 4169,
  [gr-qc/0210048].

\bibitem{Camanho:2014apa}
  X.~O.~Camanho, J.~D.~Edelstein, J.~Maldacena and A.~Zhiboedov,
  {\it ``Causality Constraints on Corrections to the Graviton Three-Point Coupling,''}
  JHEP {\bf 1602} (2016) 020
  [arXiv:1407.5597 [hep-th]].

\bibitem{Papallo:2015rna}
  G.~Papallo and H.~S.~Reall,
  {\it ``Graviton time delay and a speed limit for small black holes in Einstein-Gauss-Bonnet theory,''}
  JHEP {\bf 1511} (2015) 109,
  [arXiv:1508.05303 [gr-qc]].

\bibitem{Abbott:2016blz}
  B.~P.~Abbott {\it et al.} [LIGO Scientific and Virgo Collaborations],
  {\it ``Observation of Gravitational Waves from a Binary Black Hole Merger,''}
  Phys.\ Rev.\ Lett.\  {\bf 116} (2016)  061102,
  [arXiv:1602.03837 [gr-qc]].

\bibitem{Abbott:2016nmj}
  B.~P.~Abbott {\it et al.} [LIGO Scientific and Virgo Collaborations],
  {\it ``GW151226: Observation of Gravitational Waves from a 22-Solar-Mass 
  Binary Black Hole Coalescence,''}
  Phys.\ Rev.\ Lett.\  {\bf 116} (2016)  241103,
  [arXiv:1606.04855 [gr-qc]].

\bibitem{TheLIGOScientific:2017qsa}
  B.~P.~Abbott {\it et al.} [LIGO Scientific and Virgo Collaborations],
  {\it ``GW170817: Observation of Gravitational Waves from a Binary Neutron Star Inspiral,''}
  Phys.\ Rev.\ Lett.\  {\bf 119} (2017)  161101,
  [arXiv:1710.05832 [gr-qc]].

\bibitem{AmaroSeoane:2012je}
  P.~Amaro-Seoane {\it et al.},
  {\it ``Low-frequency gravitational-wave science with eLISA/NGO,''}
  Class.\ Quant.\ Grav.\  {\bf 29} (2012) 124016,
  [arXiv:1202.0839 [gr-qc]].

\enlargethispage{\baselineskip}

\bibitem{Loeb:2015ffa}
  A.~Loeb and D.~Maoz,
  {\it ``Using Atomic Clocks to Detect Gravitational Waves,''}
  arXiv:1501.00996 [astro-ph.IM].

\bibitem{Kolkowitz:2016wyg}
  S.~Kolkowitz, I.~Pikovski, N.~Langellier, M.~D.~Lukin, R.~L.~Walsworth and J.~Ye,
  {\it ``Gravitational wave detection with optical lattice atomic clocks,''}
  Phys.\ Rev.\ D {\bf 94} (2016) no.12,  124043,
  [arXiv:1606.01859 [physics.atom-ph]].

\bibitem{Aichelburg:1995fi}
  P.~C.~Aichelburg and H.~Balasin,
  {\it ``Symmetries of p p waves with distributional profile,''}
  Class.\ Quant.\ Grav.\  {\bf 13} (1996) 723,
  [gr-qc/9509025].

\bibitem{Ozsvath}
 I.~Ozsv{\'a}th and E.~Sch{\"u}cking, 
 {\it ``An anti-Mach metric'',}
 in `Recent Developments in General Relativity', 339, Pergamon Press, Oxford(1962).

\bibitem{Ilderton:2018lsf}
  A.~Ilderton,
  {\it ``Screw-symmetric gravitational waves: a double copy of the vortex,''}
  Phys.\ Lett.\ B {\bf 782} (2018) 22,
  [arXiv:1804.07290 [gr-qc]].


\end{thebibliography}
\end{document}